%% file: main.tex
\documentclass[fleqn,usenatbib]{mnras}

\usepackage[T1]{fontenc}
\usepackage{ae,aecompl}

\usepackage{newtxtext,newtxmath}

\usepackage{graphicx}

 \usepackage{multirow}

\newcommand{\Msun}{\ensuremath{M_{\odot}}}
\newcommand{\msun}{\ensuremath{M_{\odot}}}

\newcommand{\logL}
{\ensuremath{\log_{10}(L/L_{\odot})}}

\newcommand{\sesn}{SESN}
\newcommand{\sesne}{SESNe}

\newcommand{\SESNe}{SESNe}

\newcommand{\Ibc}{SN\,Ib/c}
\newcommand{\IIb}{SN\,IIb}

\newcommand{\posydon}{\texttt{POSYDON}}
\def\mesa{\texttt{MESA}}
\newcommand{\fig}[1]{Fig.~\ref{#1}}

\defcitealias{Zapartas+2017b}{Z17}

\newcommand{\Zt}{\citetalias{Zapartas+2017b}}

\title[Companions to stripped-envelope SNe]
{The Demographics of Binary Companions to Stripped-Envelope Supernovae: Confronting Observations with Population Synthesis}

\author[E.~Zapartas \textit{et al.}]{
E.~Zapartas,$^{1}$\thanks{Contact e-mail: \href{mailto:ezapartas@gmail.com}{ezapartas@gmail.com}}
O.~D.~Fox,$^{2}$
J.~Su,$^{3,4}$
D.~Souropanis,$^{1}$
M.~R.~Drout,$^{3,4}$
K.~A.~Rocha,$^{5,6,7}$
S.~D.~van~Dyk,$^{8}$
\newauthor
B.~F.~Williams,$^{9}$
M.~Briel,$^{10,11}$
M.~Renzo,$^{12}$
J.~J.~Andrews,$^{13,14}$
T.~Fragos,$^{10,11}$
S.~Gossage,$^{6,7}$
\newauthor
M.~U.\,Kruckow,$^{10,11}$
C.~Liotine,$^{6,7}$
S.~D.~Ryder,$^{15,16}$
P.~M.~Srivastava,$^{17,6,7}$
and E.~Teng$^{5,6,7}$
\\
$^{1}$Institute of Astrophysics, Foundation for Research and Technology, N. Plastira 100, Heraklion, 70013, Greece\\
$^{2}$Space Telescope Science Institute, 3700 San Martin Drive, Baltimore, MD 21218, USA\\
$^{3}$David A.~Dunlap Department of Astronomy and Astrophysics, University of Toronto, Toronto, ON, Canada\\
$^{4}$Observatories of the Carnegie Institution for Science, Pasadena, CA, USA\\
$^{5}$Department of Physics and Astronomy, Northwestern University, Evanston, IL, USA\\
$^{6}$Center for Interdisciplinary Exploration and Research in Astrophysics (CIERA), Northwestern University, Evanston, IL, USA\\
$^{7}$NSF\textendash Simons AI Institute for the Sky (SkAI), Chicago, IL, USA\\
$^{8}$Caltech/IPAC, Pasadena, CA, USA\\
$^{9}$Department of Astronomy, University of Washington, Seattle, WA, USA\\
$^{10}$D\'epartement d’Astronomie, Universit\'e de Gen\`eve, Versoix, Switzerland\\
$^{11}$Gravitational Wave Science Center (GWSC), Universit\'e de Gen\`eve, Geneva, Switzerland\\
$^{12}$University of Arizona, Department of Astronomy \& Steward Observatory, 933 N. Cherry Ave., Tucson, AZ 85721, USA\\
$^{13}$Department of Physics, University of Florida, Gainesville, FL, USA\\
$^{14}$Institute for Fundamental Theory, University of Florida, Gainesville, FL, USA\\
$^{15}$School of Mathematical and Physical Sciences, Macquarie University, Sydney, NSW, Australia\\
$^{16}$Astrophysics and Space Technologies Research Centre, Macquarie University, Sydney, NSW, Australia\\
$^{17}$Department of Electrical and Computer Engineering, Northwestern University, Evanston, IL, USA
}

\begin{document}
\label{firstpage}
\pagerange{\pageref{firstpage}--\pageref{lastpage}}
\maketitle

\begin{abstract}
Stripped-envelope supernovae (\sesne) mark the deaths of massive stars without  hydrogen-rich envelopes. Most \sesne\ likely originate from binary systems where a companion stripped the progenitor of its envelope. Years of \textit{HST} imaging of nearby \sesn\ sites have produced a statistically meaningful sample of constraints on surviving binary companions. 
We assemble the current sample of six companion detections and six non-detections from the literature, re-analyzing whenever needed. We then conduct the first statistical comparison with binary population-synthesis predictions, primarily based on new calculations performed with the \texttt{POSYDON} framework. 
Across a metallicity range, our models predict that $80$–$90$\% of Type Ib/c and $60$–$85$\% of IIb SNe explode with a rapidly rotating, main-sequence companion. The observed luminosity distribution favors fairly inefficient mass accretion and failed explosions of the most massive stripped stars. The companion detection fraction broadly matches predictions, given the imaging depth, but appears elevated for \IIb. In all but one non-detection, a faint, undetected companion is the most likely scenario. The red, apparently evolved companions in a few \Ibc\ may result from strong interaction with the ejecta, expected in $\sim12\%$ of them. 
Companion demographics offer a powerful, independent probe of \sesn\ progenitor systems, with the current sample disfavoring efficient accretion and supporting Wolf-Rayet non-explodability. Larger companion samples and follow-up studies will further clarify binary pathways to \sesne, serving as benchmarks for transient surveys. 
\end{abstract}

%\keywords{stripped-envelope supernovae, massive stars, binary stars}

\date{Accepted XXX. Received YYY; in original form ZZZ}
\pubyear{2025}

%%%%%%%%%%%%%%%%%%%%%%%%%%%%%%%%%%%%%%%%%%%%%%%%%%

%%%%%%%%%%%%%%%%% BODY OF PAPER %%%%%%%%%%%%%%%%%%

\section{Introduction}\label{sec:intro}

%Massive progenitors of stripped-envelope supernovae (SNe IIb, Ib, Ic) are one of the main classes of explosions of massive stars, but the origin and evolution of their progenitors is not yet fully understood. Very massive single stars are potentially one group of progenitors of these events, whereas binary stripping from a binary companion is an alternative method for a progenitor to lose its H-rich envelope.
 
% Stripped-envelope supernovae (SNe  Ib, Ic and IIb mainly) show (almost) no sign of Hydrogen in their spectra. % but the origin and evolution of their progenitors is not yet fully understood. 
% 
Stripped-envelope supernovae (\sesne), mainly SNe Ib, Ic, and IIb, but also including broad-lined Ic (Ic-BL) and Ibn events, originate from stellar progenitors with (almost) no hydrogen-rich envelope \citep[e.g.,][]{Filippenko1997, Patat+2001, Pastorello+2007}. However, the origin and evolutionary pathways of these progenitors are not yet fully understood. 
Although historically expected to arise from massive Wolf-Rayet (WR) stars, extensive compelling evidence over the past decades suggests that the vast majority of \sesne\ likely originate from binary systems. The high observed fraction of massive stars in close binaries %\citep{Kobulnicky+2007, Chini+2012, Kiminki+2012, Sana+2012, Sana+2013, Kobulnicky+2014, Dunstall+2015, Almeida+2017, Moe+2017, XX} 
\citep{Kobulnicky+2007, Sana+2012, Sana+2013, Kobulnicky+2014, Dunstall+2015, Moe+2017, Banyard+2022, Villasenor+2025} 
supports an alternative evolutionary scenario in which the hydrogen-rich envelope of the progenitor is removed via mass transfer to a companion star \citep[e.g.,][]{Blaauw1961, Podsiadlowski+1992, Woosley+1994, Nomoto+1995, Gotberg+2017} leading to a hydrogen-poor SN progenitor \citep{Podsiadlowski+1992, De-Donder+1998, Yoon+2010, Eldridge+2013, Yoon+2017, Sravan+2019, Sravan+2020, Gotberg+2017, Zapartas+2017b, Ercolino+2025}. 
 % I need to add Eldridge+ Yoon already here, or merge with below.
 
Binary channels can explain several key observational properties of \sesne: their high relative rate  \citep{Smartt+2009, Smith+2011, Li+2011,Shivvers+2017}, their low ejecta masses \citep[e.g.,][]{Drout+2011, Taddia+2015, Lyman+2016, Modjaz+2016, Liu+2016, Fremling+2016}, properties of the host environment such as age and metallicity \citep[e.g.,][]{Arcavi+2012, Williams+2014,  Modjaz+2016, Graur+2017, Williams+2018, Maund2018, Sun+2020,Sun+2022, Singleton+2025,  Williams+2025}, and the lack of luminous WR progenitors identified in pre-explosion images of nearby events \citep[although see also discussion from \citealt{Yoon+2012, Eldridge+2013, Tramper+2015, Jung+2022,  Gilkis+2025}]{Van-Dyk+2003,Maund+2005,Maund+2005a,Smartt+2009,Eldridge+2013}. %,Van-Dyk2016}. 
Even the only WR candidate progenitor of a \sesn, SN~2017ein, has been debated since its detection \citep{Van-Dyk+2018, Kilpatrick+2018b} and most probably corresponds to an unrelated stellar source \citep{Zhao+2025}. 

In addition, \sesne\ progenitors need to successfully explode, avoiding a direct collapse into a black hole \citep[BH; e.g.,][]{Burrows+1995, Fryer+Heger2000, Janka2013}. Successful explosions are found more likely for binary-stripped helium stars \citep{Schneider+2021, Patton+2020, Laplace+2021,Vartanyan+2021} compared to massive WR progenitors that tend to not explode \citep[e.g.,][]{Zapartas+2021b, Aguilera-Dena+2023}, although some WR stars may be able to produce an observable event \citep{Ott+2018,Kuroda+2018, Burrows+2023}. 

%More on \sesne\ progenitor detections?
%The only robust identification of a \sesn, (for the SN~Ib/c iPTF13bvn) is a helium giant probably stripped in a binary system \citep{Cao+2013,Groh+2013b, Bersten+2014, Fremling+2014, Eldridge+2015, Kuncarayakti+2015,Folatelli+2016, Eldridge+2016}. 

%Wolf-Rayet phenomenon? \citep{Shenar+2020, Sanders+}

%%%%%%%%%%%%%%%%%%%%%%%%%%%%%%%%%%%
%\subsection{Companions to stripped-envelope SNe}

In most binary scenarios for \sesne, the companion responsible for stripping the progenitor %, through a phase of stable or unstable mass transfer, 
is expected to be present at the time of the \sesn\ explosion.   In a previous study by \citet[][hereafter \Zt]{Zapartas+2017b} employing the parametric population synthesis code \texttt{binary\_c} \citep{Izzard+2004,Izzard+2006,Izzard+2009}, it was estimated that a main-sequence (MS) companion should be present in roughly two thirds of \sesne, with the remainder having a compact object companion, or no companion at all. %(a white-dwarf, NS, BH)
%due to prior disruption of the system or merging of the two stars \
 Earlier work by \citet{Kochanek2009} also predicted a high companion fraction, even without explicitly modelling binary interactions. Further investigations into the stellar companion population, as well as the impact of the \sesn\ explosion on such companions, have been conducted \citep{Liu+2015, Hirai+2018, Ogata+2021, Hirai2023}. 

 %Observational detections or constraints to 
Observational searches for such companions provide valuable and direct insight into the nature of \sesn\ progenitor systems, both in the case of a positive detection, or even in the event of a non-detection since they still place meaningful upper limits on the presence of a luminous companion.
%\citep[e.g.,][]{Fox+2022} 
Targeted post-SN searches for these companions in nearby \sesne\ sites %(up to tens of Mpc away; Table~\ref{tab:comp_sample}) 
are feasible using the spatial resolution of \textit{HST}, and have led to several detections or upper limits of different luminosity depths. To date, there exists twelve well-characterized companion constraints in nearby \sesne, including five \textit{HST} companion detections in post-explosion imaging, one inferred companion from a pre-explosion spectral energy distribution (SED), and six constraining non-detections. This sample and their properties are summarized in Table~\ref{tab:comp_sample} as well as \fig{fig:luminosity-distance}, and is further discussed in Sec.~\ref{sec:obs_sample} and Appendix~\ref{sec:obs_sample_app}. 
%in Type IIb \citep{}, Ib events \citep[][]{vandyk2002,maund2016,ryder2018}, or Ic events \citep[SN2013ge; 2012df][]{Fox+2022, Williams+2024}, or to deep upper limits in the existence of one \citep[1994I, 2002ap][]{van-Dyk+2016,Zapartas+2017b}.  

Albeit small, the observed sample already allows for studying and drawing conclusions not just from individual events, but for the first time from a statistical approach. In this article, we explore the sample's collective  properties, including the companion luminosity distribution and detection fraction. Combining empirical constraints with predictions from updated, detailed binary evolution and population synthesis modeling uniquely enables us to extract constraints on the uncertain physical processes governing \sesn\ progenitor evolution.  
%In this manner, we extend the previous work of \Zt, which focused on the case of SN~2002ap using parametric population synthesis results of \sesn\ progenitors. 
 %we include the comprehensive, up-to-date observational sample and an updated, detailed binary evolution and population synthesis modeling. Altogether, this work enables more robust physical insights on the physics that .
% 
As these companions are the mass gainers in binary mass transfer, our analysis of the SN companion population allows us to independently constrain the mass transfer efficiency, one of the biggest uncertainties in binary evolution. Previous studies have inferred efficiency values spanning a broad range for similar types of systems, including semi-detached binaries \citep[e.g.,][]{de-Mink+2007, Mennekens+2017, Sen+2022}, %to $\gtrsim 30\%$ in Be X-ray binaries \citep{Vinciguerra+2020},
and Be X-ray binaries \citep{Vinciguerra+2020, Rocha+2024}, with high efficiencies found in Be+sdOB systems \citep{Pols2007, Schootemeijer+2018b,Lechien+2025}, % consistent with case studies such as in $\phi$~Persei \citep{}, 
compared with low ones in WR binaries \citep{Petrovic+2005, Shao+2016, Nuijten+2025}. Constraining this parameter is critical, as it shapes binary evolution outcomes  \citep[e.g.,][]{Renzo+2023} and sets the initial conditions for X-ray binaries and compact-object mergers \citep[e.g.,][]{van-den-Heuvel+De-Loore1973, Abbott+2016, Bavera+2021}. In addition, our analysis allows us to probe the uncertain explodability of massive stripped cores, independently constrained by the detection rate of luminous companions.

The remainder of the manuscript is organized as follows. In Section~\ref{sec:obs_sample} we describe the observational sample of binary companions and their properties.  Section~\ref{sec:pop_synth} summarizes the {\posydon} and auxiliary population-synthesis calculations and model caveats. In Section~\ref{sec:pie_chart} we present the predicted occurrence rates and evolutionary states of companions, while Section~\ref{sec:luminosity_cumulative} compares those predictions with the observed luminosity distribution.  The detectable companion fraction is studied in Section~\ref{sec:bin_frac_Z}, as well as the companions' positions in the Hertzsprung-Russell diagram (HRD; Sec.~\ref{sec:HRD}) and the possibility of interaction with  SN ejecta (Sec.~\ref{sec:ejecta_interaction_companion}). We discuss the astrophysical implications and observational caveats in Section~\ref{sec:discussion}, summarizing our main conclusions in Section~\ref{sec:conclusions}.

%%%%%%%%%%%%%%%%%%%%%%%%%%%%%%%%%%%%%%%%%%%%%%%%%%%%%%%%%%
\section{Method}\label{sec:method}
%%%%%%%%%%%%%%%%%%%%%%%%%%%%%%%%%%%%%%%%%%%%%%%%%%%%%%%%%%

We begin by outlining the combined observational and theoretical framework of our study. In Sec.~\ref{sec:obs_sample}, we compile the current sample of detected and constrained non-detections of companions to \sesne. In Sec.~\ref{sec:pop_synth}, we construct a suite of binary population synthesis models spanning a range of metallicities. Our aim is to use these models to predict the distribution of \sesn\ endpoints and their possible binary companions and compare them with observations, thereby constraining key physical processes in the evolution towards these events. % such as mass ratio and accretion efficiency.

\subsection{Observational sample of binary companions to \sesne}\label{sec:obs_sample}

The current observational census of companions searches with \emph{HST} imaging in nearby \sesne\ consists of 6 detections and 6 non-detections. We highlight that non-detections, particularly deep upper limits on the luminosity of a possible stellar companion, are also very useful as they suggest either a faint stellar companion, a compact object, or the absence of a companion at the moment of explosion. In most parts of our analysis in this study, unless mentioned otherwise, we employ only the \emph{direct} observational limits on a putative companion—namely, detections or flux upper limits measured at the exact SN position, not potential model-dependent inference in the literature about the most probable progenitor and companion system. Likewise, we exclude from the main sample limits obtained in older supernova remnants \citep{Dincel+2015, Kochanek2018,Kerzendorf+2019, Dincel+2024}, as well as indirect hints of binarity such as the periodic modulation attributed to a companion in the light curve of SN2022jli \citep{Moore+2023,Chen+2024}, although they are discussed in Sec.~\ref{sec:remnants} and \ref{sec:ejecta_interaction_companion}, respectively.

As we will primarily rely on measurements (or upper limits) of the companion luminosities, we divide the sample into three groups based on the type of information available in the literature as well as whether or not a candidate companion is detected:

\emph{Detections with constraints on the companion luminosity in the literature:} In general, when a companion detection has been reported and the HST magnitudes have been modeled to constrain the companion’s luminosity and temperature, we take these values directly from the literature. This is the case for four objects in our sample: SN2019yvr, SN2006jc, SN2001ig, and SN2011dh. For further information on each object, including which measurements we adopt when multiple are available, see Appendix~\ref{sec:obs_sample_app}.

\emph{Detections with new constraints on the companion luminosity computed here:} For two systems in our sample with claimed companion detections in the literature, we reanalyze them here in order to compute a constraint on their luminosity. First, SN2013ge was a Type Ib/c SN with a candidate companion detection in post-explosion HST imaging reported by \cite{Fox+2022}. While they found the companion to be consistent with a B5I star with an absolute magnitude of F275W $= -6.5$ mag (Vega), they do not explicitly provide a constraint on the companion’s luminosity. We therefore take the photometry from \cite{Fox+2022} and perform $\chi^2$ fitting with a set of Kurucz model atmospheres \citep{1993Kurucz} to derive a best-fit temperature and luminosity. Second, SN1993J is a well-studied Type IIb SN, with multiple studies examining a putative companion star \citep{Maund+2004,Maund+2009,Fox+2014}. The most recent analysis was performed by \cite{Fox+2014} who found evidence for a hot B-star (T$_{\rm{eff}} \sim 24000^{+3000}_{-5000}$K) through joint analysis of an HST near UV spectrum and photometry. However, as above, no explicit luminosity constraint for these models was provided. We therefore take the best-fit spectral model and normalization provided by \cite{Fox+2014} and integrate to derive a luminosity for the companion star. For further information on the reanalysis of both SN2013ge and SN1993J, see Appendix~\ref{sec:obs_sample_app}.

\emph{Systems with Upper Limits on the Presence of a Companion:} Our sample contains six non-detections of SN companions based on deep post-explosion HST imaging: SN2015G, SN2008ax, SN1994I, SN2017gax, SN2002ap, and SN2012fh. While in some cases, companion luminosity upper limits are provided in the literature, we chose to re-assess all systems with non-detections to ensure consistent analysis. In particular, we note that any stellar luminosity limit is formally temperature dependent\footnote{For example, if upper limits are available primarily at optical wavelengths, then constraints on the stellar luminosity will be weaker for both very hot and cool stars, whose spectra peak in the UV and IR, respectively.}. For each object, we therefore take the published HST upper limits and determine what types of companions they allow as a function of temperature. Specifically, we take a set of \cite{1993Kurucz} model atmospheres with temperatures between $T = 3500-36000$ K, scale them to luminosities ranging from $\log_{10}(L/L_{\odot}) = 2-7$, apply appropriate distance and reddening for each object, and perform synthetic HST photometry for comparison with the observed upper limits (for further information see \citealt{Williams+2025} and Su et al., in prep.). For much of the analysis below, we adopt the luminosity upper limits for these objects that correspond to the location of the MS. However, in Appendix~\ref{sec:obs_sample_app} we provide additional context on the types of companions that are ruled out as a function of temperature.

The inferred properties of the companions and the SN progenitor systems included in our sample are summarized in Table~\ref{tab:comp_sample} and each event is further discussed in Appendix~\ref{sec:obs_sample_app}. In particular, in Table~\ref{tab:comp_sample} we list the luminosities and temperatures for each system (for systems with upper limits, the temperatures and luminosities quoted in the Table correspond to the location of the MS)
as well as the distances and extinction values used to derive them . We also provide the SN subtype (which spans a wide range), the timing of the HST imaging relative to explosion (all are post-explosion with the exception of SN2019yvr), and estimate of the mass of the progenitor at the time of explosion. For the latter value, these are typically derived from light curve modelling (see Appendix~\ref{sec:obs_sample_app} for further details on a case-by-case basis). Finally, we also provide context on the metallicity of the SN host galaxy, when available. 

\begin{figure}
    \centering
    \includegraphics[width=0.99\linewidth]{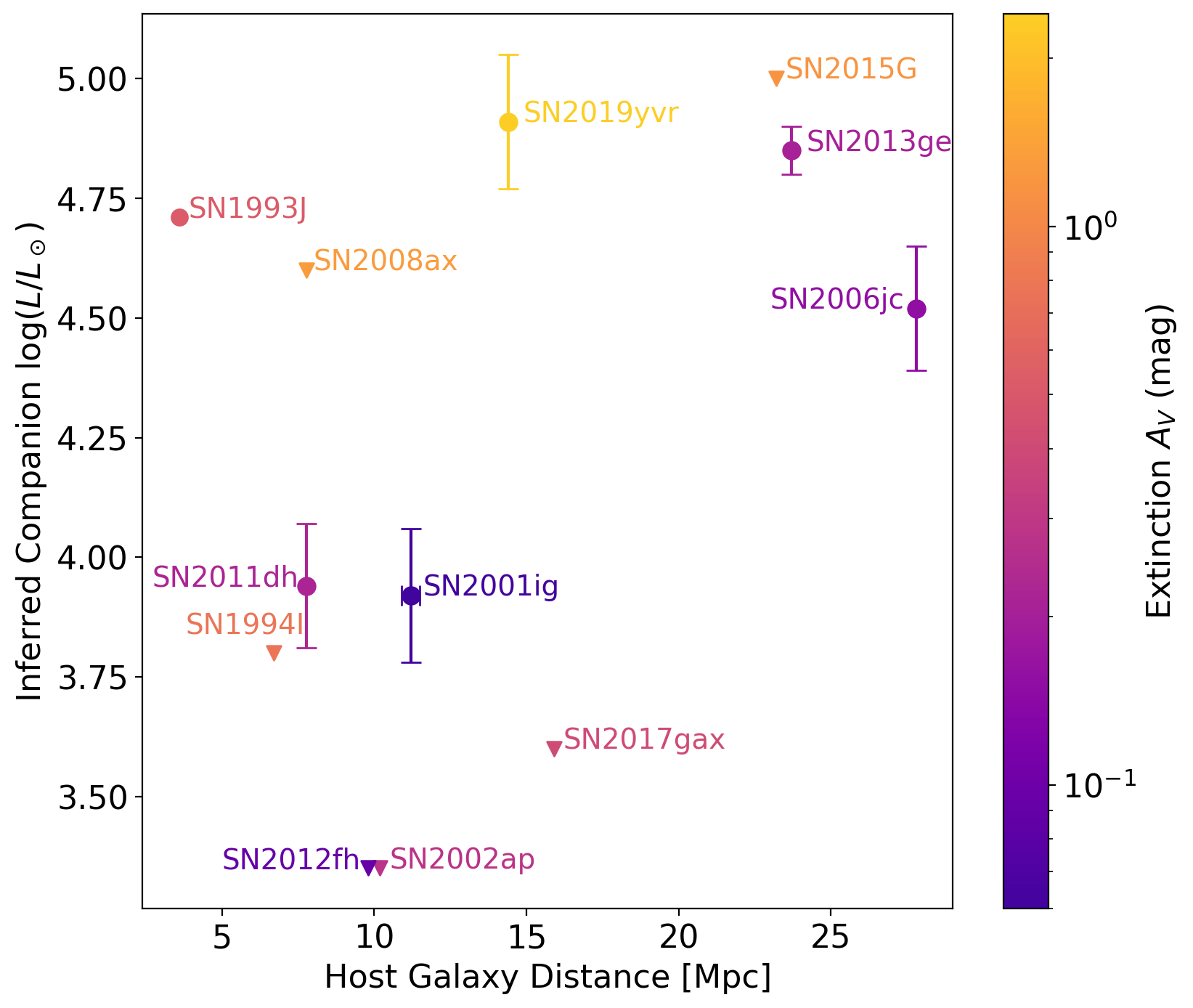}
    \caption{Luminosities inferred for the SN companions in our observational sample (see Section~\ref{sec:obs_sample} and Appendix~\ref{sec:obs_sample_app}) versus the distance to the SN host galaxies.  Upper limits for non-detections are plotted as triangles.  The color scheme indicates the extinction of the individual host galaxies.\label{fig:luminosity-distance}}
\end{figure}

In Figure~\ref{fig:luminosity-distance}, we plot the luminosity constraint for each companion in our sample versus the distance to the galaxy in which the SN exploded. Overall, the proposed companion detections are all found at luminosities of $4 \lesssim \log(L/L_{\odot}) \lesssim 5$ and were identified in galaxies with distances less than 30 Mpc. However, for several very nearby events (d $\lesssim 10$ Mpc), deep upper limits on the presence of a MS companion ($\log_{10}(L/L_{\odot}) \lesssim 3.35$) have also been possible.

%The inferred properties of the companions and the SN progenitor systems are summarized in Table~\ref{tab:comp_sample}, and each event is further discussed in Appendix~\ref{sec:obs_sample_app}.

\subsection{Population synthesis of binary companions next to stripped-envelope supernovae }\label{sec:pop_synth}

We model the evolution of massive binaries with the open-source and publicly available \posydon\footnote{\href{https://posydon.org/}{posydon.org}. Exact commit used in this work is \href{https://github.com/POSYDON-code/POSYDON/tree/Dimitris_WD_mergers}{891c5897}.} code \citep{Fragos+2023, Andrews+2024}.  
\posydon\ couples large grids of detailed single– and binary–star models, computed with \mesa\ \citep{Paxton+2011, Paxton+2013, Paxton+2015, Paxton+2018, Paxton+2019, Jermyn+2023}, to a rapid binary–population engine.  
The current \posydon~v2 release\footnote{\posydon~Data Release 2 grids are archived in the POSYDON Zenodo community at \href{https://zenodo.org/records/15194708}{https://zenodo.org/records/15194708}} spans eight metallicities from $10^{-4}$ to $2\,Z_\odot$ \citep[with $Z_\odot = 0.0142$;][]{Asplund+2009}; in this study we focus on $Z \ge 0.1\,Z_\odot$ because these metallicities encompass the majority of stripped-envelope supernovae (SESNe) discovered within $\lesssim 30$\,Mpc, where companion searches are currently practical \fig{fig:luminosity-distance}.  
For each metallicity, $Z/Z_{\odot} = \{0.1, 0.2, 0.45, 1, 2\}$, we evolve $2 \cdot 10^{5}$ stellar systems, adopting a binary fraction of $0.6$ in a period range [0.75-3170] days. These binary systems follow a \citet{Sana+2013} initial period distribution when $M_1>15\msun$  and extending it 
down to $0.75$\,d with a logarithmically uniform tail, as in \citet{Bavera+2021}. For lower mass (B-type) systems we assume a logarithmically uniform distribution across the
entire period range. Binary primary (or single) star masses follow a \citet{Kroupa+2001} initial-mass function from $4$ to $250\,\mathrm{M_\odot}$, with secondaries drawn from a flat mass-ratio distribution across $q=M_{2}/M_{1}\in[0.05,0.99]$ \citep[][]{Sana+2012}, although note that the \posydon~v2 grids do not extend to $M_2 < 0.5 \msun$.  %with $M_{2}\ge0.35\,\mathrm{M_\odot}$ 
Binaries start in circular orbits and with tidally aligned spins. Initial-final stellar and binary property interpolation within the grids follows Section~7 of \citet{Fragos+2023}.   

% A key source of uncertainty in predicting the stellar companion's mass and luminosity at the moment of \sesn\ is the efficiency of mass accretion during binary transfer. \posydon’s treatment, of the fairly non-conservative accretion, limited by the stellar spin of the accretor \citep{Fragos+2023} is physically motivated but still uncertain. As we currently lack alternative binary \mesa\ grids in \posydon\ with varied mass transfer efficiencies %(although they are under development), 
% in our analysis we also compare where relevant with the  default model of \Zt\ (where mass gain is limited by the thermal timescale of the accretor), as well as with their alternative models assuming fixed values of $\beta \equiv M_{\rm accreted}/M_{\rm transferred} = 0$, $30\%$, and $100\%$. A discussion and our conclusions on this uncertainty of binary evolution are presented in \Sec.~\ref{sec:conserv}.  

% Common-envelope (CE) evolution is treated with the classical $\alpha_{\mathrm{CE}}$–$\lambda_{\mathrm{CE}}$ prescription \citep{Webbink1984, Livio+1988}.  Our fiducial model adopts $\alpha_{\mathrm{CE}}=1$ and computes $\lambda_{\mathrm{CE}}$ directly from the \mesa\ profiles; a variant with $\alpha_{\mathrm{CE}}=5$ is explored in Section~\ref{sec:ce}.  Radiation-driven winds follow the \mesa\ {\tt Dutch} scheme with the addition of LBV-like winds for beyong the Humphresy-Davidosn limit \citep{Humphreys+Davidson1979,McDonald+2022}; their impact is revisited in Section~\ref{sec:variations_caveats}.

The \sesn\ progenitor populations are the same as those of the Souropanis et al. (in prep.)  %[][hereafter S25a]
upcoming study. We follow \citet{Andrews+2024} assumptions unless otherwise noted, summarizing the key processes here.  
In contrast to previous parametric population-synthesis studies (such as \Zt), the progenitor's detailed surface information within \posydon\ allows us to distinguish among different \sesn\ subtypes. A progenitor is tagged as a \sesn\ if the ejected hydrogen mass satisfies $M_{\mathrm{H,ej}} < 0.5\,\mathrm{M_\odot}$. Type~IIb systems meet $0.033\,\mathrm{M_\odot} \le M_{\mathrm{H,ej}} \le 0.5\,\mathrm{M_\odot}$, whereas Types~Ib and Ic are grouped together as \Ibc\ when $M_{\mathrm{H,ej}} < 0.033\,\mathrm{M_\odot}$ \citep{Hachinger+2012, Gilkis+2022}, due to the uncertainties both in the progenitor structure from modeling (Sec.~\ref{sec:variations_caveats}) and in the requirement to avoid helium signatures \citep[e.g.,][]{Yoon2015,Dessart+2020}. 
Although the sample also includes some companions for rare subtypes, such as  SNe Ic-BL or Ibn, we do not attempt to distinguish them in this study. Properly capturing these events would require incorporating additional physics, such as central-engine activity and strong circumstellar interaction, for which the key ingredients remain uncertain and beyond the capabilities of our present models. We further discuss these limitations for specific SN subtypes in Sec.~\ref{sec:subtypes}.

 %Properly capturing these events would require incorporating additional physics—such as central-engine activity and strong circumstellar interaction—for which the key ingredients remain uncertain and beyond the capabilities of our present models. A dedicated exploration of these sub-types is therefore deferred to future studies.  The limitations introduced by this choice, and the way our SN classification could map onto the specific supernova sub-types represented in the observed companion sample, are examined in detail in Sec.~\ref{sec:subtypes}.

 We include electron-capture SNe, with the lowest core mass able to explode defined according to  \citet{Tauris+2015}. 
We define successful core–collapse explosions as those that result in non-zero ejecta, forming either a neutron star (NS), or a BH after fallback.  As all our \mesa\ runs stop at core carbon depletion, in our default \posydon\ model, we estimate the collapse outcomes following the mapping to preSN structural parameters by \citet{Patton+2020} and the explodability criterion by \citet{Ertl+2016}. In our default model, successful \sesne\ together with BH formation from fallback are found to be negligible and therefore omitted.  %which maps the carbon–oxygen core mass and its central carbon abundance 
%Each stripped stellar candidate with lead to either a successful \sesn\ explosion with a NS formation or a BH without an assumed transient (events with fallback are found less important in this prescription and are omitted). 
For comparison, we also explore two alternative engines, the N20-calibrated prescription of \citet{Sukhbold+2016} and the ``delayed" prescription of \citet{Fryer+2012}, as discussed in Section~\ref{sec:non_expl}. We also refer to \citet{Zapartas+2021b} for a  comparison of SN explodability on single WR \sesn\ progenitors. We note that the explodability of \Zt\ models that are also shown in this study is assessed following the ``rapid’’ prescription of \citet{Fryer+2012} with non-zero ejecta, as these models do not retain sufficient information on the central abundance or core structure to apply the criterion of \citet{Patton+2020}.  Natal kicks are drawn from Maxwellian distributions with dispersion $\sigma = 265$\,km\,s$^{-1}$ for canonical iron CCSNe \citep[][although see also recent analysis of \citealt{Disberg+2025}]{Hobbs+2005}, and $\sigma = 20$\,km\,s$^{-1}$ for ECSNe \citep{Arzoumanian+2002, Giacobbo+2019}; BH kicks are down-scaled proportionally to the remnant mass.  A variation adopting a reduced normal NS kick dispersion of $\sigma = 60$\,km\,s$^{-1}$ is also discussed (Sec.~\ref{sec:ce_kicks}). 

\posydon, built upon detailed stellar structure models, provides an improved treatment of the mass-transfer phase compared to previous population synthesis codes, including its stability as well as the effects of rotation, tidal interactions, and spin-up of accreting companions. The models adopt default wind prescriptions, primarily following the \mesa\ ``Dutch'' scheme; a discussion on the possible impact of different wind prescriptions can be found in Sec.~\ref{sec:variations_caveats}. 
Unstable mass transfer \citep[as defined in section 4.2.4 of][]{Fragos+2023} triggers common-envelope evolution (CE), treated with the $\alpha_{\rm CE}$–$\lambda_{\rm CE}$ energy-balance formalism \citep{Webbink1984, Livio+1988}. We set $\alpha_{\rm CE}=1$, compute $\lambda_{\rm CE}$ from the donor’s detailed structure, and place the core–envelope boundary where $X_{\rm H}<0.3$ (assuming that the remaining post-CE envelope is removed during a subsequent, instantaneous, stable mass transfer episode; referred to as ``two\_phases\_stableMT" in \citealt{Andrews+2024});  we also tested a variation with more efficient CE ejection  with $\alpha_{\rm CE}=5$ discussed in Section~\ref{sec:ce_kicks}. 
A binary merger product, if one component has left the MS, is approximated with a single star with a helium core mass equal to the sum of the helium core masses of both pre-merger stars (which is considered zero for a MS star) with a central helium abundance equal to the mass-weighted average in the pre-merger cores (not trying to match the total mass as in default \citealt{Andrews+2024}). This way we try to better capture the core’s explodability parameters, while constraining less the envelope mass (which is anyway uncertain for merger products) and thus the resulting SN type. We further discuss this limitation in Sec.~\ref{sec:variations_caveats}.

A key source of uncertainty in predicting the stellar companion's mass and luminosity at the moment of \sesn\ explosion is the efficiency of mass accretion, $\beta \equiv \Delta M_\mathrm{acc} / \Delta{M}_\mathrm{d,RLO}$, the ratio of mass accreted to mass transferred by the donor through Roche lobe overflow %M_{\rm accreted}/M_{\rm transferred}$, during binary transfer 
\citepalias[as discussed in][and quantitatively shown in our analysis too]{Zapartas+2017b}. \posydon’s treatment, limited by the stellar spin of the accretor \citep{Fragos+2023} is physically motivated \citep{Langer1998, Heger+2000} but still uncertain. It tends to lead to very low average mass transfer efficiencies (with $\langle \bar{\beta}_{\rm rot} \rangle \sim 4\%$; calculated in detail in Appendix~\ref{sec:beta}). As we currently lack alternative binary \mesa\ grids in \posydon\ with varied mass transfer efficiencies \citep[potentially based on dedicated studies that follow the stellar-accretion disk angular momentum coupling, e.g., ][]{Popham+1991,Paczynski1991,  Ichikawa+1994, Dittmann2021} %(although they are under development), 
in our analysis, we also compare where relevant with the default model of \Zt\ (where mass gain is limited by the thermal timescale of the accretor), as well as with their alternative models assuming fixed values of $\beta = 0$, $30\%$, and $100\%$. A discussion and our conclusions on this uncertainty of binary evolution are presented in Sec.~\ref{sec:conserv}.

 %Additional model assumptions, along with their impact on the predicted companions to \sesne, are reviewed in Souropanis et al. (in prep.) and Section~\ref{sec:variations_caveats}.

\subsubsection{Main modeling limitations}\label{sec:variations_caveats}

Here, we briefly discuss the main modeling limitations in our analysis, which we identify as the treatment of mergers and the wind mass-loss rate. Beyond these, we place our results in the context of uncertainties in the binary mass accretion efficiency (Sec.\ref{sec:binary_physics}), explodability (Sec.\ref{sec:non_expl}), CE and SN kicks (Sec.~\ref{sec:ce_kicks}),  as well as the possible  physical origin of \sesn\ subtypes in our sample (Sec.~\ref{sec:subtypes}).

Our treatment of binary mergers is highly simplified, especially for non-MS mergers, matching essentially the merger product to a single stellar track. Modeling such systems in detail lies beyond the scope of this work. Most dedicated works on merger models focus mostly on Type II SN progenitors \citep[e.g.,][referring to the latter for an extensive discussion on the uncertainties of simulating merger structures in 1D stellar evolution models]{Menon+2019, Farrell+2020a, Chatzopoulos+2020, Schneider+2024, Patton+2025}. 
While this approximation may, in principle, affect the predicted fraction of \sesn\ originating from progenitors without companions, we find that most such merger products that have to get stripped through their own winds as WR stars, form cores that are too massive to explode under standard assumptions \citep[e.g.,][]{Zapartas+2021b}, mitigating the impact of this uncertainty. For more optimistic explosion criteria that allow for the explodability of massive helium cores, the contribution of isolated merger remnants to \sesne\ could increase. %Nevertheless, our main conclusion—that most \sesn\ progenitors arise from mergers involving main-sequence companions—remains robust; even in the limiting case where all models are assumed to explode, \citet{Zapartas+2017} found that roughly two-thirds originate from main-sequence mergers.

Uncertainty in stellar wind assumptions \citep[e.g.,][]{Smith2014} should typically be less important for the companions themselves, as they are usually MS stars with weak winds \citep[e.g.,][although they may impact their further evolution and explodability; \citealt{Renzo+2017}]{Vink+2000, Gormaz-Matamala+2021}, except perhaps in very luminous systems \citep{Sabhahit+2021, Josiek+2024}. Instead, winds are most critical during the evolution of the \sesn\ progenitor, i.e., the object whose companion we aim to identify. Mass loss at earlier stages \citep[e.g.,][]{Vink+2000, Gormaz-Matamala+2021}, and especially during the RSG phase \citep{de-Jager+1988, Kee+2021, Beasor+2020, Yang+2023, Antoniadis+2024, Decin+2024} or brief eruptive phases with enhanced winds \citep{Bonanos+2024, Cheng+2024}, can determine whether the hydrogen envelope is removed before explosion, enabling a star to become a \sesn\ candidate, irrespective of its single or binary origin \citep[e.g.,][]{Meynet+2015, Beasor+2021, Zapartas+2025}. The metallicity dependence of such winds is also a key factor \citep{Vink+2001, Antoniadis+2025}. Finally, winds from stripped stars, both massive WR \citep[e.g.,][]{Nugis+2000, Sander+2019} and low-mass binary-stripped ones \citep[e.g.,][]{Vink2017, Woosley2019, Sander+2020, Gotberg+2023}, are relevant for distinguishing between different \sesn\ subtypes.

 % Langer, Sanyal, expansion of M>20 Msun stars, MLT++, so RLOF. COre convective Overshooting for luminosity (apart from mass transfer efficiency). In any case Ic-Bl and Ibn discussed below.  

%reinforcing its minor role compared to stable mass transfer.

%\begin{table*} 
% \centering
\input{summary_comp_detected2.tex}
% \end{table*}

%%%%%%%%%%%%%%%%%%%%%%%%%%%%%%%%%%%%%%%%%%%%%%%%%%%%%%%%%%
%\section{Results}
%%%%%%%%%%%%%%%%%%%%%%%%%%%%%%%%%%%%%%%%%%%%%%%%%%%%%%%%%%

 \begin{figure*}  
\begin{center}
 %\includegraphics[width=0.95\linewidth]
 %{plots/companions next to stripped - envelope from all scenarios (including implosions into BH)}\\
 %{plots/companions next to stripped - envelope from all scenarios (including implosions into BH)log_yaxisFalse}\\
  %\includegraphics[width=0.95\linewidth]{plots/companions next to stripped - envelope from everything (not excluding implosions into BH)} % same as above but linear y axis)
  %\vspace{-0.1 cm}
   %\includegraphics[width=0.95\linewidth]{plots/companions next to stripped - envelope SNe from all scenario (but including only successfull explosions)}
   \includegraphics[width=0.99\linewidth]{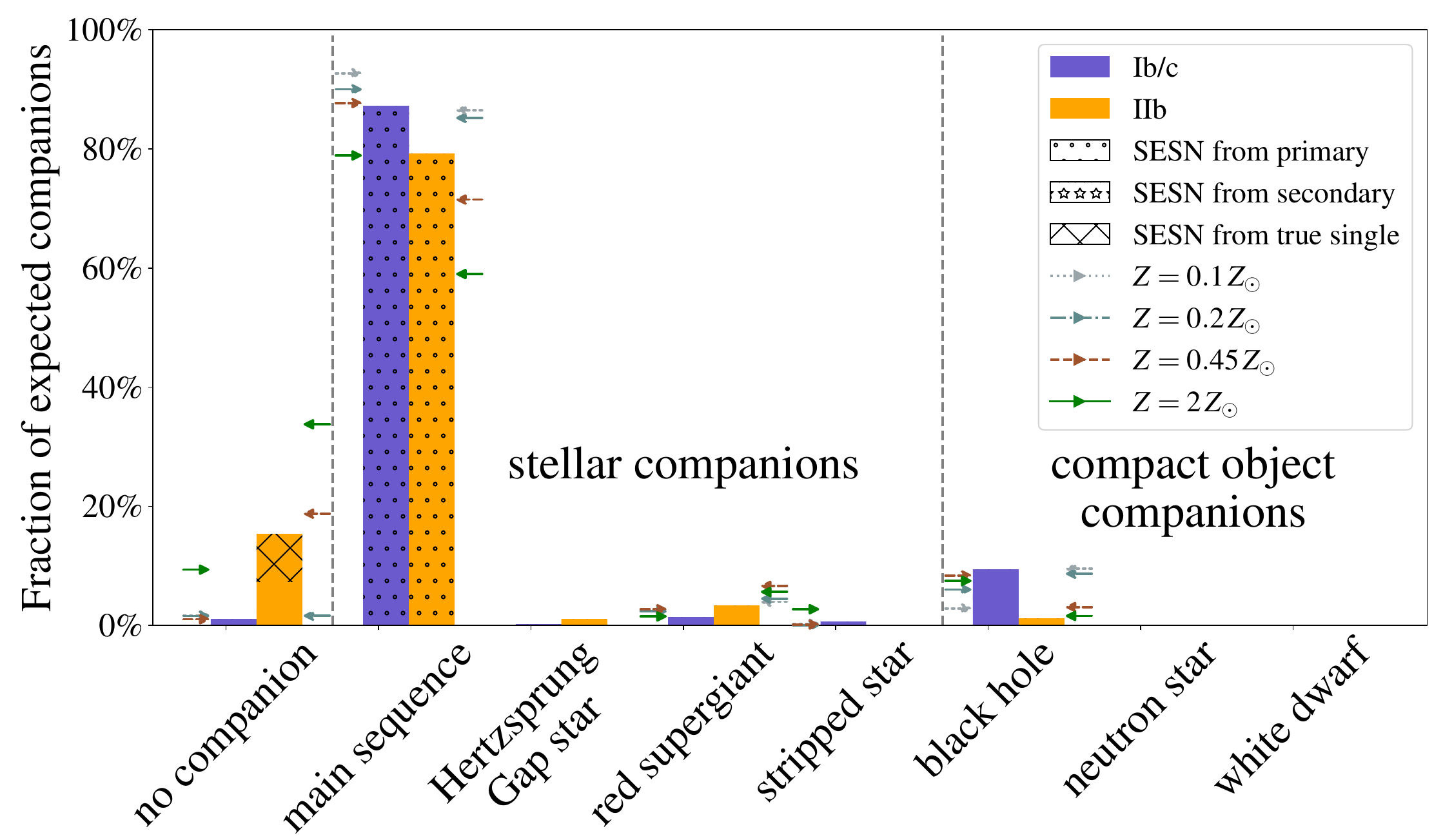}
 \caption{ 
 %Expected evolutionary phases of possible companions next to type Ib/c (cyan or blue) and IIb (yellow or orange) SNe at the moment of explosion. Top panel is assuming all core-collapse lead to an explosion, bottom includes only the successful explosions according to \citep{Patton+2020}. 
 %Compare to pie chart of \Zp\ (which also assumed that everything explodes).
 Expected relative frequencies of companion states to type \Ibc\ (blue) and \IIb\ (orange) SNe at the moment of explosion. We differentiate between the cases the \sesn\ progenitor is the primary star in a binary (dotted hashes), the secondary star (star hashes), or a true born-single star (X-pattern hashes). The bars are for $Z_\odot$, and with arrows next to each non-negligible bar we show the relative frequencies for $Z = 0.1, 0.2, 0.45,$ and $2\cdot Z_\odot$. \label{fig:pie_chart}
 %Top panel is assuming all core-collapse lead to an explosion, bottom includes only the successful explosions according to \citep{Patton+2020}. 
 %\MZ{Maybe differentiate among primary or secondary hashes and a different symbol specifically for mergers. For example for IIb, it would nice to see that the many with no-companions are indeed mergers, not just primaries}
 }
  \end{center}
\end{figure*}

\section{Results}
\label{sec:results}

\subsection{Predicted occurrence and evolutionary phase of companions}\label{sec:pie_chart}

We first calculate the expected companions next to \sesne\ based on our population synthesis results. Considering only successful explosions, the distribution of companions next to \sesne\ is illustrated in \fig{fig:pie_chart}. The majority of \Ibc\ events, around 87\%, retain a MS companion, predominantly from binary systems where the initially more massive primary star (dotted hashes) has been significantly stripped due to binary interactions. The fraction is so high due mainly to two reasons. The presence of the companion at explosion is the key reason for stripping the stellar progenitor, which would not be massive and luminous enough to eject its envelope through its own mass loss, in order to produce a \sesn. In addition, the most massive WR stars, candidates for \sesne, collapse directly into BHs without producing a \sesn\ \citep{Zapartas+2021b}. The latter is the main reason for the increased fraction of MS companions compared to the $\sim 2/3$ in \Zt, where BH implosions with no transient was not taken into account. 

The small fraction of \Ibc\ without a companion ($\sim2\%$) originate predominantly from binary mergers
 (dotted or star hashes, depending on which star initiated the merging). %\sesne\ merger scenarios initiated from secondary stars originate from close to equal mass systems where the secondary evolves first due to mass exchange. 
 In principle, \Ibc\ with no companions can also originate from massive WR secondaries that get isolated after being ejected from a disrupted binary, but these are found to be negligible, mostly leading to Type II SNe \citep{Zapartas+2019, Zapartas+2021a,  Wagg+2025}. The fraction of \sesne\ with no companions increases with metallicity, as winds become more effective at stripping and indirectly lowering the core mass, making the progenitors prone to explode, reaching \Ibc\ with no companion at 7\% for $2\cdot Z_\odot$.

For \IIb, the fraction with MS companions is slightly lower at $\sim79\%$. The fraction decreases as a function of metallicity,  since binary stripping \citep{Gotberg+2017} and its subsequent wind mass loss \citep{Yoon2017,Vink2017,Woosley2019, Drout+2023} increasingly remove any remaining thin H-rich layer on top. SNe IIb are also better candidates for harboring no companion at their SN site compared to \Ibc, either from single stars ($\sim7\%$, X-pattern hashes) or merger events ($\sim8\%$; initiated from both primaries and secondaries in roughly equal proportions), even at $Z_\odot$. %This result reflects the fact that isolated \IIb\ progenitors form lower core masses and are more likely to explode successfully, compared to the massive, fully stripped WR progenitor candidates of \Ibc\ \citep{Zapartas+2021b}. 
%Thus, the relative abundance of isolated \IIb\ reflects explosion efficiency, not progenitor frequency.  
Isolated \IIb\ progenitors tend to form lower core masses and explode more easily than the massive WR progenitors of \Ibc\ \citep[e.g.,][]{Zapartas+2021b}. Thus, the relative abundance compared to isolated \Ibc\ reflects explosion efficiency of partially stripped \IIb\ progenitors, not higher formation frequency (although these conclusions are subject to the merger treatment and explodability uncertainties, Sec.~\ref{sec:non_expl} and \ref{sec:variations_caveats}). 
At $2\cdot Z_\odot$, the no companion scenario reaches up to 32\% of all \IIb.

%For everything exploding, no implosions: around 2\% of all \Ibc\ (dotted pattern in the cyan column). % This is much different to 11.6\% of disrupted \sesne\ for Zsolar in \Zt. The main reason is the lower rate of disruptions found in \posydon,  due to more BH formation in the prior SN, that are expected to have have low or no kicks.

 \begin{figure*}  
% \vspace*{-2.0 cm}
\begin{center}
\includegraphics[width=0.99\linewidth]{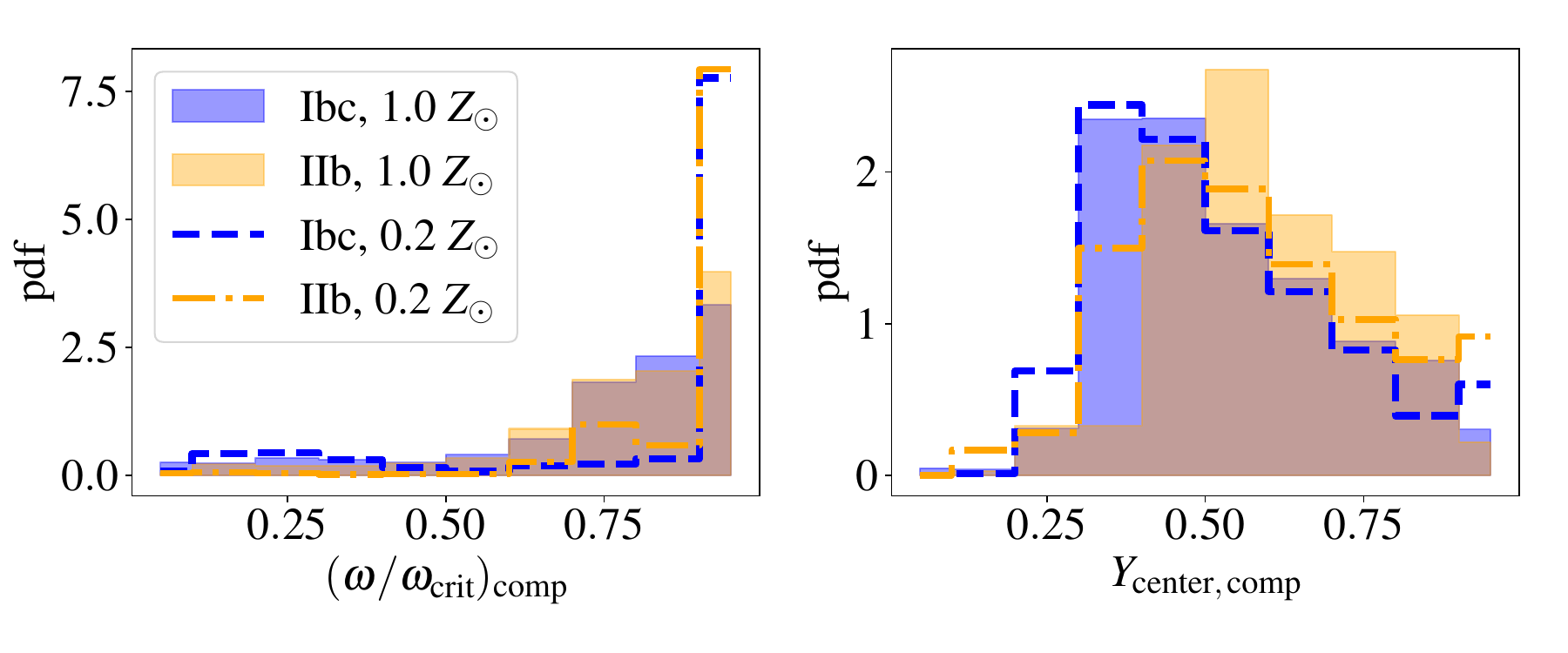}
 \caption{
Normalized probability density distributions of surface critical rotation (left) and central helium mass fraction (right) of companions next to \Ibc\ (blue) and \IIb\ (orange) progenitors at $1.0 Z_{\odot}$ (shaded) and $0.2 Z_{\odot}$ (lines) at the moment of explosion. \label{fig:centerhe_omega}
%\MZ{Only MS or all stellar companions?}
}
  \end{center}
\end{figure*}

Almost all our MS companions to \sesne\ have experienced a prior stable mass accretion phase. 
% %
This phase spins up the companion \citep{Packet1981, de-Mink+2013, Renzo+Gotberg2021},  having an average spin at the moment of the \sesn\ of $\omega/\omega_{\rm crit}\sim 0.84$ at $Z_\odot$ ($0.92$ and $0.77$ for $0.1$ and $2\,Z_\odot$, respectively) until the point of the progenitor's explosion as a \sesn\ (left panel of \fig{fig:centerhe_omega}). %This could potentially make them appear as Be stars \citep{Rivinius+2013, Rocha+2024}. 
Since most companions are still on the MS at the first explosion, they avoid the efficient spin down taking place during the post-MS expansion, and thus remain fast rotating at the time of the \sesn. At low metallicities, the companions are expected even closer to critical velocity at explosion, mostly due to weaker winds. Even at high metallicities where winds would tend to slow down the companion, only a few percent of the companions are found with a negligible spin, with only $5\%$ found with $\omega/\omega_{\rm crit}< 0.3$ at $2\,Z_\odot$. 
Scenarios of successful CE ejection after an unstable mass transfer phase  contribute only marginally to our \sesn\ progenitors %\citep[of the order of few \% at high metallicities, although increasing at lower ones;][]{Souropanis+2025} 
(of the order of few \% at high metallicities, although increasing at lower ones) 
and thus also on the companion sample. This differs from earlier population synthesis studies, as our binary models (that treat  in detail the mass transfer phase and stellar structure) predict fewer unstable mass transfer phases \citep[e.g.,][]{Gallegos-Garcia+2021}, and rare CE survival (Sec.~\ref{sec:ce_kicks}), typically leading to mergers that most likely eventually explode as  Type II SNe.  

 The predicted MS companions are found at various stages of their MS evolution. \Ibc\ companions are typically found in their early MS at the moment of explosion, with an average helium mass fraction at the center of $Y_\mathrm{center,\, comp}=0.54$ for $Z_{\odot}$ (right panel of \fig{fig:centerhe_omega}) %(corresponding to roughly 37\% of the fractional MS lifetime), 
and a broad distribution extending to their terminal-age main sequence (TAMS). \IIb\ companions are slightly more evolved, with an average   $Y_\mathrm{center,\, comp}=0.60$. % corresponding to approximately the midpoint of their MS lifetime. 
This is because they tend to originate systems that have more equal mass ratios than \Ibc\ \citep{Claeys+2011, Yoon+2017, Sravan+2018}, which result in more similar timescales with the \sesn\ progenitor. The central helium abundances at lower metallicities are similar for \Ibc\ and only slightly lower for \IIb, with the median $Y_\mathrm{center,\, comp}$ for the latter being $ $ at $0.2 Z_{\odot}$. As almost all these companions have accreted mass during the progenitor's stripping, their central abundances have been influenced by rejuvenation \citep{Neo+1977,Hellings1983,Hellings1984,Braun+Langer1995, Renzo+Gotberg2021,  Renzo+2023,Wagg+2024}. 

% %
% These MS companions are found at various stages of their MS evolution. Companions of \Ibc\ have spent an average of $\sim37\%$ of their MS lifetime, independent of metallicity but with a broad distribution until their terminal-age main sequence (TAMS). Companions next to \IIb\ are a bit more evolved, having already spent $49\%$ of their MS lifetime by the time of explosion, on average. This more evolved state occurs because \IIb\ tend to originate from less extreme mass ratio systems \citep{Claeys+2011, Yoon+2017, Sravan+2018, Souropanis+2025}. Thus the companion has a comparable timescale with the \sesn\ progenitor and has evolved more than in the case of \Ibc\ companions. %\citet{Claeys+2011} IIb showed that q<0.7-0.8 ends up in contact and that is the reason that \citet{Yoon+2017} limited themselves to q=0.9.

In addition to the dominant scenario of MS companions, about 8\% (2\%) of \Ibc\ (\IIb) progenitors have a compact object companion next to them (right part of \fig{fig:pie_chart}), predominantly a black hole (BH). These \sesne\ originate from systems where the initially less massive secondary remained bound to a BH following the primary’s earlier collapse. The fraction of SESNe with compact object companions slightly decreases for \Ibc\ at low metallicities, in contrast to \IIb\ that increases and reach $10\%$ at low metallicities, due to the tighter orbits and weaker winds that favour compact object companion next to partially-stripped progenitors. This fraction is higher than \Zt, where $\lesssim 2.8\%$ of SESNe at solar metallicity had a compact companion. The shift in number of SESNe with compact objects in \posydon\ suggests that most systems that would have been disrupted instead remain bound after the collapse of the initially more massive star into a BH, due to the lower assumed kicks for them. BH formation is favoured by the updated stellar physics of massive stars in \posydon, such as the higher convective overshooting appropriate for them \citep[e.g., ][]{Bavera+2023}. 

\sesne\ with NS companions are extremely rare ($<0.5\%$), as they are typically disrupted following the earlier explosion of the primary star. Even if the NS remains bound, such systems are more likely to trigger CE when the secondary initiates mass transfer and results in a merger event. Most \sesn\ with NS companions would most likely be ultra-stripped SNe \citep{Tauris+2015} with their rate consistent with previous predictions of this channel \citep[0.1-1\% of core-collapse SNe;][]{Tauris+2013}.  %very important as possible progenitors of double neutron star mergers, but they are a extreme minority channels in the  
Similarly, \sesne\ with white dwarf companions are disfavored by our default \posydon\ model assumption of low mass-transfer efficiency onto the secondary, ultimately limiting the secondary mass and typically resulting in a second white dwarf.

A small number of evolutionary channels with stellar companions more evolved than TAMS are also found, with roughly 2\%  (8\%) of \Ibc\ (\IIb) progenitors having predominantly red supergiant (RSG) companions. Again \IIb\ are more likely to harbor them, as they originate from systems with more equal mass ratio so that both stars evolve on similar timescales.%, with an increase in  the RSG companion rate to $\sim 7\%$ for $2\cdot Z_\odot$.  
We also find that $\lesssim1\%$ of \Ibc\ have stripped helium star companions, with a slight increase at higher metallicities. Although stripped stars are frequently anticipated from binary evolution scenarios, they generally constitute the progenitors of \sesne\ rather than their companions. The rate of occurrence of evolved companions is further discussed in Sec.~\ref{sec:bin_frac_Z}.

% Compact object companions, predominantly BHs, are found in $\sim10\%$ of \Ibc, and $\sim2\%$ (0.5\%) of \Ibc\ (\IIb) have giant companions. Stripped helium star companions account for about 1\% (4\%) of \Ibc\ (\IIb). Overall, these trends remain broadly consistent with \Zp, although differences arise for \sesne\ originating from secondaries, driven by the lower disruption rates predicted by \posydon, which favor BH formation in earlier SNe with reduced kicks, allowing systems to more often remain bound.

\begin{figure} 
% \vspace*{-2.0 cm}
\begin{center}
 \includegraphics[width=0.999\linewidth]
%{plots/cumulative_companions_pureTrue_upperlimitinTrue21_TypeIIbFalse_onlyObservationpointsFalse}
 {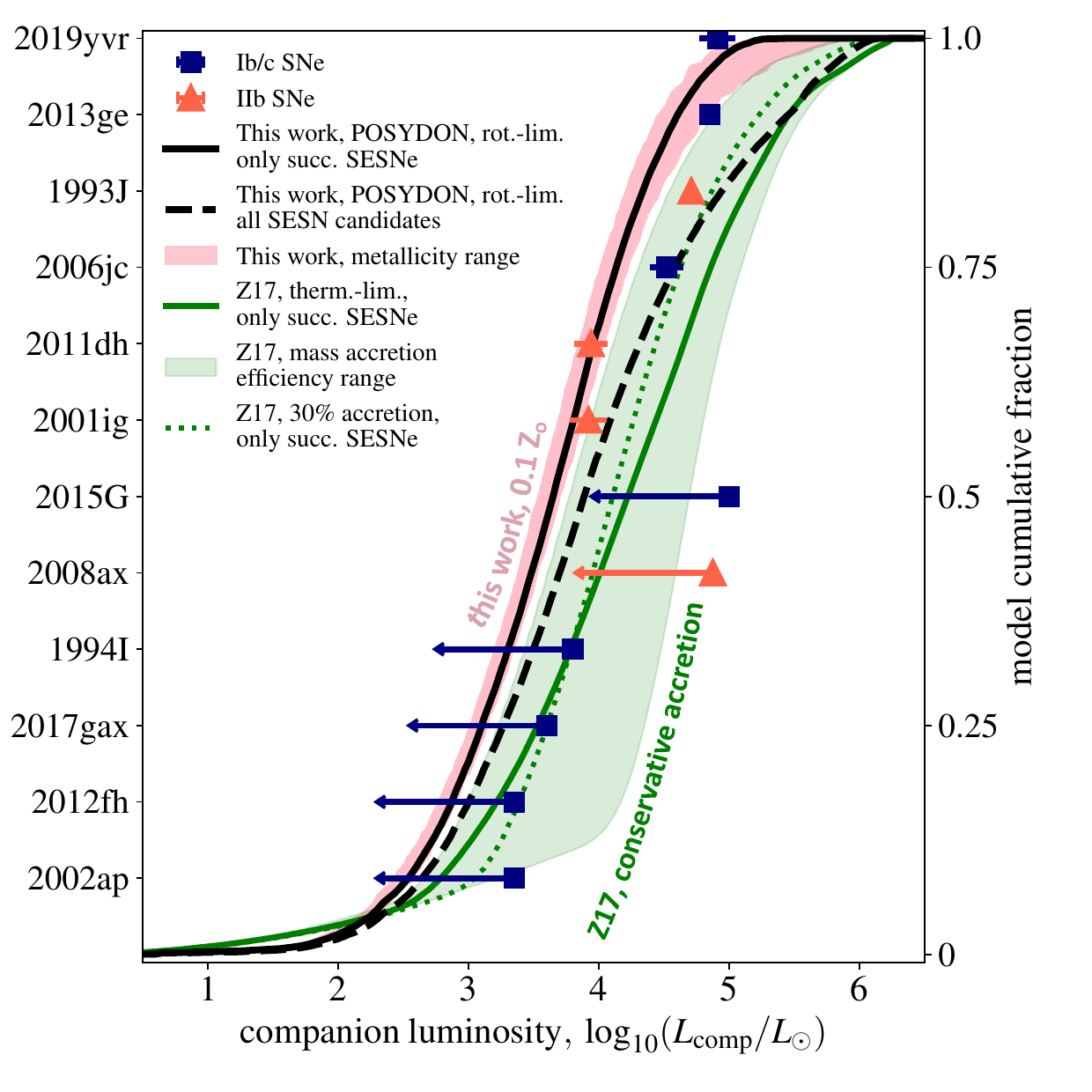} 
% \vspace*{-1.0 cm}
 \caption{Cumulative distribution of the detected luminosities, or upper limit in case of non-detections (shown at the bottom), of binary companions to \sesne\ (blue for \Ibc, red for \IIb). We also show the predicted luminosity distribution based on the default \posydon\ model in this study, keeping only the successful explosions (black solid line), and assuming all \sesn\ candidate progenitors explode (black dashed line), for $Z_{\odot}$. The pink shaded region shows the effect of metallicity for only successful explosions, ranging from $0.1$ to $2\,Z_{\odot}$. For comparison we also show the results of \Zt\ for $Z_{\odot}$ for different mass accretion efficiencies: the range between completely non-conservative and fully conservative mass accretion (green shaded region, left and right edge, respectively); the model which assumes a fixed $\beta=30\%$ (green dashed line); and the default thermally-limited \Zt\ model (green solid line), accepting in all of them only successful \sesne. 
  \label{fig:cumulative}
 %We also show the case of fully conservative mass transfer for \Zt\ including all \sesne\ progenitors (green dashed line). 
 %The model assuming high mass accretion efficiency, and the one assuming all massive \sesn\ progenitors explode, seem even by eye to overpredict luminous companions.
 }
 \end{center}
\end{figure}

\subsection{Luminosity distribution of stellar   companions}\label{sec:luminosity_cumulative}
 
%The distribution of companions is well-characterized by their luminosities (L$_{\rm comp}$) and effective temperatures ($T_{\rm eff}$). These parameters are more directly connected to the evolutionary state of the companion and its position in the Hertzsprung-Russell diagram (HRD). 
Our analysis primarily focuses on the luminosity of the
companion stars, whereas their effective temperature ($T_{\rm eff}$), more connected on the evolutionary state of the
companions, and their position in the HRD is discussed separately in Sec.~\ref{sec:HRD}. 
We deliberately refrain from focusing on companion mass estimates
%due to their greater uncertainties. Specifically, 
because translating observed luminosities and HRD positions (or equivalently \emph{HST} SEDs) into mass estimates is highly model-dependent and thus less reliable. However, we do discuss the predicted companion's mass distribution in Appendix~\ref{sec:mass_distr}. 

 In \fig{fig:cumulative} we show the cumulative distribution of the observed luminosities of stellar (ie., non-compact object) binary companions next to \sesne\ (as detailed in Table~\ref{tab:comp_sample}) together with the theoretically predicted ones. 
 %At this stage, we strictly consider observational limits for possible companions, without incorporating additional inferences based on progenitor or host galaxy characteristics. 
 Although we differentiate between the observed \Ibc\ (blue points) and \IIb\ (orange points), we analyze them collectively here. 
 % 
%This type of graph resembles that of the main complementary core-collapse class, i.e., Type II SN progenitors \citep[e.g.,][]{Smartt+2009}, though here we focus on binary companions rather than progenitors. 
%
%Given our results in Sec.~\ref{sec:pie_chart} and further quantified in Sec.~\ref{sec:upper_limits}, we anticipate that most cases will involve faint, stellar, most likely MS companions. Consequently, we place all upper limits at the lower end of the cumulative distribution (although this affects only the visually the graph and not the Bayesian analysis discussed below in Sec.~\ref{sec:better_statistics_cumulative}). % for further examination in Sec.\ref{sec:upper_limits}. 
%
% Our results from our models in Sec.~\ref{sec:pie_chart} %Sec.~\ref{sec:upper_limits} 
% provide some new insight into the interpretation of the observed upper limits. Since isolated progenitors are rare, most non-detections still suggest the likely presence of a faint, stellar, MS companion (and we further quantify this in Sec.~\ref{sec:upper_limits}). For this reason we justify placement of all upper limits at the lower end of the cumulative distribution (although this is only a visual choice, and does not influence the Bayesian model comparison below). 

Given the rarity of isolated progenitors (Sec.~\ref{sec:pie_chart}), most non-detections likely correspond to faint MS companions (further quantified and discussed in Sec.~\ref{sec:upper_limits}). We therefore place all upper limits at the lower end of the cumulative distribution, a choice for visual purposes only that does not affect the Bayesian model comparison below anyway.

We compare the observed luminosity distribution with our population synthesis models. %, following various assumptions. 
The default \posydon\ model at solar metallicity, including only successful \sesne, is shown by the solid black line and already appears, by eye, to be in decent agreement with observations,  predicting most companions in the range of $\logL = 2.5-5$. 
% for further examination in Sec.\ref{sec:upper_limits}. 
% The observed luminosity distribution clearly shows 6 detections at luminosities between $\log_{10}(L/L_\odot) \sim 4$ and $5$, as well as 6 non-detections mostly putting constraints as low as  $\log_{10}(L/L_\odot) \sim 3.5$, although two upper limits are shallower.

% I think Variations at the faint end are likely artificial, influenced by the limited modeling input physics, particularly the lack of self-consistent internal structure feedback (IF) in these metallicity variations.

%A key source of uncertainty in predicting companion luminosities is the efficiency of mass accretion during binary transfer. As we currently lack full conservative binary grids in \posydon\ with varied mass transfer efficiencies (although they are under development), 
%We also investigate the effect of other physical assumptions, such  the accretion efficiency onto the companion, $\beta$, and the explodability of the SN progenitors. % by comparing our results with a range of assumptions adopted in models from \Zt\ (green shaded area and green lines), 
We explore the impact of various properties on the shape of our distribution.  For metallicity, we plot the cumulative distributions across a range from $Z = 0.1\,Z_\odot$ to $Z = 2\,Z_\odot$ (pink shaded region). The overall impact is not strong, with higher metallicities resulting in slightly more luminous companions, mostly at the bright end of the distribution. This again corroborates that most stripping is caused by binary interactions rather than wind mass loss, the latter being much more sensitive to metallicity.

We perform a more quantitative Bayesian comparison among different stellar and binary physics assumptions below. 
 For each model assumed, ~$M$, we build a smooth probability–density function $p_M(L)$ of the companion luminosities, $L$,  %of the model-produced  catalogue
using a Gaussian kernel–density estimator (KDE).  The
bandwidth is set by the
Silverman rule \citep{Silverman2018}
%\footnote{$h_\mathrm{Silverman}=1.06\,\hat\sigma\,n^{-1/5}$ for one dimension; with $n=13\,391$ \textsc{POSYDON} points and $\hat\sigma\sim0.6$\,dex we obtain $h\simeq0.13$\,dex.}  
which balances bias and variance for a sample of ${\sim}10^{4}$ entries.  The
intrinsic cumulative–distribution function is then
%$F_M(L)=\int_{-\infty}^{L}p_M(L')\,\mathrm dL'$.
$F_M(L)=\int_{0}^{L}p_M(L')\,\mathrm dL'$.

Each observed \sesn\ event, $i$, contributes to either a detection at
$L_i$ or an upper limit $L_{\lim,i}$.
In the case of upper limits, we allow for the possibility of no stellar companion at the moment of explosion, meaning either a compact object or no companion at all. We define as $f_{\rm bin}$  the intrinsic probability for having a stellar companion. Although $f_{\rm bin}$ is found $\gtrsim 0.8$ for our default model (\fig{fig:pie_chart}), in this analysis we allow $f_{\rm bin}$ to be a free parameter and vary the whole range, calculating the likelihood of each model for different (fixed) 
$f_{\rm bin} = \{1.0,\,0.95,\,0.9,\,0.85,\,0.7,\,0.5,\,0.25,\,0.10\}$. The likelihood for a single SN can therefore be written as:

\begin{equation}
\label{eq:single_like}
\mathcal L_i \;=\;
\begin{cases}
\displaystyle
  f_{\rm bin}\cdot\,p_M(L_i),
  & \text{for detection},\\[10pt]
\displaystyle
  \underbrace{(1-f_{\rm bin})}_{\text{no stellar comp.}}
  +\;
  \underbrace{f_{\rm bin}\cdot\,F_M(L_{\lim,i})}_{\text{faint stellar comp.}},
  & \text{for upper limit}.
\end{cases}
\end{equation}

Observational uncertainties of detections ($\sigma_i\simeq0.05$--$0.10$\,dex) are relatively 
small compared to the KDE bandwidth. Tests with an explicit Gaussian
convolution of the observed uncertainties indeed leave $\ln\mathcal L$ unchanged at the $<0.01$ level and
are therefore neglected. The joint log–likelihood of a given model $M$ for the entire sample is written as:

\begin{equation}
\label{eq:logL_total}
\begin{aligned}
\ln\mathcal L_M\!\bigl(f_{\rm bin}\bigr) \;=\;
&\sum_{i\in\text{detections}}
 \ln\!\Bigl[f_{\rm bin}\,p_M\!\bigl(L_i\bigr)\Bigr] \\[6pt]
+&\sum_{i\in\text{upper limits}}
 \ln\!\Bigl[(1-f_{\rm bin})
            +f_{\rm bin}\,F_M\!\bigl(L_{\lim,i}\bigr)\Bigr] .
\end{aligned}
\end{equation}

\input{likelihoods_table3.tex}

The results for selected models are shown in Table~\ref{tab:likelihoods}. We adopt as a reference the highest likelihood  for the \posydon\ default model (at $Z_\odot$ and when only including successful \sesne) when $f_{\rm bin}=1.0$ (highlighted bold in Table~\ref{tab:likelihoods}).  Bayes factors, $\mathcal B=\exp\!\bigl[\ln\mathcal L-\ln\mathcal L_{\rm ref}\bigr]$, are quoted relative to it to indicate which model better fits the observational luminosity sample ($\mathcal B$ values closer to unity are equally consistent with the reference model). 

For assumed values of $f_{\rm bin} \gtrsim 0.7$, the Bayes factors exhibit only minor variation within any given model, indicating that there is little variation within each model at such high binary fractions. However, all models show a steady decline in likelihood toward lower $f_{\rm bin}$, suggesting lower fractions of \sesne\ with companions is disfavoured. %This behavior is consistent across all models. %although a few display local maxima in likelihood around $f_{\rm bin} \sim 0.5$–$0.7$, but never at $f_{\rm bin} \leq 0.5$. 
A more detailed discussion of the binary fraction is presented in Sec.~\ref{sec:bin_frac_Z}. 

The \posydon\ model including the explosion of all SESNe candidates (dashed black line) diverges from our default model especially for luminous companions, as these tend to be next to massive stripped progenitors that are more likely to implode into a BH. It thus results in slightly poorer agreement compared to the default, with $\mathcal B \sim 0.4-0.5$ at best (not shown at Table~\ref{tab:likelihoods}).

%  (all other model variations, of explodability and metallicity, perform worse and therefore omitted from the table but discussed below).
%and explodability (metallicity effects remain subdominant for the luminosity distribution and are therefore omitted). %%We adopt as a reference the highest likelihood for the \posydon\ default model (accepting only successful \sesne) when $f_{\rm bin}=1.0$ (bold in Table~\ref{tab:likelihoods})  which best reproduces the observed luminosity distribution. 
%The highest likelihood (i.e., the model which best fit the results) is found for the default \posydon\  model (at $Z_\odot$ and when only including successful \sesne) when $f_{\rm bin}=1.0$ (highlighted bold in Table~\ref{tab:likelihoods}). We adopt this as our reference model, with Bayes factors, $\mathcal B=\exp\!\bigl[\ln\mathcal L-\ln\mathcal L_{\rm ref}\bigr]$, quoted relative to it as a comparison of which model better fit the observational luminosity sample ($\mathcal B$ values closer to unity are equally consistent with the reference model to observations). %For each model, 
%

%

 %Even the cases including only successful explosions but thermally limited accretion have low likelihoods ($\mathcal B = 0.1-0.2$ at best). %, but approaching unity for $f_{\rm bin} \sim 0.85$.
% 

Focusing on different assumptions in mass transfer efficiency (one of the main uncertainties for the luminosity of binary companions), we find that the fixed accretion efficiency ($\beta = 0\%$, left edge of the green shading) %(defined as model variation 1 in that work), 
%when considering only successful \sesne, 
performs almost equally well with the default for $f_{\rm bin}>0.85$ ($\mathcal B \sim  0.9-0.95$). Similarly, the thermally-limited accretion model of \Zt\ (green solid line, default model at that work, with accretion scaling on average with stellar mass; \citealt{Schneider+2015}) reaches $\mathcal B \sim 1$ for slightly lower $f_{\rm bin}=0.7$.  The model with an accretion efficiency $\beta = 30\%$ (green dashed line) also performs greatly, having even slightly higher Bayes factor than our default \posydon\  model for $f_{\rm bin}=0.7-0.85$ (up to $\mathcal B = 1.24$). % (not shown at Table~\ref{tab:likelihoods}) for slightly lower $f_{\rm bin}=0.7$. 
In other words, all these low- to moderately-conservative models perform quite well, with a slightly more conservative mass-transfer assumption compensated by a lower $f_{\rm bin}$ value. %Similarly, a model with an accretion efficiency $\beta = 30\%$ reaches $\mathcal B = 0.5$ (not shown at Table~\ref{tab:likelihoods}) for slightly lower $f_{\rm bin}=0.7$. In other words, the more conservative a mass transfer model is, a lower $f_{\rm bin}$ value is needed to be consistent with the non-detections. For the thermally-limited accretion model of \Zt\  we find $\mathcal B = 0.1-0.2$ at best, indicating a weaker support than the other models above. %  However, given the uncertainties in our models and observations, as well as the Bayes factor close to unity among the favoured models, we refrain from both a detailed comparison of these differences to avoid an over-interpretation of the inferred mass accretion efficiency. 
In contrast, the fully conservative \Zt\ model ($\beta = 100\%$, corresponding to the right edge of the green shaded region) results in systematically higher companion luminosities and a sharp rise at $\log_{10}(L/L_\odot) \gtrsim 4$). %(3), assuming fully conservative mass transfer 
It is the least favoured among the options examined, with  $\mathcal B \le 0.26$. % and  in some cases $\ll 1$. %(e.g., $10^{-3}$ for $f_{\rm bin} = 1$). 
This is because it tends to overpredict the number of luminous companions and struggles to statistically account for the number of deep non-detections, as can also be inferred visually from \fig{fig:cumulative} (right edge of green shading). The disagreement would become even more pronounced if all \sesne\ candidates are assumed to explode. % (green dashed line).

\subsection{Companion detection fraction with metallicity}\label{sec:bin_frac_Z}

%In this subsection, we compare the predicted binary fractions of visible companions next to \sesne\ across a range of metallicities, comparing them with observational constraints derived from our sample. 
%
%From Appendix~\ref{sec:better_statistics_cumulative} 
The results of Sec.~\ref{sec:luminosity_cumulative} show that all plausible best fitting models to our data have a higher likelihood for binary fraction, $f_{\rm bin}\ge 0.7$ (which is considered a free parameter of that analysis). 
%and most peak between $f_{\rm bin} = 0.7$ and $1$. 
The comparison cannot confidently differentiate among high $f_{\rm bin}$ values, but all models have an extreme drop in likelihood for $f_{\rm bin}<0.7$. Here, we consider the fraction of detectable binary companions, $f_{\rm det}$, as a function of both metallicity and imaging depth.

%the fraction of the predicted and observed fraction of detections, $f_{\rm det}$, as a function of metallicity. %for two particular factors: metallicity and observation depths. % considering different visible stellar companions. % observed binary fractions as a function of metallicity and of our the depth of our observational surveys. 

\begin{figure}  
% plotting_metallicity_companions.ipynb
\begin{center}
 \includegraphics[width=0.99\linewidth]%{plots/annotated_HRD_companions_poster.pdf}
%{plots/binary_fraction_metallicity_companions_sequential_shading_MS}\\
{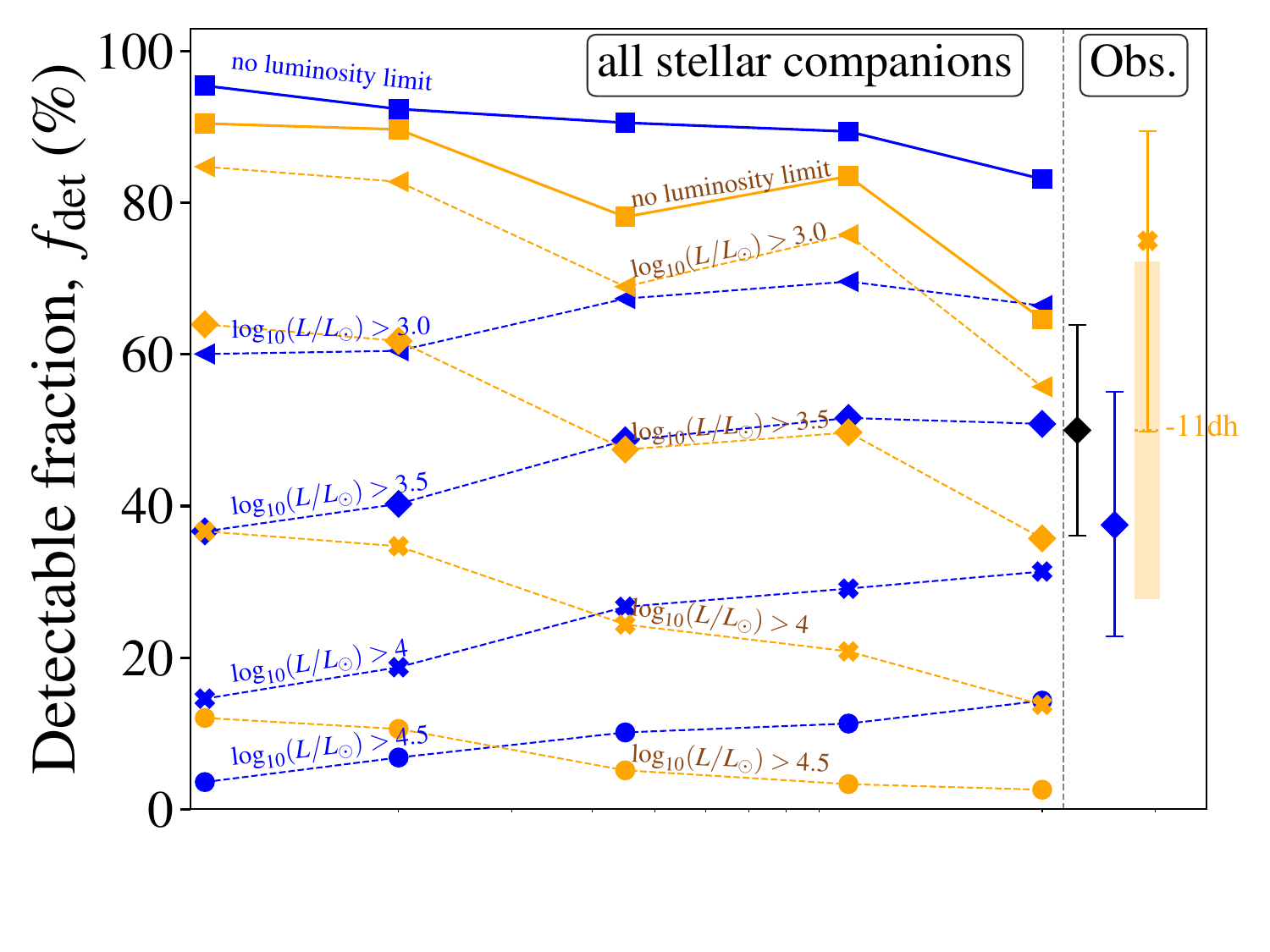}\\
 \vspace*{-0.9 cm}
 \includegraphics[width=0.99\linewidth]%{plots/annotated_HRD_companions_poster.pdf}
{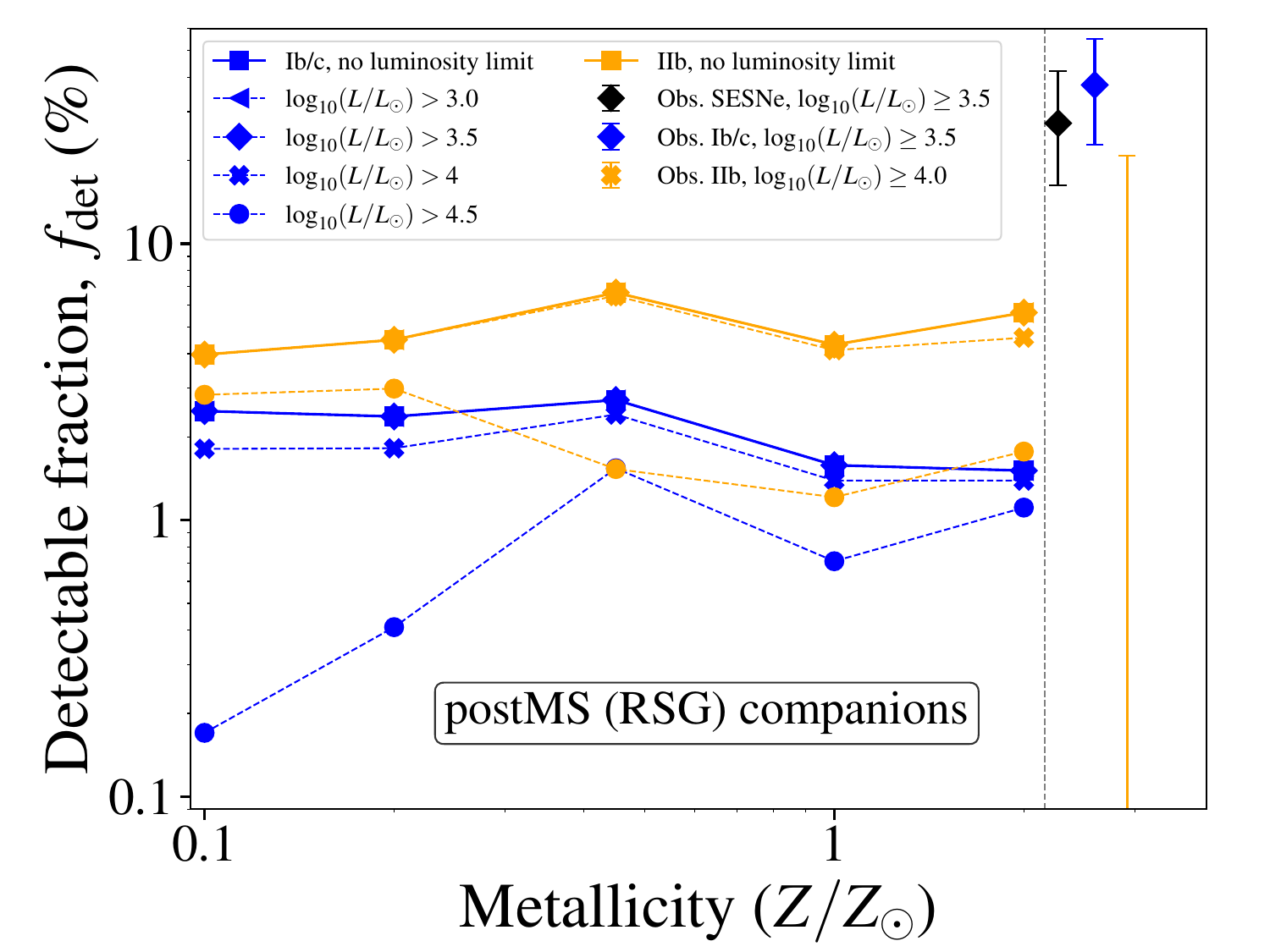}
 \caption{
  {\bf Top:} Fraction of detectable binary companions next to \Ibc\ (blue) and \IIb\ (orange), as a function of metallicity. The fractions decrease depending on the luminosity threshold of the searches, from infinitely deep searches (top line with squares) to increasingly shallower luminosity thresholds (dashed lines of different symbols). On the right of each panel, we show the fraction of detected companions for all \sesne\ (black), \Ibc\ (blue), and \IIb\ (orange),  for selected luminosity depths, corresponding approximately to the observational surveys for each type. In each observed fraction, we show the binomial 1-sigma uncertainty due to our small sample (errorbar), as well as the recalculation of the fraction with no companion in 2011dh for \IIb  (orange shaded rectangle). {\bf Bottom:} Same as above, but only showing the fraction of expected post-MS companions (mostly RSGs). Note the difference in the y-axis scale. 
 \label{fig:bin_frac_Z}
 %We also show the fraction of all stripped companions only in \Ibc\ (grey), with none expected in \IIb.
 }
  \end{center}
\end{figure}

In \fig{fig:bin_frac_Z} we show $f_{\rm det}$ for both \Ibc\ and IIb (blue and orange, respectively) for the default \posydon\ model, as a function of metallicity and for a variety of search sensitivities. The top lines of the top panel correspond to sufficiently deep searches that detect all low luminosity companions. They have values that are the same as presented in Sec.~\ref{sec:pie_chart}. This result highlights, unsurprisingly, that the detectable fraction of binary companions is the equivalent of the predicted fraction ($f_{\rm det}\equiv f_{\rm bin}$). %This would imply that with infinitely deep observational surveys, we would expect to see a companion in almost all cases. 
%At solar metallicity, the fractions for \Ibc\ and \IIb\ are about 87\% and 79\%, respectively (top blue/orange square points for $Z_{\odot}$), as discussed earlier in Sec.\ref{sec:pie_chart}. 
We find only a modest decline in $f_{\rm det}$ with increasing metallicity, primarily due to slightly more efficient stellar winds enabling the formation of isolated, individual WR progenitors. 
%Hence, the vast majority of \sesne\ across all typical host environments is predicted to have stellar companions (most probably a MS one).

%
%A notable observational outlier is the case not of a \sesn\ but of the CasA SN remnant, discussed separately in Sec.~\ref{sec:remnants}, which lacks any detectable companion despite very stringent luminosity limits.

%Magnitude limits of the images in the searches naturally reduce the companion detection fraction. 

Dashed lines of different symbols show the effects of increasing the luminosity threshold (i.e., having shallower depth) of the detectability limit. By excluding faint companions with $\log_{10}(L/L_\odot) < 3.0$, the predicted  fraction drops to about $f_{\rm det} \sim 50-60\%$ across the considered metallicities for \Ibc, although it does remain above 70\% for \IIb. Further restricting ourselves to only more luminous companions with $\log_{10}(L/L_\odot) > 3.5$ reduces $f_{\rm det}$ to around 40\% for \Ibc, or as low as 25\% at sub-solar metallicities. 

A couple interesting overall trends emerge. For \Ibc, there is a decrease in the detectable fraction at lower metallicities. These progenitors originate from systems with lower initial primary masses, and subsequently with less luminous companions, requiring a second mass transfer episode to lose their thin hydrogen-rich envelope  (Souropanis et al., in prep.).
%due to an increased proportion of low-luminosity MS companions, which are 
The less luminous companions are inherently more challenging to detect observationally.  Conversely, \IIb\ increases its detectable fraction at lower metallicities %from  $f_{\rm det}\sim50\%$ with $\logL>3.5$ at $Z_\odot$ to 70\% at sub-SMC metallicities (which becomes even $\gtrsim 80\%$ for $\logL>3.0$). 
because their progenitors originate from massive stars stripped in binaries across a wide range of initial orbital periods (Souropanis et al., in prep.). 
%Reduced binary stripping efficiency \citep{Gotberg+2017} and diminished wind-driven mass loss of the remnant \citep{Yoon2017} at low metallicities allow \IIb\ progenitors to retain a thin hydrogen-rich envelope, avoiding \Ibc. 
Consequently, the more massive progenitors of \IIb\ typically lead to correspondingly more massive and luminous companions. 

Our current observational sample includes 6 binary companion detections (i.e., not counting upper limits) out of 12 \sesne\ (Table~\ref{tab:comp_sample}). This %(with 15G) else 10 
translates to an observational fraction of $f_{\rm det}=50^{+14}_{-14}\%$ (black diamond on the right, at the top panel of \fig{fig:bin_frac_Z}). The uncertainty of this fraction due to our small sample is calculated as the ``Jeffreys'' binomial confidence interval for small number statistics  \citep[assuming a $1\sigma$ confidence of $0.6827$;][error bars in the observed fractions]{Cameron2011}. The detected fraction for \Ibc\ is about $37.5^{+18}_{-15}\%$ (3 out of 8), while the fraction for \IIb\ is roughly $f_{\rm det}=75^{+14.5}_{-25}\%$  (3 out of 4). For these detected fractions we treat all events together, although they may originate in different metallicity environments, due to our already small sample and the overall measurement uncertainties on the host metallicities (Table~\ref{tab:comp_sample}). However, this assumption still allows us to monitor trends and comparison between the subclasses.
%
 %Anyway, the observed \sesne\ companion fraction rises to 7 out of 11 (black errorbar) if we assume a possible indirect detection in case of hypothetical deeper luminosity limits for SN1994I and SN2002ap, which are inferred to have low-mass, undetected companions based on progenitor mass and ejecta properties \citep{Van-Dyk2016,Zapartas+2017b}. Similarly, fraction of \Ibc\ increases to 50\% if we assume a possible companion existence for SN 2015G (blue error bar), which has a much shallower luminosity limit than $\logL\sim 3.5$. Further analysis of how we treat upper limits on possible companions is shown in Sec.~\ref{sec:upper_limits}. 

Given our luminosity limits, which tend to be roughly $\logL\sim 3.5$ for most \sesne\ (see Table~\ref{tab:comp_sample}), our observational results for \Ibc\ (and even SESNe overall) are broadly consistent with theoretical predictions in \fig{fig:bin_frac_Z}.
In contrast, the detection fraction for \IIb\ is a bit high (apart from the cases of low metallicity) %the detection fraction for \IIb\ is approximately $f_{\rm det}=75^{+14.5}_{-25}\%$  (3 out of 4). 
given the faintest \IIb\ detection is at $\logL\sim 4$ (and the one non-detection has very shallow limits). %it appears that we detect more companions for \IIb\ than expected. 
In fact, if SN 2015G (which has shallow limits) also harbors a luminous companion, $f_{\rm det}$ can even reach $>80\%$. %$100\%^{0\%}_{-20\%}$. 
Even if we consider the possibility that the post-SN source in 2011dh is not a companion but instead is dominated by a light echo \citep{Maund2019} or CSM interaction \citep{Maund+2015}, $f_{\rm det}=50^{+22}_{-22}\%$ (2 out of 4, orange shaded rectangle) would still be on the high side. Excluding SN2011dh from our sample altogether, due to its uncertain nature, would have even less impact on the detected fraction than considering it a non-detection. Although limited by the overall small sample size, %this high fraction is only consistent with the predicted ones for \IIb\ at low metallicities, 
the high $f_{\rm det}$ for \IIb\ may be interpreted as indirect evidence that \IIb\ are favoured in those low-metallicity environments \citep[e.g.,][Souropanis et al., in prep.]{Anderson+2015} or may leave room for channels towards more luminous companions in \IIb. For example, interaction of the SN ejecta with a companion has been proposed to yield a more luminous and redder star (Sec.~\ref{sec:ejecta_interaction_companion}). However, as we in that Section, such a scenario is unlikely for \IIb, and anyway  detections in our sample are more consistent with bluer MS stars.

%\MZ{Chen+2023 discuss the possibility of 11dh}

% Even the lowest value of the observed $f_{\rm bin}$ seems on the high side  and inconsistent with the low predicted companion fraction for $\logL\gtrsim4.0$ (which is the rough limit for \IIb\ surveys only, orange points in \fig{fig:cumulative}). Although we cannot extract any strong conclusions from the small numbers of \IIb\ only, given at face value, this discrepancy may potentially  point that most 

% to higher mass transfer efficiency for \IIb\ which would favour more massive and luminous companions and thus more predicted detections too (further discussed in XX).
% \MZ{Although this contradicts the conclusion for less conservative result based on the luminosity distribution of all \sesne\ ( discussed in Sec.~\ref{sec:luminosity_cumulative}}
% \MZ{It also needs further investigation of a physical reason for higher mass transfer efficiency mostly for \IIb\ binary progenitor systems, which may be caused by the fact that \IIb\ occur predominantly at lower metallicities compared to \IIb\ \citep{Sravan+2018,Souropanis+2025}. Also different range of initial progenitors masses (at low Z) or periods (for high Z).}

The lower panel of \fig{fig:bin_frac_Z} is the same as the upper, but includes only post-MS companions, which are expected to be predominantly RSGs. The predicted fraction of RSG companions remains low (<10\%) across all metallicities. At higher metallicities evolved RSG companions can be as frequent as 8\% of all \IIb, reflecting a progenitor origin from binaries with near-equal initial masses (and evolutionary timescales) so that a companion is more likely to have evolved off the MS by the time the primary explodes.  These companions are almost always of high luminosity ($\logL\gtrsim 4$), as RSGs have higher luminosities compared to MS stars of the same mass. %At lower metallicities, similar fractions of $\sim 5\%$ of evolved companions are predicted for both \Ibc\ and \IIb, but with a wider range of luminosities for the former.
%and relatively long orbital periods (several hundred to thousands of days). Such systems undergo predominantly Case~BC or C mass transfer \citep{Yoon+2017, Sravan+2019, Souropanis+2025}, facilitating the evolution of the secondary off-MS by the time the primary explodes, due to similar evolutionary timescales. This companions are almost always of high luminosity ($\logL\gtrsim 4$), as RSGs have higher luminosities compared to MS stars of the same mass. At lower metallicities, similar fractions of $\sim 5\%$ of evolved companions are predicted for both \Ibc\ and \IIb, but with a wider range of luminosities for the former.
%No evolved companion has been detected in the 4 \IIb\ events (orange errorbar, $<21\%$). 

Observationally, red companions have not been detected for \IIb, but have been identified in possibly three \Ibc: 2006jc, 2013ge, and SN2019yvr (via pre-explosion analysis). If these scenarios correspond to post-MS companions, it would result in a detected fraction of post-MS companions of about 15–40\% for all \sesne\ (black errorbars) and even higher specifically for \Ibc\ (blue errorbars). %This detected fraction may be even higher (black and blue shading) due to a potential third \Ibc\ (SN2013ge) with a companion potentially evolved off the MS. 
This fraction significantly exceeds theoretical expectations. It is worth noting, however, that none of the detected sources are actually found at the RSG region of the HR diagram, but instead are within the Hertzsprung Gap (discussed later in Sec.~\ref{sec:HRD}). Intriguingly, misclassification may at least partially explain this discrepancy, as interactions of SN ejecta with a MS companion could make it appear redder and mistakenly identified as evolved. We discuss this further in Sec.~\ref{sec:ejecta_interaction_companion}.

% Finally, stripped helium star companions are predicted exclusively for \Ibc\ (WHY?), with fractions decreasing from about 2\% at twice solar metallicity to below 1\% at sub-solar metallicities. These companions are expected to be typically luminous, averaging around $\logL \sim \gtrsim 4.5$. Although frequently anticipated from binary evolution scenarios, stripped helium stars generally constitute the progenitors of \sesne\ rather than their companions. Furthermore, their significant UV brightness coupled with their optical faintness poses additional challenges for their detection \citep{Gotberg+2018}, though recent observational efforts have started addressing these difficulties \citep{Drout+2023}. %The extremely low predicted fraction aligns with the current absence of observational detections of stripped helium companions in any \sesn, even in cases of very deep targeted searches such as the investigation of the Type IIb SN remnant Cas~A \citep{Kerzendorf+2019}.

%%%%%%%%%%%%%%%%%%%%%%%%%%%%%%%%%%%%%%%%%%%%%%%%%%%%%%
\subsection{Interpreting companion non-detections} \label{sec:upper_limits}

 \begin{figure}  
% \vspace*{-2.0 cm}
\begin{center}
\includegraphics[width=0.999\linewidth]%{plots/prob_for_faint_companion_Ibc_includingBHimplosions_inclnocompanions_POSYDON}
{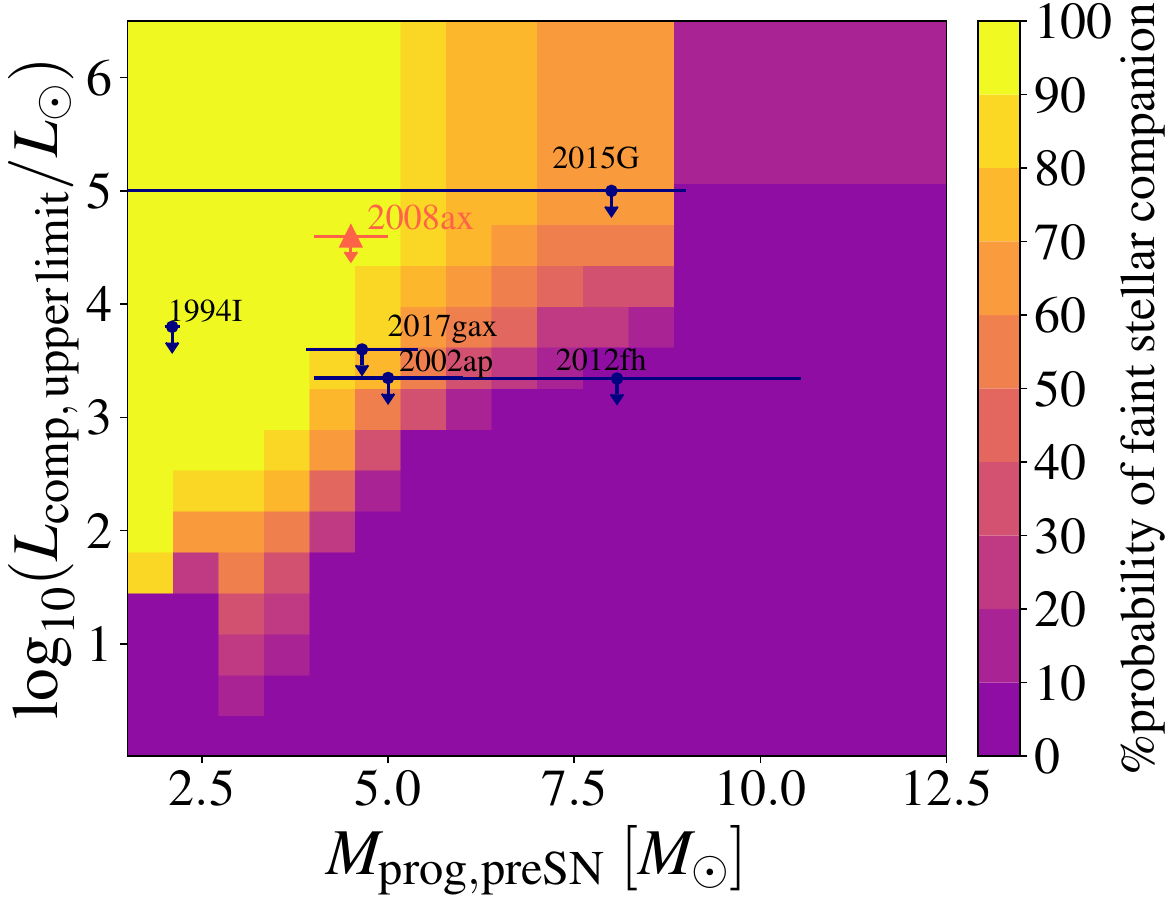}
 \caption{
Probability of presence of a stellar companion for our default \posydon\ model for $Z_\odot$, as a function of the mass of the stripped progenitor, depending on the luminosity depth of non-detections in a \sesn\ site.  Points represent the empirical non-detections to \Ibc\ (blue) and \IIb\ (red). For all non-detections in our sample, a faint stellar companion is still very likely present, with the exception of SN2012fh. 
\label{fig:prob_nondetection}
%\MZ{12fh is lower limit in Mej, 15G is upper limit}
}
  \end{center}
\end{figure}

%Here we discuss the interpretation of companion non-detections in conjunction with constraints on the progenitors of \sesne. 
Figure~\ref{fig:prob_nondetection} shows the probability for the presence of a faint, undetected companion as a function of the imaging luminosity limit and the progenitor primary star mass (typically inferred from the SN light curves or from progenitor imaging). Deeper luminosity limits reduce the likelihood of an undetected faint companion.  Additionally, progenitors with higher masses (towards the right of the plot) are inherently less reliant on binary interactions for envelope stripping or tend to harbor more massive companions, so shallower companion limits can be still very constraining.

Following our fiducial $Z_\odot$ model, we predict a faint stellar companion is likely present at the SN site in most non-detections. For example, in the case of the Type Ic SN1994I, the likelihood of an existing stellar companion remains extremely high (at $\sim98\%$) despite the deep observational limits. This fact reflects the necessity of binary interaction to strip its low-mass progenitor, consistent with \citet{Nomoto+1994, van-Dyk+2016}.  Similarly high likelihoods are found for the Type Ib SN2008ax ($\sim92\%$) and the Ibn SN2015G ($\sim91\%$).% Although in these cases the limits are relatively shallow, the theoretical predictions still confirm that binary stripping via a companion is expected to be the main evolutionary scenario for most \sesne.  
The Ib/c SN2017gax also retains a high probability, $\sim80\%$, consistent with Su et al.  (in prep.).  A more moderate likelihood of $\sim65\%$ is obtained for the Ic‑BL SN2002ap, still supporting a binary evolutionary origin, albeit with a bit less certainty compared to the conclusions drawn by \Zt. 

An exception is SN\,2012fh, which likely originated from a progenitor of $5.6-10.56\Msun$, where deep observational limits for a companion become very constraining. As we show also in \citet{Williams+2025}, leveraging multiple observational constraints and a similar theoretical framework, the presence of any faint stellar companion is strongly disfavored below $9\%$. This calls for alternative scenarios, most probably involving a compact object companion (undetectable by \emph{HST} imaging) or the absence of one. The latter scenario includes mostly merger products, as true single-star WR progenitors remain unconfirmed \citep[e.g.,][]{Eldridge+2013,Van-Dyk+2018,Kilpatrick+2018b} and \sesne\ from prior binary disruptions are found to be negligible in our study  \citep[e.g., see also discussion in][]{Williams+2025}.

%Notably, for progenitor masses $M_{\mathrm{prog}} \lesssim 8,M_{\odot}$, the results remain essentially unchanged irrespective of the explodability prescriptions for \sesne, as these lower-mass progenitors are predominantly expected to explode anyway. %If BH implosions are included, 

% providing a theoretical framework to interpret current or future companion searches.

Interpreting these probabilities in the context of our population synthesis models provides some direct insight and constraints on possible mass transfer scenarios. It turns out that the probability of a surviving companion is highly sensitive to assumptions regarding mass transfer efficiency. More conservative mass transfer models predict more massive and luminous companions, thereby allowing for stellar companions to be ruled out even with shallower luminosity limits. 
%for ruling out the presence of a companion. 
Our default \posydon~model population implemented here leads to almost non-conservative mass transfer (Appendix~\ref{sec:beta}), so these probabilities should be interpreted as upper bounds.
%Consequently, the probabilities derived from our default \posydon\ population characterized by predominantly non-conservative mass transfer (see Section~\ref{sec:pop_synth}) should be viewed as upper bounds. 
 %
When employing \Zt\ population models, which incorporate thermally-limited or fully conservative mass accretion scenarios (\fig{fig:prob_nondetection_var}), the existence of nearly all stellar companions are effectively ruled out by current observational limits. Specifically, in fully conservative scenarios, the likelihood of a faint companion drops to a few percent for nearly all non-detections, except SN2017gax, which retains a non-negligible probability around 10\%, and SN2008ax %and SN2015G, 
which remain at $\sim 50\%$ due to shallower limits. 
%
%However, we emphasize that most observed \sesne, particularly those with low inferred ejecta masses such as SN1994I, SN2017gax, and SN2008ax, strongly favor  binary interactions with companions to explain their stripped progenitor envelopes. 
However, we emphasize that most observed \sesne, particularly those with low inferred ejecta masses such as SN1994I, SN2017gax, and SN2008ax, strongly favour binary interactions to explain their stripped progenitor envelopes, because if they were isolated progenitors their winds would be too weak to self-strip them in the first place.  
Thus, the assumption of fully conservative mass transfer ($\beta = 100$\%) strongly contradicts the observational evidence of 6, mostly deep, non-detections, as it effectively would exclude  the presence of nearly all expected stellar companions. This inconsistency strongly indicates that highly conservative mass transfer is improbable for the majority of \sesne, aligning with our conclusions in Sec.~\ref{sec:luminosity_cumulative} and discussed in Sec.~\ref{sec:conserv}. %This conclusion can be made from a population perspective approach 
Nevertheless, we advise caution before extending this conclusion to each individual \sesn. % high-ejecta-mass events like SN~2012fh. % high-ejecta-mass events like SN~2012fh.

% The above analysis pertains primarily to individual cases. However, adopting a broader population perspective, we emphasize that most observed \sesne, particularly those with low ejecta masses such as SN1994I, SN2017gax, and SN2008ax, strongly require binary interactions with companions to explain their stripped progenitor envelopes. When we apply the conservative mass-transfer assumptions \Zt\ models to the same \sesn\ sample, almost all visible companions would be excluded by our observational upper limits, leading to a contradiction. This inconsistency strongly indicates that highly conservative mass transfer is improbable for the majority of \sesne, aligning with the conclusions presented in Sec.~\ref{sec:luminosity_cumulative}. Nevertheless, caution is warranted when making such inferences based solely on individual, high-ejecta-mass events like SN~2012fh.

%%%%%%%%%%%%%%%%%%%%%%%%%%%%%%%%%%%%%%%%%%%%%%%%%%%%%%%%%%
\subsection{Position at the Hertzsprung-Russell Diagram }\label{sec:HRD}

\begin{figure}  
% \vspace*{-2.0 cm}
\begin{center}
 \includegraphics[width=0.99\linewidth]%{plots/annotated_HRD_companions_poster.pdf}
 {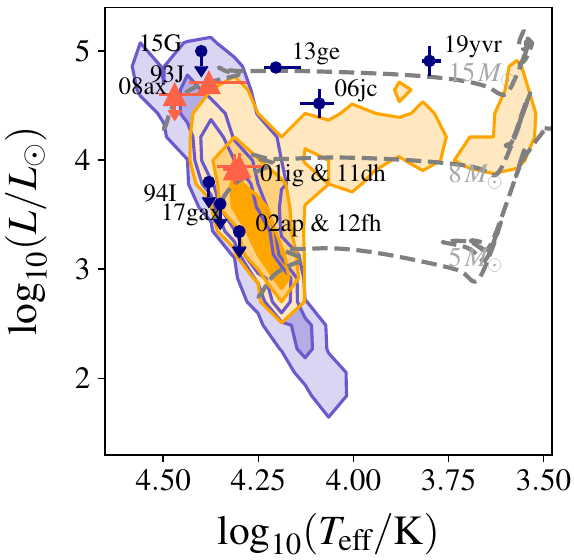}
 \caption{
Hertzsprung-Russell diagram showing detected or constraints on companion stars (points), overlaid on population synthesis predictions (contours). Blue represents companions next to \Ibc, and orange to \IIb. Contour levels denote the enclosed probability densities at 50\%, 95\%, and 99\% of $Z_\odot$. Note that the companions to SN2011dh and SN2001ig, as well as the upper limits of SN2002ap and 2012fh almost coincide. \label{fig:HRD_comp}
   }
  \end{center}
\end{figure}

\fig{fig:HRD_comp} shows the distribution of companions predicted by the \posydon~models within the HRD, with a majority of companion expected to reside on their MS. 
\Ibc\ companions are predicted to span a broad luminosity range on the MS region on the HRD from $\logL = 1.5$ to 5.2 (blue contours), whereas \IIb\ companions exhibit a more constrained distribution between $\logL = 2.5$ and 4.8 (orange contours).

We note that, for each empirical non-detection, the excluded region in the HRD is determined by both luminosity and effective temperature (\fig{fig:rule-out}), although here we present only the luminosity upper limit corresponding to the MS regime.  The three detected \IIb\ companions (SN1993J, SN2011dh and SN2001ig; red triangles) are found to reside in the upper-luminosity regime of the predicted MS companion population. 
%In contrast, detected MS companions of \Ibc\ are generally fainter, with many events limited to upper luminosity bounds. %This is because they originate from higher periods (at solar metallicity) and do case BC or case C and may merge more easily of it is extreme q. Type Ib/c are coming from caseAB or case B and can survive at more extreme q. \citet{Claeys+2011} IIb showed that q<0.7-0.8 ends up in contact and that is the reason that \citet{Yoon+2017} limited themselves to q=0.9.
%
However, for the three \Ibc\ detections, there is provisional evidence that they are  redder than expected for MS stars. %, apparently having evolved off the MS. 
The coolest companion is of SN2019yvr, though this is based on a tentative pre-explosion detection (the only one in our sample) of a composite SED from the progenitor and the companion. \citet{Jung+2022} suggest that the progenitor dominates the observed SED, so confirmation of the companion’s properties requires waiting until the SN has sufficiently faded. SN2006jc is found in the Hertzsprung Gap, and SN2013ge lies near or just beyond the TAMS phase of massive single stars. 

Evolved, post-MS, companions are a rare prediction of our models, occurring in fewer than 4\% of companions at all metallicities, and almost always occurring during their RSG phase, not in a transition through the Hertzsprung Gap (Sec.~\ref{sec:pie_chart}). In addition, such evolved companions would have been more likely in \IIb\ systems ($8\%$ of all companions; Fig.~\ref{fig:bin_frac_Z}), as they originate from binaries with closer to equal masses (particularly at solar metallicity), increasing the likelihood of a post-MS companion (orange contours).   %Further analyses providing two-dimensional constraints with HST in luminosity and effective temperature is discussed in Sec.~\ref{sec:HRD} and has been further investigated in the studies of SN2015G \citep{Sun+2020}, SN2017gax \citep{Su+2025}, as well as generally for our broader sample in Su et al. in prep.  %\citep{Su+2025b}.
Although their expected high rotational (\fig{fig:centerhe_omega}) may help explain the borderline case  of SN2013ge, it cannot explain the other two events. 
 For the two post-explosion detections, an intriguing possibility is that a luminous, redder-than-MS source may be observed for years to decades following the SN, due to interaction between the SN ejecta and the MS companion that is driven out of thermal equilibrium, becoming more luminous and redder for a timescale of years to even more than a decade \citep{Hirai+2015, Liu+2015, Hirai+2018, Ogata+2021, Chen+2023, Hirai2023}.% a possibility we further discuss below.

\subsection{Possibility of interaction of supernova ejecta with a surviving stellar companion}\label{sec:ejecta_interaction_companion}

 \begin{figure}  
% \vspace*{-2.0 cm}
\begin{center}
 \includegraphics[width=.99\linewidth]{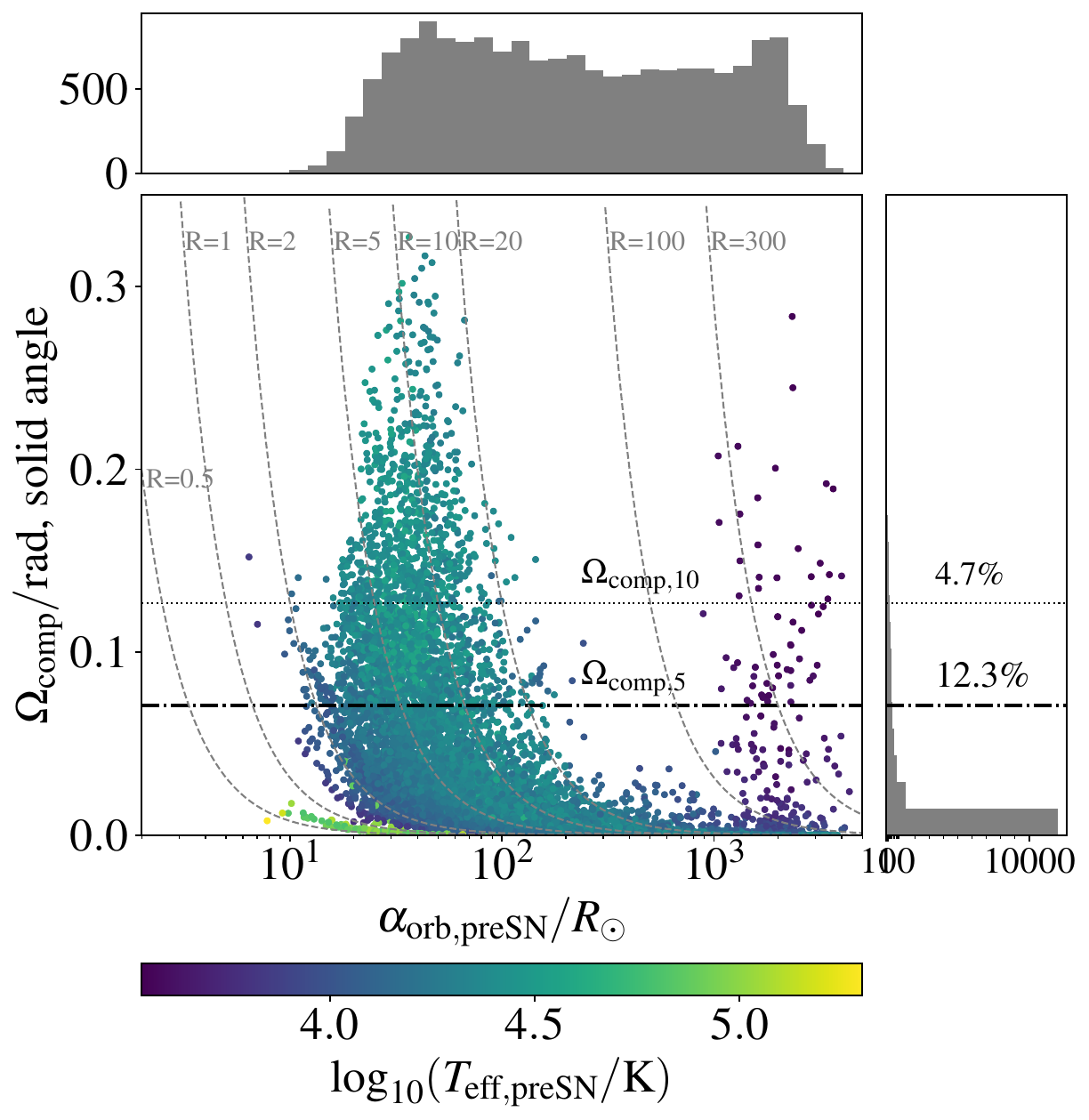}
 \caption{
Solid angle of the stellar companion $\Omega_{\rm comp}$ as a function of \sesn\ preSN orbital separation,  for $Z_\odot$. Grey dashed curves represent constant companion radii, in solar units. We also show rough solid angle thresholds above which the most applicable model of \citet{Ogata+2021} would reside on the cool side of their TAMS, for more than $\sim 5$ (dot-dashed line) and $\sim 10$ (dotted line) years after SN (see text), as well as their occurrence rate on the right-hand distribution. \label{fig:solid_angle}
%The calculated solid angles for the a_sep=[] and Rcomp = 6 (of Ogata+2021) are [1.2566,\ 0.2894,\ 0.1269,\ 0.0711,\ 0.0315]
}
  \end{center}
\end{figure}

We explore the possibility that a SN will impact a companion by calculating 
%The companions of \sesne\ are expected to be influenced by the interaction with SN ejecta, including getting stripped of mass from their surface, having ejecta mass polluting their surface abundances, experiencing a momentum kick, and most importantly here, injecting energy which temporarily brings them out of thermal equilibrium. 
%
%To estimate the possibility of SN interaction with a surviving companion, we show in \fig{fig:solid_angle} 
the solid angle of the surviving companion from the reference point of the exploding star, 
\begin{equation}
    \Omega_{\rm comp} = 2\pi \left(1 - \sqrt{1 - \left(\frac{R_{\rm comp}}{\alpha_{\rm orb,preSN}}\right)^2} \right),
\end{equation}
%$\Omega_{\rm comp} = 2\pi \left(1 - \sqrt{1 - \left(\frac{R_{\rm comp}}{\alpha_{\rm orb,preSN}}\right)^2} \right)$. $R_{\rm comp}$ 
%
where $R_{\rm comp}$ is the companion's radius at the moment of explosion, and %, accounting for both spin-up due to mass accretion and its stellar evolution and rejuvenation up to that point (see Sec.~\ref{sec:HRD}), and %, which is commonly at their early MS but with a range up to TAMS. 
$\alpha_{\rm orb,preSN}$ is the orbital separation at the moment of explosion.  
\fig{fig:solid_angle} shows the results. The effect of the ejecta-companion interaction (ECI) with the SN ejecta would roughly increase with $\Omega_{\rm comp}$ \citep{Hirai+2015, Hirai+2018, Ogata+2021, Chen+2023, Hirai2023}. 
To provide a rough threshold for the solid angle required to push the companion out of thermal equilibrium, causing it to appear more luminous and cooler for a significant timescale to be observable, we use as an example reference the 10 \msun model of \citet{Ogata+2021}. This model has a pre-interaction luminosity of $\logL \sim 4$, typical of our MS companion population and near the peak of the companion mass distribution (see \fig{fig:mass_distr}). They also adopt a SN ejecta mass of $M_{\rm ej} = 2,\Msun$, representative of typical \SESNe. 
Their models show that a companion stays cooler than its TAMS temperature for longer than a  typical SN fade time of $\sim 5$ years only when the binary separation is $\lesssim 40~R_\odot$, %(circles on the dashed lines in their Figure~4),  
corresponding to a solid angle $\Omega_{\rm comp,5}$ of 0.071 (for their example $6~ R_\odot$ model), 
which we adopt here as a rough lower limit for observability of such an effect. We also include a more extreme ECI case with a separation of $\sim 30~R_\odot$, where the companion subtends a larger solid angle of $\Omega_{\rm comp,10} \sim 0.127$ and returns to near-MS $T_{\rm eff}$ roughly 10 years after the SN (although note that \citet{Chen+2023} require even larger solid angles for decade-long signatures). 
%If we have picked lower companion mass and radius, corresponding 
%
%These two solid angle thresholds, 0.071 and 0.127, correspond to ECI signature durations of $\sim 6.4$ and $11.4$ years, respectively, based on \citet[][Eq.~5]{Hirai2023} for a $10~M_\odot$ companion and explosion energy of $10^{51}$~erg. However \citet{Chen+2023}, who modeled the ECI observables for the inferred binary configuration of SN2020oi \citep{Gagliano+2022}, suggest that even larger solid angles may be required for decade-long signatures. 

 %The vast majority reside at sufficiently large separations that ECI is unlikely: approximately 95.3\% of companions lie below the $\Omega_{\rm comp,10}$ threshold. This limit roughly maps to a ratio $\alpha_{\rm orb,preSN}/R_{\rm comp} \sim 5$, consistent with the findings of \citet{Liu+2015}, who report that 95\% of systems fall below this cross-sectional threshold. 
%
We find that 12.3\% of all companions have $\Omega_{\rm comp} > \Omega_{\rm comp,5}$, meaning they could exhibit ECI-induced observables for at least five years post-explosion, according to our simplified model. The fraction of companions with possible decade-long effects (i.e.,  $\Omega_{\rm comp} > \Omega_{\rm comp,10}$) is 4.7\%, consistent with the findings of \citet{Liu+2015}. These fractions increase slightly at lower metallicities, owing to reduced wind-driven orbital widening, which dominates over the reduced companion radii at lower $Z$ \citep[e.g.,][]{Xin+2022}. %, thus generally results in tighter pre-SN configurations. 
For example, at $0.1~Z_\odot$ we find that 15\% and 9.3\% of companions exceed the $\Omega_{\rm comp,5}$ and $\Omega_{\rm comp,10}$ thresholds, respectively. Companions that are ECI candidates are typically MS stars of $10$–$30~\Msun$ with pre-SN radii of $5-10 \, R_\odot$, often located in the later stages of their MS evolution, sometimes near the TAMS, and orbiting at pre-SN separations of $15 \lesssim \alpha_\mathrm{orb, preSN}/R_{\odot} \lesssim 100$. Exceptions include 
%hot, compact helium star companions with $R_{\rm comp} < 1 \, R_\odot$ that exceed the rough ECI threshold due to being very close to the SN progenitor typically after of a successful double common-envelope ejection. as well as a small number of 
RSG companions in wide orbits, due to their radial expansion after core helium exhaustion. Across the entire population the maximum solid angle attained is $\Omega_{\rm comp} \sim 0.35$.

The fraction of companions exceeding the ECI solid angle thresholds is significantly lower for \IIb\ systems at $Z_\odot$, with only $\sim 3\%$ and 1\% surpassing the $\Omega_{\rm comp,5}$ and $\Omega_{\rm comp,10}$ thresholds, respectively (and involving mostly already giant progenitors at wider orbits). This is primarily because \IIb\ progenitors at high metallicity tend to originate from wide initial orbits, resulting in correspondingly wide pre-SN separations \citep{Yoon+2017, Sravan+2019}. At lower metallicities, where \IIb\ systems are more prevalent, their more massive progenitors still experience some orbital widening, though reduced due to weaker winds, leading to only modest increases in solid angle fractions. We therefore argue that \IIb\ systems are not the best candidates for detecting ECI signatures, being roughly four times less likely  than \Ibc\ to exceed the observable ECI threshold compared.

Overall, the theoretical model presented here suggests that $\sim12\%$ of our companions are expected to experience ECI effects long enough to be observable at least for a few years post-explosion. That corresponds, observationally, to about one ore even two objects in our sample of 11 post-explosion companion searches.  %, not including 2019yvr 
%(and of 7 of \Ibc\ specifically) has experienced this. 
The Type Ibn SN~2006jc \citep[as suggested also by][]{Sun+2020, Ogata+2021} or the Type Ib/c SN~2013ge are the obvious candidates, as their companions lie within the Hertzsprung Gap. Follow-up observations at their SN sites could confirm or refute this scenario. SN2019yvr cannot be explained through ECI as it was inferred to be red and evolved even before the explosion. %In contrast, if the existence of a cool companion is confirmed for 2019yvr \citep[although see][]{Jung+2022}, it cannot be explained through ECI as it was inferred to be red and evolved even before the explosion. 

The ECI-induced expansion of a companion may cause it to fill its Roche lobe and interact with the newly formed compact object. This appears to be the case for SN~2022jli, as an explanation to its  periodic light curve variability \citep{Moore+2023,Chen+2024}. It is thus crucial to search for a potential companion in the coming years as the SN fades, while it may still be in the ECI-inflated state, to confirm and further study this phenomenon.

\section{Discussion}\label{sec:discussion}

\subsection{Constraints on the physics of \sesn\ progenitors}\label{sec:binary_physics}

Our analysis in Sec.~\ref{sec:results} provides direct constraints on the physics of \sesn\  progenitors and the associated binary star evolution. Our observational sample disfavors key aspects of the physical scenarios proposed to produce \Ibc\ and \IIb. Below, we discuss several of the foremost implications.

\subsubsection {Evidence against highly conservative mass accretion}\label{sec:conserv}

%Our analysis (Sec.~\ref{sec:luminosity_cumulative}, and also Sec.~\ref{sec:upper_limits}) 
In Sec ~\ref{sec:luminosity_cumulative} and ~\ref{sec:upper_limits} we show that the observational sample is found in strong tension with fully conservative mass transfer, %and points towards fairly inefficient mass accretion onto the companions, 
as these models consistently overpredict luminous companions and are challenged by the number of non-detections. %next to \sesne. 
%, more than found in the observed sample. 
In contrast, low or moderately conservative models, such as the rotationally-limited scenario of \posydon\ (with $\langle \bar{\beta}_{\rm rot} \rangle \sim 4\%$; Appendix~\ref{sec:beta}),  the fixed $\beta = 0$ and especially the 30\% adopted in \Zt, provide a better fit to the data within uncertainties, particularly when assuming that not all massive WR stars successfully explode. Still, the high detection fraction of companions in our \IIb\ sample (Sec.~\ref{sec:bin_frac_Z}), and their the high luminosity (such as SN1993J, \fig{fig:HRD_comp}) may indicate higher mass transfer efficiencies particularly for these progenitor systems. Given the uncertainties in all binary models, the sensitivity of the luminosity distribution to model assumptions, and the limited observational sample and its own uncertainties, we refrain from tightly constraining a best-fit accretion efficiency. For example, if the SN2019yvr companion is ruled out by future observations, and those of SN2006jc and SN2013ge are found to be brightened by ECI, the cumulative distribution shifts toward lower luminosities, favouring lower mass accretion efficiencies. 

Event-by-event modelling of individual \sesne, though less powerful than our statistical population approach, converges on the same picture: companions usually accrete only a modest fraction of the overflowing envelope. %Events with stronger constraints generally support low mass-transfer efficiency. 
SN1994I, 2002ap, 2011dh, 2013ge, and 2017gax each favour highly non-conservative transfer to explain either the binary system's properties or the companion's non-detection \citep[][Su et al., in prep. and \citetalias{Zapartas+2017b}]{van-Dyk+2016, Maund2019, Fox+2022}. SN2001ig also suggests low efficiency, based on the HRD position, possible lack of rejuvenation, and the presence of CSM \citep{Ryder+2018}. For SN1993J, revised photometry and models yield a lower inferred companion mass of $\sim$15$\Msun$ (\fig{fig:HRD_comp}), removing the need for the high accretion required to reach the $22~\Msun$ value inferred by \citet{Maund+2004}. Non-accreting scenarios are also supported by \citet{Hirai2023} for SN2006jc, although potentially involving common envelope. \citet{Sun+2020} explored a near-equal-mass scenario for it, allowing a range of efficiencies tied to two possible SN ages inferences. We favour the younger age based on the SN's high ejecta mass, which indirectly supports low-efficiency transfer. SN2012fh likely lacks a stellar companion and thus offers little leverage \citep[][and Sec.~\ref{sec:upper_limits}]{Williams+2025}, while the remaining events provide no meaningful constraints.

Our results align more closely with the low to moderately efficient accretion inferred for massive WR binaries \citep[][probing higher masses compared by the typical binary \sesne\ progenitors]{Petrovic+2005, Shao+2016, Nuijten+2025}, rather than with highly conservative scenarios in low-mass stripped donors \citep{Pols2007, Schootemeijer+2018b, Lechien+2025}, below the threshold for core collapse. %\citep{Podsiadlowski+2004, Tauris+2015}. 
Thus, our results may point towards a mass-dependent accretion efficiency, which decreases for higher donor masses.

% Recent observations have identified binary companions next to the first empirical detection of stripped stars, with typical companion masses in the range of 1–4~M$_\odot$ \citep{Drout+2023,Gotberg+2023}, lower than our resulting mass distribution (\fig{fig:mass_distr}). However, their findings are largely biased toward low-mass stripped stars that are not expected to explode as core-collapse \sesne. In addition, their observational selection favors systems with low-mass, faint companions, otherwise the stripped star would not be observable as the companion would dominate the system’s luminosity \citep{Gotberg+2018}. 

\subsubsection{Non-explodability of massive stripped-envelope cores}\label{sec:non_expl}

%Model variants in which \emph{all} systems explode are disfavoured (Sec.~\ref{sec:luminosity_cumulative}; dashed lines in \fig{fig:cumulative}). 

%Together with constraints on the mass accretion efficiency, permitting the most massive stripped helium cores, expected to be observable as luminous WR stars, to collapse without a bright transient improves the fit for all $f_{\rm bin}>0.5$. 
Sec.~\ref{sec:bin_frac_Z} discusses the high fraction of targets found to have a companion, which we interpret to argue against the ability of the most massive stripped helium cores, such as WR stars, to explode. A successfully exploding WR population would have resulted in a significant fraction of born-single, merger, or ejected \sesn\ progenitors that lack any companion at explosion. On top of that, massive progenitors that are in binaries at explosion tend to have more luminous stellar companions  (e.g., the gap between the two black lines in the \posydon\ models in the luminosity distribution; \fig{fig:cumulative}), which we do not observe in a high fraction (Sec.~\ref{sec:luminosity_cumulative}). These results are consistent with the lack of WR progenitors in the literature \citep[e.g.,][although see \citealt{Gal-Yam+2022}]{Eldridge+2013} as well as empirical evidence for failed explosions of hydrogen‐poor massive cores \citep{De+2024}. 
%They are also in the same direction as the conclusion in \citet{Zapartas+2021b}, where similar single-star models as in this study have been used.   In summary companion searches, both individually and approached as a population in this work, provide an additional argument towards the significance of binary progenitors for \sesne.
%

In contrast, \sesn\ progenitors stripped in binaries typically form lower‐mass helium cores that tend to explode, with binary interaction reshaping their final  stellar structure potentially further enhancing explodability \citep{Laplace+2021, Vartanyan+2021}. We capture these effects in our default \posydon\  model via the parametrized prescription of \citet{Patton+2020} coupled with \citet{Ertl+2016} explosion criteria, as shown in \citet{Schneider+2021, Patton+2022}. 
%

%To investigate the uncertainty in the explodability of massive stars, 
To investigate these effects further, we go on to vary the SN prescriptions in our \posydon\ models. Changing our assumption of explodability to one based on final helium core mass following \citet{Sukhbold+2016} (calibrated according to the ``N20'' engine) leads to more optimistic explosion outcomes for \IIb\ at $Z_{\odot}$, consistent with \citet{Zapartas+2021b}. In this scenario, nearly half of \IIb\ originate from massive, single-star WR progenitors, reducing the contribution of MS companions to about $40\%$, although has little effect on \Ibc. This may help explain individual cases such as the Type IIb CasA remnant, which seems to have no stellar companion (see Sec.~\ref{sec:remnants}), but overall has little impact in our interpretation because \IIb\ are more important at lower metallicities \citep{Kelly+2012, Sravan+2019}.
%the overall impact on \sesn\ is limited, as the most massive WR candidate progenitors of \Ibc\ still fail to explode, so companion predictions remain largely unchanged. 

At low metallicity, weaker winds suppress self-stripping and when stripped cores do form, they often collapse into BHs even under the more optimistic SN prescription of \citet{Sukhbold+2016}. A notable effect is the increase in BH companions next to exploding secondaries in \IIb\ systems, rising from negligible at $Z_\odot$ to $\gtrsim 10\%$ at $0.1\,Z_{\odot}$ and up to $20\%$ with \citet{Sukhbold+2016}. %\other{Why...?}. 
The \citet{Fryer+2012} model has milder effects, aligning more with the results of companions following \citet{Patton+2020}. Still, we emphasize again that BH formation does not necessarily proceed without observable signatures, even in the case of very compact or high-mass progenitors, as sustained accretion can keep neutrino luminosity high  \citep{Ott+2018,Kuroda+2018, Burrows+2023}. At the same direction, newer, more optimistic models, such as of \citet{Maltsev+2025}, need to be explored as they may further increase the occurrence of single WR exploding \sesne.

%Our results are consistent with not detecting WR as progenitors of \sesne\ \citep[][although see \citealt{Gal-Yam+2022}]{Eldridge+2013} as well as empirical evidence for failed explosions of hydrogen‐poor massive cores \citep{De+2024}. They are also in the same direction as the conclusion in \citet{Zapartas+2021b}, where similar single-star models as in this study have been used.   In summary companion searches, both individually and approached as a population in this work, provide an additional argument towards the significance of binary progenitors for \sesne.

%%%%%%%%%%%%%%%%%%%%%%%%%%%
\subsubsection{Common envelope scenarios and SN kicks}\label{sec:ce_kicks}

Models for our default binary assumptions also show that a CE channel does not contribute significantly to the \sesne\ population. This is because CE is found to be subdominant compared to stable mass-transfer, and if it is triggered, these systems are less likely to survive. 
Given the predominance of stable mass transfer, a natural correlation between the stripped progenitor mass and the companion luminosity emerges (e.g., the trend can be seen at \fig{fig:prob_nondetection}). Extreme mass ratios are less likely, as they would trigger CE and likely mergers.

CE produces only \Ibc\ in our \posydon\ simulations because, by definition, it strips a star deep enough to expose its hydrogen-deficient core \citep[i.e., surface hydrogen mass fraction $<1\%$;][see Sec.~\ref{sec:method}, although see also \citealt{Wei+2024}]{Fragos+2023}. 
At $Z_\odot$, CE contributes merely $\simeq 3\%$ to the predominant channel of  \Ibc\ with stellar companions. This increases to $\sim6\%$ of \Ibc\ for $0.1 Z_\odot$, although we note the decreased likelihood of this subtype at lower metallicity (Souropanis et al., in prep.).  Surviving companions in the CE scenario tend to be faint ($\log L/\mathrm{L_\odot} \lesssim 2.5$), with solar-type companions being extremely rare. %As CE is quite inefficient in producing \sesn\ progenitors we only 
A variant of the fiducial model with a higher $\alpha_{\rm CE}=5$, compared to the default value of unity, yields only a modest rise in low-luminosity companions at $Z_\odot$: from 8\% to 9.5\% below $\log L = 2.5$, and from 2.3\% to 3.3\% below $\log L = 2$. Even with $\alpha_{\rm CE}=5$, the fraction of \sesn\ from secondaries with compact object companions (almost exclusively BHs) remains low, with CE still accounting for only $\simeq 15\%$ of these events. The rest arise from stable mass transfer.

SN kicks affect only the second explosion in a system, most likely whether the secondary will stay bound and get stripped by the remaining compact object companion. Assuming a reduced NS kick dispersion of $\sigma = 60$\,km\,s$^{-1}$ only slightly increases the possible NS companions for \Ibc\ (to a few \%). Even if the system survives disruption it will most likely not avoid merging when reverse mass transfer is triggered, due to the extreme mass ratio of the companion with the NS.   
%For $Z_{\odot}$ a slight increase of BH companions for both \Ibc\ ($>10\%$) and \IIb\ ($\sim 5\%$) again in the expense of MS companions and no companions, but the main conclusions stay the same. 
Potentially a combination of low kicks and high $\alpha_{\rm CE}$ would increase the number of NS companions significantly. These systems could be progenitors of ultra-stripped SNe \citep{Tauris+2015} and double neutron stars \citep[e.g.,][]{Andrews+2019}.

%%%%%%%%%%%%%%%%%%%%%%%%%%%%%%%%%%%%%%%%%%%
\subsection{Companions of specific SN subtypes}\label{sec:subtypes}

In this work we grouped all \sesne\ together, separating them into \IIb\ and Type Ib/c categories. However, many of the observed events exhibit distinctive observational features that place them into more specific subtypes. 
In particular, SN2006jc and SN2015G, are classified as Type Ibn due to clear signatures of interaction with circumstellar material (CSM). The physical origin of this CSM remains uncertain. In our analysis, we do not model or account for mass loss from the progenitor system and do not evolve the systems up to the time of explosion, as our simulations are halted at core carbon depletion. Recent studies suggest that Type Ibn SNe originate from binary progenitors undergoing Roche-lobe overflow at the time of explosion, either with a stellar or a compact-object companion \citep{Wu+2022, Ercolino+2025, Moriya+2025}. % implying the presence of a companion, possibly embedded within the CSM. 
If this late mass loss episode occurs within years to decades prior to core collapse, some of our \sesne\ could appear observationally as Type Ibn events. Given that SN2006jc is also the main candidate for ECI, its progenitor system warrants particular attention. Alternatively, if the mass transfer occurs earlier, they may instead exhibit a transitional behavior from Type Ib to Type II, as observed in SN~2014C \citep{Margutti+2017}. %\other{@Dimitris How many at RLOF at our final POSYDON models?}

On the other hand, SN2002ap is a Type Ic-BL, a subtype commonly linked to rapidly rotating collapsing cores, eventually forming a BH. SNe Ic-BL are both observationally \citep[e.g.,][]{Galama+1998,Hjorth+2003,Modjaz+2016} and theoretically \citep{MacFadyen+Woosley1999,Woosley+2006a, Gottlieb+2024} %Central engine? \citep{Rodriguez+2024, Soker+}. 
associated with long gamma-ray bursts.  
In such cases, our framework, which assumes successful supernova explosions occur only in conjunction with NS formation, may not be directly applicable and may exclude these \sesne\ and their potential companions. However, binary interactions, such as mass accretion \citep[e.g.,][]{Cantiello+2007} and tidal spin-up \citep[e.g.,][]{Bavera+2022}, may still be essential for preserving the high angular momentum required at core collapse. The formation channels leading to such events, including the potential for alternative evolutionary pathways, have been investigated with \posydon\ in \citet{Briel+2025}.

% While this study focuses on \sesne, we note that no companions have been identified to date in association with Type II SNe, despite several targeted searches in nearby SNe \citep[e.g.,][]{Qin+2024,  Bell+Kilpatrick2024} and remnants \citep[SN~1987A and Crab;][]{Kochanek2018}. This observational outcome remains consistent with the predictions of \citet{Zapartas+2019}, which suggest that most Type II SNe are not expected to retain a companion at the time of explosion. However, non-detections do not rule out a possible binary origin of a Type II SN, as many such systems may have undergone binary interactions that resulted in either mergers or binary-ejected stellar progenitors \citep{Podsiadlowski+1992, Zapartas+2019,  Bostroem+2023, Wagg+2025}. 
% %
% Nevertheless, several scenarios still allow for the presence of a surviving companion even for Type II SNe. These include systems with very wide separations or former triple systems that evolve without significant interaction, potentially retaining a distant, non-interacting companion. Additionally, partially stripped yet still hydrogen-rich progenitors have been proposed as plausible channels for SNe exhibiting short-plateau Type II-P or Type II-L light curves and spectral features \citep{Morozova+2015,  Eldridge+2018, Ercolino+2024, Dessart+2024}, in which case a stellar companion responsible for the partial stripping would be expected.

While this study focuses on \sesne, we note that no companions have been identified to date for Type II SNe, despite several targeted searches in nearby events and remnants \citep[e.g.,][]{Qin+2024, Bell+Kilpatrick2024, Kochanek2018}. This aligns with predictions by \citet{Zapartas+2019}, who suggest most Type II SNe lack close companions at explosion, without precluding a prior binary history. Surviving companions may be identified even for Type II SNe if they are in very wide orbits or in prior triple progenitor systems \citep[this should be the cases of companions found around RSGs; e.g.,][]{Patrick+2025}, or alternative in partially stripped hydrogen-rich progenitors linked to short-plateau Type II-P or II-L SNe \citep{Morozova+2015, Eldridge+2018, Ercolino+2024, Dessart+2024}.

\subsection{Companion fate after the supernova}\label{sec:remnants}

%The typical high stellar spin of this companion may give rise to Be phenomena for them \citep{}.
Most companions next to \sesne are still early in their evolution and their typically high stellar spin (\fig{fig:centerhe_omega}) could potentially cause them to appear as Be stars \citep{Rivinius+2013, Rocha+2024}. % may give rise to Be phenomena for them. 
After explosion most of them are expected to become unbound from the stellar system \citep{Blaauw1961,Eldridge+2011,Renzo+2019,Wagg+2025},  with velocities ranging from a few to several tens of $\mathrm{km/s}$. %\MZ{How many of them get disrupted in POSYDON?} 
These objects would be classified as walkaway or runaway mass accretors, such as the case of $\zeta$ Ophiouchi \citep{Neuhauser+2020,Renzo+Gotberg2021}.
 While such velocities do not displace the companion from the SN center for an extragalactic source in the first years to decades after explosion, they become relevant when attempting to locate it within an expanding %galactic 
 SN remnant thousands of years post-explosion. 

 If some \sesn\ form magnetars, then we would expect a small fraction of companions to remain bound in orbit to them \citep{Chrimes+2022}. However \citet{Sherman+2024} investigated the occurrence of both bound and unbound companions linked back to magnetars and found fewer than predicted by binary population models, suggesting that many of these systems may undergo merging prior to explosion \citep[e.g.,][]{de-Mink+2014}. 

If a the star remains bound after the explosion, the system could evolve into an intermediate- or high-mass X-ray binary \citep[e.g.,][]{van-den-Heuvel+De-Loore1973, Tauris+van-den-Heuvel2006, Misra+2020}. However, staying bound is expected in only the minority of the systems \citep{Eldridge+2011, Renzo+2019}, as found also empirically in nearby SN remnants \citep{Kochanek+2019}. Most bound systems will most likely harbor a BH, with the possibility of the system to undergo future reverse mass transfer, leading to a scenario in which the secondary star explodes as a \sesn\ while orbiting a BH companion. This configuration accounts for approximately 10\% of \sesne\ in our study (\fig{fig:pie_chart}), and represents possible progenitor systems for future gravitational wave sources, mostly avoiding CE evolution \citep{van-den-Heuvel+2017}.
 
Multiple searches exists for surviving companions in galactic SN remnants \citep[e.g.,][]{Hoogerwerf+2001, Tetzlaff+2014, Dincel+2015,Boubert+2017}. Several candidate companions have been identified and traced back to the explosion point, although much of the information about the original SN, including its Type, is unknown. Some of these candidate companions are B- and O-type stars, which is expected from the modeled mass distribution that ranges from 3-20 \Msun (Fig.~\ref{fig:mass_distr}), but few of them are inferred to have lower masses. If these stars are in fact the true surviving companions, this would imply that \sesne\ channels involving CE, following unstable mass transfer from a progenitor in an extreme mass ratio binary, are more prominent than predicted in our fiducial scenario, or alternatively, that multiple stellar systems are involved in the production of \sesne\ \citep[e.g.,][]{Moe+2017}. 

In contrast, \citet{Dincel+2024} explored a sample of 12 SNRs, with at least 7 being of core-collapse nature, and found no definitive OB companion.  
Even though the majority of these remnants may be hydrogen-rich Type II SNe (whose progenitors most likely explode in isolation, Sec.~\ref{sec:subtypes}), the absence of any massive runaway in even a few remnants potentially associated with \sesne\ remains on the low side. 
However, the small-number statistics preclude any firm conclusions, and the authors discuss various possible observational and physical biases that could hinder such identifications.

One final important case study is that of CasA, which has been classified as a \IIb\ from light echoes \citep{Rest+2008}. Searches have ruled out the presence of a luminous binary companion down to very deep detection limits \citep{Kochanek2018,Kerzendorf+2019}. \citet{Kerzendorf+2019} suggested the most likely scenarios were a compact object companion or no companion at all. %, providing a key constraint. 
It is interesting to note that, in the absence of a MS companion, our fiducial model results for \IIb\ at $Z_\odot$ 
(\fig{fig:pie_chart}) favor either the merger scenario (dotted and star hashes) or the single-star channel (X-pattern hashes). The model predicts a much lower probability for a compact object companion because, as previously mentioned, binary mass stripping followed by winds at $Z_\odot$ typically leads to \Ibc\ progenitors rather than \IIb. 

It is worth noting that the relative importance of the allowed channels remains sensitive to uncertainties in the merging process, wind mass-loss rates, and progenitor explodability, making any conclusions tentative. A more optimistic prescription for successful SN explosions could further favor scenarios of isolated progenitors (see Sec.~\ref{sec:non_expl}). 
Further empirical constraints may be possible from ongoing studies of CasA’s geometry and nucleosynthesis \citep[e.g.,][]{Orlando+2022,Sakai+2024,Milisavljevic+2024,Bamba+2025}.

%It is worth considering that if a fraction of the \sesn\ population forms magnetars, then we would expect a small fraction of companions to be  bound in orbit to them \citep{Chrimes+2022}. However \citet{Sherman+2024} investigated the occurrence of both bound and unbound companions linked back to magnetars and found fewer than predicted by binary population models, suggesting that many of these systems may undergo merging prior to explosion \citep[e.g.,][]{de-Mink+2014}. 

%If a the star remains bound after the explosion, the system could evolve into an intermediate- or high-mass X-ray binary \citep[e.g.,][]{van-den-Heuvel+De-Loore1973, Tauris+van-den-Heuvel2006, Misra+2020}. However, this is expected in only the minority of the systems \citep{Eldridge+201, Renzo+2019}, as found also empirically in nearby SN remnants \citep{Kochanek+2019}. Most bound systems will most likely harbor a BH, with the possibility of the system to undergo future reverse mass transfer, leading to a scenario in which the secondary star explodes as a \sesn\ while orbiting a BH companion. This configuration accounts for approximately 10\% of \sesne\ in our study (\fig{fig:pie_chart}), and represents possible progenitor systems for future gravitational wave sources, mostly avoiding CE evolution \citep{van-den-Heuvel+2017}.

%%%%%%%%%
\subsection{Observational caveats}\label{sec:obs_caveats}

\subsubsection{Cluster or chance alignment pollution in our observational sample detections}

%\MZ{Important only for cases of detection} \other{Below is text from @Ori in an HST proposal + our 13ge paper, I think on the optimistic side needed in a proposal} \MZ{it is also too HST specific (makes sense) but we probably need to make it more generic and looking into the future missions too.}
There is always an underlying question about whether one can be sure that a detected source is, in fact, the surviving companion. One possibility is that some of the putative companions in our sample may actually be clusters or even just a handful of stars within the cluster. Without analyzing each case individually, we consider a couple of general examples. The observations presented in this paper are obtained with HST/WFC3, which has a pixel-scale resolution of $0\arcsec.0396$ pixel$^{-1}$. For a typical distance in our sample of 20 Mpc (\fig{fig:luminosity-distance}),  
%(a reasonable upper limit given the distance distribution in our sample), 
this corresponds to a physical scale of %print,np.tan(0.0396/206265)*20.e6
$\sim$3.8 pc pixel$^{-1}$ (at smaller distances, the physical scale per pixel decreases, and vice versa).  RU 141, which can be used as a representative compact cluster or association, % Ru 141 
has a physical radius of $\sim$7 pc~and has a single dominant star \citep{Camargo+2009}, which would imply a roughly $\sim$30\% chance of that star landing on a pixel associated with a putative companion at 20 Mpc. NGC 2004 (20 Myr and $R_{\rm eff} \approx 6$~pc) is one of the worst case scenarios with $\sim$10 stars brighter than $M_{\rm F275W} > -5$  \citep{Niederhofer+2015}.  At a distance of 20 Mpc, such a cluster would be spread over $\sim$4 pixels, implying the potential for up to 3 bright stars per pixel.  However, in both cases, the SEDs show that because some of the cluster stars are evolved off the MS, any composite would likely appear a bit redder than a single star. Second, a typical OB association is often spread over a much larger area than the examples given above \citep[$\sim$10-20 pc;][]{Smith+2010}.  Third, %and perhaps most important, 
most of the targets in our sample are at distances $\lesssim$20 Mpc, which greatly decreases the likelihood of either scenario above. We therefore conclude the likelihood of a chance alignment for any of our putative companions is $\ll10\%$.  This may still be something to be cautious about for a single event, but it becomes less significant when considering a population, as in this paper. It is worth noting that overall, searches for surviving companions won't improve much with JWST. The NIRCam pixel scale is comparable to WFC3/UVIS. Furthermore, the SEDs of bright, blue OB-type stars tend to decrease significantly in the infrared. Any benefit of JWST's bigger aperture is lost at longer wavelengths, except perhaps in the case of a reddened companion. A future detailed study is needed to explore this phase space.
%

%If a detected source is more likely an unresolved cluster, we can still derive properties of the local stellar population's age and mass, since any undetected companion must be among the most massive stars in the cluster (assuming a similar mass as the SN progenitor).  Even in cases where we inadvertently interpret several stars as a single star, we would only overestimate the age of a massive star by 2$\times$ \citep{Girardi+2010}.  Thus, our most conservative approach would be to assume that inferred ages are lower limits. 
%
%\other{@Ori, should I add the distance in Table 1, to argue here that the closer detections are more certain, whereas that far away ones can be clusters?}
%\MZ{Bayesian analysis from Boubert+}
%https://ui.adsabs.harvard.edu/abs/2017A%26A...606A..14B/abstract
%\other{@other Should we discuss also the possibility of a random stellar source in chance alignment Say something about the density of the region? Maybe mention that JWST's potential due to higher spatial resolution , for future proposals?}

%Chance alignment is an issue but it is mostly important "per event" but its significant should diminish at population level which is the focus of this paper, becoming irrelevant. 

%
Nonetheless, let's consider here the possibility that some of the putative companions in our sample are, in fact,  chance alignments. This would mean that any actual companion for that event would be even fainter or non-existent all-together. In the former case, the results would shift the overall distribution in \fig{fig:cumulative} towards lower luminosities, supporting channels of even less conservative mass transfer or CE scenarios.

\subsubsection{Shock interaction and light echoes}

Another possibility is that a detected source is actually the result of ongoing interaction between the forward shock and a dense circumstellar medium (CSM) or an optically reflected light echo. For most of the targets in our sample, the color, brightness, and shape of the SED were used to distinguish between these possibilities. For example, CSM interaction typically produces a flat SED, while a stellar companion, especially a blue star, shows a sharp decline beyond 3000–4000~\AA. However, in some cases, there is cause for ambiguity, such as the Type IIb SN 2011dh \citep{Maund+2015}. In such a scenario, an optical spectrum was able to rule out the presence of H$\alpha$~emission from a shock and a potential light echo contribution was removed, leaving only the analysis of the resulting companion \citet{Maund2019}.

%The SED can help disentangle some of these components  two contributions since CSM interaction typically produces a flat SED, while a stellar companion, especially a blue star, shows a sharp decline beyond 3000–4000~\AA. At later epochs, \citet{Maund2019} rule out suggests that in this case we see the contribution of both a companion and light echo onto CSM instead of CSM interaction. Unlike a stellar source, CSM interaction (or light echo?)\other{@Schuyler, Ori?} is expected to evolve over timescale of years, a behaviour that would only be seen in companions in the presence of strong ECI effect (see Sec.\ref{sec:ejecta_interaction_companion}). Thus, the colours and their time evolution are key to interpreting companion detections. \other{@Ori: can we quantify how certain we are that our list in Table 1 has only stellar sources?}

\subsubsection{Dust obscuration and misinterpretation of CSM-interaction}

Post-supernova dust formation can significantly affect the detectability of a surviving companion. While dust formation is particularly relevant for Type II events, where high levels of dust production are well established \citep[e.g.,][]{Sarangi+2015,  Shahbandeh+2023}, \sesne\ have traditionally been expected to produce little dust, owing to their lower ejecta masses and higher expansion velocities. However, some recent results suggest the possibility of dust in this subclass \citep{Szalai+2019,Tinyanont25}. Future JWST studies will better provide an understanding of the dust yield in this subclass, with detailed observations of the Type Ib SN~2024ahv and Type Ic SN~2023dbc already underway \citep{Shanbandeh+25dbc,Shahbandeh+25ahv}. Alternatively, other studies suggest that the presence of UV emission from a companion could potentially suppress any dust formation in the primary explosion \citep{Kochanek2017}.
% Analytical studies further suggest that the presence of a companion could potentially suppress dust formation in its vicinity due to UV radiation, thus increasing its own visibility \citep{Kochanek2017}. This limiting factor of extinction can be even better addressed with future \emph{JWST} search of companions in nearby SN sites. 
%Shahbandeh+    observations

If we consider the possibility that some of the upper limits in our sample are due to dust obscuration, the overall distribution in \fig{fig:cumulative} would shift upward towards higher luminosities (potentially allowing for higher mass accretion scenarios) and the empirical  detection fraction (\fig{fig:bin_frac_Z}) would be even higher.
% https://ui.adsabs.harvard.edu/abs/2025AAS...24545806S/abstract
%https://ui.adsabs.harvard.edu/abs/2024sros.confE.178S/abstract

%%%%%%%%%%%%%%%%%%%%%%%%%%%%%%%%%%%%%%%%%%%%%%%%%%%%%%%%

\section{Conclusions}\label{sec:conclusions}

We have conducted a comprehensive analysis of binary companions to nearby stripped-envelope supernovae (\sesne), assembling the six companion detections with \textit{HST} and six non-detections (mostly with deep upper luminosity limits) reported in the literature or obtained through our own re-analysis. Confronting this observational sample with state-of-the-art, detailed population-synthesis modelling, we characterize the statistical properties of the companion ensemble and thereby deepen our understanding of the progenitor systems and uncertain binary  pathways that produce \sesne.

%\begin{itemize}

% \item   Across the full metallicity range $0.1$–$2\,Z_\odot$, about $85$–$90\%$ of SN~Ib/c and $70$–$80\%$ of SN~IIb explode next to a visible companion, predominantly a main-sequence one.  These stars are stripped through prior \emph{stable} 
% %but \emph{inefficient} 
% mass transfer that spins them up to near-critical rotation, with mean $\langle\omega/\omega_{\mathrm{crit}}\rangle\simeq0.84$ at $Z_\odot$ and $\simeq0.92$ at $0.1\,Z_\odot$ at the time of explosion.  Companions in SN~IIb systems are, on average, more luminous and slightly more evolved than those in SN~Ib/c, %reflecting their origin in (i) higher mass-ratio binaries at solar metallicity or (ii) generally more massive binaries at sub-solar metallicity.  
% Common-envelope channels contribute negligibly under our default assumptions, so very low-mass, low-luminosity companions are intrinsically rare.

% \item   The remaining events follow two minority pathways.  First, the “no-companion’’ channel—driven by mergers or single-star winds—accounts for $15$–$20\%$ of SN~IIb at $Z\!>\!Z_\odot$ but for only $\lesssim7\%$ of SN~Ib/c, with Cassiopeia A remnant likely being such a companion-free IIb example.  Second, about $8\%$ of SN~Ib/c and up to $10\%$ of SN~IIb retain a non-detectable compact object.
% %, predominantly a black hole, making them plausible descendants of high-mass X-ray binaries and promising antecedents of gravitational-wave sources.

%\item   
Across  a range of metallicities of $0.1$–$2\,Z_\odot$, about $80$–$90\%$ of SN~Ib/c and $60$–$85\%$ of SN~IIb are predicted to explode next to a stellar companion (\fig{fig:pie_chart}). These companions are almost always at various stages of their main-sequence phase, that stripped the progenitor through earlier stable mass transfer and rotate close to their critical limit up to the moment of \sesn (\fig{fig:centerhe_omega}). %\bigl($\langle\omega/\omega_{\mathrm{crit}}\rangle\!\simeq\!0.84$ at $Z_\odot$, $\simeq0.92$ at $0.1\,Z_\odot$\bigr); 
The companions in \IIb\ are, on average, predicted to be more luminous and slightly more evolved than those of \Ibc. 
%Common-envelope channels contribute only a minority of stellar companions (with only a slightly higher contribution for compact object companions). This trend persists across different common-envelope and natal kick assumptions. Consequently, intrinsically faint ($\logL \lesssim 2.5$), low-mass companions are expected to be rare  ($<10\%$).  
Common-envelope channels contribute only a minority of stellar companions, and thus intrinsically faint ($\logL \lesssim 2.5$), low-mass companions are expected to be rare  ($<10\%$).  
At high metallicities, however, mergers or true single stars stripped through winds can produce \IIb\ with no companions in $\sim20\%$ of cases, with the Cassiopeia A remnant being a likely example, compared to only $\lesssim7\%$ among SN~Ib/c. Finally, about $8\%$ of SN~Ib/c and up to $10\%$ of SN~IIb explode beside an undetectable compact object, predominantly a black hole, with SN~2012fh standing out as a plausible case.

The luminosity distribution of the six detections and six upper limits is well explained by models assuming  low- and moderate-accretion efficiency onto the companion during mass transfer  (\fig{fig:cumulative}), with metallicity playing a smaller role. It disfavors very efficient mass accretion, which would overpredict bright companions. %Low- and moderate accretion efficiency best fit the observed luminosities, with metallicity playing a smaller role. 
The observed detection fraction is broadly consistent with model expectations at the achieved luminosity depth (\fig{fig:bin_frac_Z}), especially for \Ibc, and together with the luminosity function, it favors the non-explosion of the most massive Wolf–Rayet stars, that would tend to have either no companion, or a luminous companion next to them.  However, the high detection rate for \IIb\ is only marginally consistent with their expected binary fraction at low metallicity. A faint stellar companion remains the most probable scenario in every non-detection, except in the case of SN~2012fh  (\fig{fig:prob_nondetection}).

Three \Ibc\ show companions apparently cooler than expected for main-sequence stars (\fig{fig:HRD_comp}), even though evolved companions, predominantly RSGs, should be rare for \Ibc\ ($\lesssim4\,\%$), and only slightly more common for \IIb\ ($\lesssim8\,\%$).  This apparent excess may reflect post-SN bloating from ejecta–companion interaction (ECI), expected in $\sim12\%$ of \sesne\ with a companion  to produce observable effects after the SN has faded (\fig{fig:solid_angle}), mostly in SN~Ib/c systems (and only in $\sim3\%$ of SN~IIb). The companions of SN~2006jc and SN~2013ge are already strong candidates for this scenario, and SN~2022jli is promising to reveal similar signatures as the SN fades.
 %Hot stripped-helium companions remain an exotic channel, confined to $\lesssim1\,\%$ of SN~Ib/c (and virtually absent in SN~IIb), and will require deep ultraviolet follow-up to detect.
%\end{itemize}

Overall, our results underscore the value of systematic companion searches as an independent probe of binary interactions that shape \sesn\ progenitors.  The empirical companion demographics already place stringent constraints on key uncertain aspects of binary evolution, such as mass transfer efficiency and explodability of stripped progenitors.  Future observational campaigns, particularly increasing the companion sample, as well as follow-up studies for companion confirmations (especially on possible candidates of ejecta–companion interaction), are essential for deepening our insight into the rich subtype diversity and physical complexity inherent in \sesne\ progenitor pathways, across varying metallicity regimes. A well-characterized sample of nearby \sesne\ and their companions can function as a crucial benchmark for interpreting the diverse and distant populations from transient surveys, such as \emph{ZTF} and \emph{LSST}, where direct companion, progenitor or host-galaxy detections will largely be out of reach. 

%%%%%%%%%%%%%%%%%%%%%%%%%%%%%%%%%%%%%%%%%%%%%%%%%%%%%%%%%%
\section*{Acknowledgments}

We acknowledge useful interactions with Jennifer Andrews, Azalee Bostroem, Alex Filippenko, Patrick Kelly, Danny Milisavljevic, Selma de Mink, Nathan Smith, and Weikang Zheng.  
EZ thanks Thibault Lechien, Norbert Langer, Andrea Ercolino, Charlie Kilpatrick and Konstantinos Kovlakas for fruitful relevant discussions. % Extra observers?
EZ and DS acknowledge support from the Hellenic Foundation for Research and Innovation (H.F.R.I.) under the “3rd Call for H.F.R.I. Research Projects to support Post-Doctoral Researchers” (Project No: 7933). 
MMB was supported by the Boninchi Foundation, the Swiss National Science Foundation (project number CRSII5\_21349), and the Swiss Government Excellence Scholarship No. 2024.0234. 
MR acknowledges support from NASA (ATP: 80NSSC24K0932). 
JJA acknowledges support for Program number (JWST-AR-04369.001-A) provided through a grant from the STScI under NASA contract NAS5-03127.  MK was supported by the Swiss National Science Foundation Professorship grant (PI Fragos, project number PP00P2 176868). SG, CL, PMS and ET were supported by the Gordon and Betty Moore Foundation (PI Kalogera, project numbers GBMF8477 and GBMF12341). 
This work is based on observations made with the NASA/ESA {\it Hubble Space Telescope}, obtained at the Space Telescope Science Institute (STScI), which is operated by the Association of Universities for Research in Astronomy, Inc., under NASA contract NAS 5-26555. Support was provided by NASA through grants GO-14075, GO-16165, and GO-17203 from STScI. 

%%%%%%%%%%%%%%%%%%%%%%%%%%%%%%%%%%%%%%%%%%%%%%%%%%%%%%%%%%%%%%

\bibliographystyle{mnras}
%\bibliography{my_bib,bib_overleaf,bibliography_old}
\bibliography{my_new_bib,bib_overleaf_new}

\appendix

\section{Individual supernovae in our observational sample}\label{sec:obs_sample_app}

%\other{@Jenny, @Ori, @Maria would you be willing to check this list of companions constraints in Table 1 together one by one? Would it be possible to write a few sentences for each event too (this can go to the Appendix). I already have gathered various in our spreadsheet  sheet 2:}
%https://docs.google.com/spreadsheets/d/1vRtjGzRkuTzIIvaAegyqKiuYPbMxEDYCbgNXzeuji1o/edit?gid=1210109300#gid=1210109300

%\other{Question, for @Ori,  @Schuyler and other: What do we do with 16coi, 11dh?}
%\other{Updated decision on which companion data we use: - 16coi not included, but may mention that we are about to show a detection(!?) about it (@Maria @Jenny you can phrase it as you want). - 16gkg, 13bvb are not here yet, but may be mentioned as examples of how the population will grow in the near future. -11dh will go in, with the updated Maund2019 inferred limits!}

Our sample comprises in total 12 events with \emph{HST} imaging (see Table~\ref{tab:comp_sample}). These objects span the full zoo of \sesne\ sub‑types, from IIb through Ib, Ic, Ic-BL, and Ibn, and populate host environments from sub‑solar to roughly solar metallicity.  The \emph{HST} images were acquired between $\sim$1.8 yr and $\sim$23 yr after explosion, ensuring that any surviving non‑degenerate companion has now emerged from the fading SN signal, with the exception of the pre-SN imaging for SN2019yvr. Here, we provide additional information and context for each event in our sample and the origin of the numbers summarized in Table~\ref{tab:comp_sample}. % For SN2019yvr an \emph{HST} frame obtained one year \emph{before} core collapse provides the relevant constraint.)

\subsection{Companion Detections}

Six \sesne\ in our sample have luminous companion detections in the literature (these are shown in the upper part of Table~\ref{tab:comp_sample}).

{\bf SN1993J:} For Type IIb SN1993J, the well‑known hot B‑type star companion remains the touch‑stone example of mass transfer in an interacting binary \citep{Maund+2004,Fox+2014}.  The presence of a binary companion was first proposed based on the presence of excess UV/B-band flux in observations of the SN progenitor \citep[see e.g.][]{vanDyk2002}. Obtaining imaging and spectroscopy of the location of SN1993J, 10 years after the explosion, \cite{Maund+2004} identified absorption lines consistent with a companion star with an effective temperature of $\log_{10}(T_{\rm{eff}}/K)=4.3 \pm 0.1$ and luminosity $\log_{10}(L/L_\odot)=5 \pm 0.3$. However, observations at this epoch were still dominated by light from the SN. Later, \cite{Maund+2009} confirmed the presence of a B-supergiant companion, and they predicted the SN will have faded sufficiently by 2012 for more direct observations of the companion star. 

\cite{Fox+2014} subsequently obtained an HST UV spectrum of the location of SN1993J in 2012. Modeling the spectrum/photometry as a combination of SN light and a stellar companion, they find a best-fit with a hot B star with temperature $\sim 24000^{+3000}_{-5000}$K. However, as noted in Section~\ref{sec:obs_sample}, \cite{Fox+2014} do not explicitly quote the inferred luminosity for their best-fit models. As a re-analysis, we extracted the best-fit Kurucz stellar model found by \cite{Fox+2014} (with temperature T=24 kK, log (g) = 3.5, and solar metallicity) and scaled it to 22.9 Vega mag in the F438W band. After correcting for the distance of 3.6 Mpc, we found the final luminosity to be $\log_{10}(L/L_\odot) = 4.71$.   

By comparing the light curve of SN\,1993J to a set of explosion models, \cite{Woosley+1994} infer the final helium core mass at the time of explosion was 4.0$\pm0.5$ \Msun. This is consistent with other estimates of the ejecta mass \citep[e.g.][]{Lyman+2016,Prentice+2016} and we adopt this value in Table~\ref{tab:comp_sample}.

{\bf SN2001ig:} A comparable proposed companion star is also recovered for Type IIb SN SN2001ig. 
%; its F‑type companion ($\log L/L_\odot\approx3.9$; $\log T_{\rm eff}/{\rm K}\approx4.31$) sits in a low‑metallicity environment \citep{Ryder+2006, Ryder+2018} \MZ{expand on helium nebular?}. %and is surrounded by a helium‑rich nebula that strengthens the binary‑stripping interpretation . 
The first study to search for the SN2001ig companion was made by \cite{Ryder+2006}, who used ground-based imaging and found a source consistent with a late B to late F star. 14 years post-explosion, \cite{Ryder+2018} searched the site with HST/WFC3 UV imaging and found a source consistent with a MS star of effective temperature $19,000 < T_{\textrm{eff}} < 22,000$ K and luminosity $\log_{10}(L/L_\odot) = 3.92 \pm 0.14$. We adopt \cite{Ryder+2018}’s results in our analysis. 

SN2001ig is also notable in showing several indirect pieces of evidence for binary interaction. In particular, the radio light curve for SN2001ig showed clear modulations (indicative of changes in the density of the circumstellar material) with a $\sim$150 day period. \cite{Ryder2004} argue this could be caused by wind collisions within an eccentric binary system. In addition, an optical spectrum obtained 6 years post-explosion shows narrow HeII  emission, potentially indicative of a nebula formed via binary interaction \citep{Ryder+2018}.
%We do not re-analyze SN2001ig as both temperature and luminosity are reported in literature. 
%With BPASS models, Ryder et al. (2018) found the progenitor core mass to be $3.5\,M_\odot$. 
%

To the best of our knowledge, no light curve analysis is available for SN2001ig that constrains its ejecta mass. However, \cite{Silverman2009} perform modelling of a late-time spectrum and find that $\sim1.15$ \Msun\ of material is located at velocities below 4300 km/s (effectively setting a lower bound on the ejecta mass). In addition, by comparing the properties of the candidate companion (along with the Type IIb classification for the SN) to a suite of BPASS models \citep{Eldridge+2009}, \citet{Ryder+2018} favor a progenitor with a final mass of 3.5 \Msun. We therefore adopt a final progenitor mass of 3.5 $\pm$ 1 \Msun\ in Table~\ref{tab:comp_sample}.

{\bf SN2011dh:} The Type IIb SN2011dh also shows a luminous point source at the location of the SN in post-explosion imaging, interpreted as a main‑sequence companion to a low-mass progenitor \citep{Bersten+2012,Benvenuto+2013}. While earlier studies raised the possibility of contamination from circumstellar interaction or light echoes \citep{Folatelli+2014,Maund+2015}, the most recent analysis by \citet{Maund2019} supports a composite interpretation in which the detected flux arises from both a companion and extended light echo, with the companion parameters inferred after subtracting the latter’s contribution. They identified a $\sim 9-10 \Msun$ BSG with temperature of $\log_{10}(T_{\rm{eff}}/K)=4.30 \pm 0.06$ and luminosity $\log_{10}(L/L_\odot)=3.94 \pm 0.13$. We adopt these values from \cite{Maund2019} in our analysis.
% \MZ{for 11dh addition} \citep{Bersten+2012} 2-3 Msun progenitor, \citep{Benvenuto+2013} 4 Msun prognitor. \citep{Benvenuto+2013} massive companion, possibly observed \citep{Folatelli+2014} but CSM interaction in \citep{Maund2015}. Most recent companion \citep{Maund2019}, which argues that we dont see CSM interaction in the end , but most probably we see light echoes on CSM that exists more far away. Assuming this possiblity and subtracting its effect he infers a (slightly lower luminosity than initially reported at \citet{Folatelli+2014}). % companion there of logL=3.94±0.13 and logTeff=4.3±0.06 ..
%
%We do not reanalyze SN2011dh due to the complexity of the light echo. 

\citet{Bersten+2012} presented hydrodynamical modelling of the light curve of SN\,2011dh. They found that it was consistent with the explosion of a progenitor star with a final mass in the range of 3--4 \Msun\. This is consistent with other estimates of the ejecta mass \citep[]{Lyman+2016,Prentice+2016}, and we adopt this value in Table~\ref{tab:comp_sample}.

{\bf SN2006jc:} \cite{Maund+2016} detected a faint source that could be the surviving companion of Type Ibn SN2006jc, with some ambiguities if the source could be a light echo or late-time CSM interactions. With newer observations, \cite{Sun+2020} confirmed the detection is indeed a star. SED fitting yields an effective temperature of $\log_{10}(T_{\rm{eff}}/K)=4.09_{-0.04}^{+0.05}$ and luminosity $\log_{10}(L/L_\odot)=4.52_{-0.13}^{+0.13}$, thus located on the Hertzsprung Gap. We adopt the values from \cite{Sun+2020} in our analysis.

Assessing the pre-SN mass of Type Ibn SN is particularly challenging, as their light curves are heavily impacted by interaction with a dense CSM. \cite{Tominaga2008} produce a theoretical model for SN2006jc, performing evolution of the progenitor star, hydrodynamics and nucleosynthesis calculations of the subsequent SN explosion, and calculations of a bolometric light curve. They favor a model with a pre-SN progenitor mass of $\sim$6.9 \Msun. \citet{Maeda2022} also present a light curve model for Type Ibn SN, in which they model the interaction of the SN ejecta with a dense CSM. Applying their model to SN2006jc, they favor and ejecta mass in the range of 2--3 \Msun, depending on the epoch of explosion (although larger ejecta masses are still allowed). In Table~\ref{tab:comp_sample} we therefore adopt a range of possible pre-explosion progenitor masses for SN2006jc of 4.4-6.9 \Msun. 

{\bf SN2013ge:} SN2013ge was a Type Ib/c SN (early time spectra revealed weak He lines in an event that would have otherwise been classified as a Type Ic SN) that showed two peaks in its UV light curve, indicative of either an inflated progenitor envelope or an asymmetric explosion \citep{Drout2016}. Late‑time HST imaging of the explosion site of SN2013ge revealed a blue source that \cite{Fox+2022} find is consistent with a B5 I supergiant star that is redward of the MS. However, as noted in Section~\ref{sec:obs_sample}, they do not explicitly provide a luminosity constraint for the candidate companion. We therefore take the published photometry for the companion from \cite{Fox+2022} and performed new $\chi^2$-fitting with a grid of \cite{Kurucz1992} model atmospheres. Adopting the same distance and reddening values as \cite{Fox+2022}, we find best-fit stellar parameters (with 1$\sigma$ uncertainty) of $\log_{10}(L/L_{\odot}) = 4.85\pm0.05$ and $T_{\rm{eff}} = 16^{ +1}_{ - 2}$kK. These are consistent with the results of \cite{Fox+2022} and confirm that the candidate companion is located in the Hertzsprung Gap.

By modelling the bolometric light curve of SN2013ge, \citet{Drout2016} found and ejecta mass of $\sim$2--3 \Msun. Assuming a remnant mass of 1.4 \Msun, we therefore adopt a pre-SN progenitor mass of $3.9\pm0.5$ \Msun\ in Table~\ref{tab:comp_sample}.

\begin{figure*} 
% \vspace*{-2.0 cm}
\begin{center}
\includegraphics[width=\textwidth]{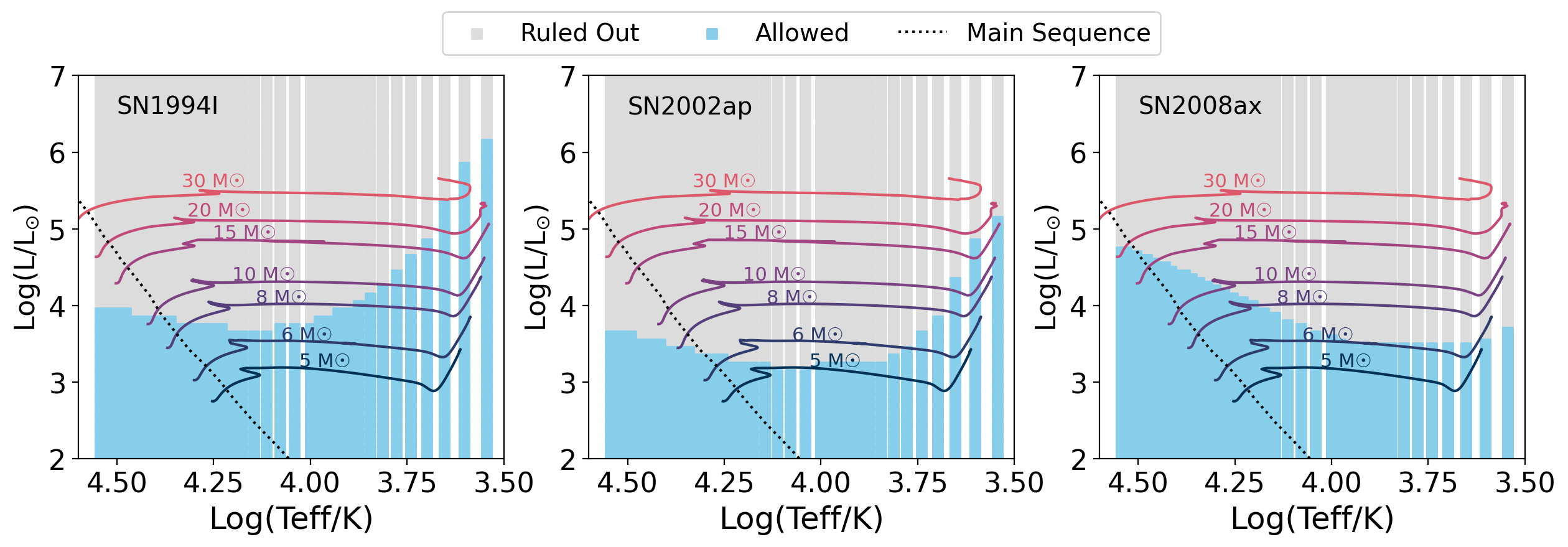}
\caption{Types of surviving companions that are ruled out (grey) or allowed (blue) based on our reanalysis of post-explosion upper limits from the explosion sites of SN1994I (left), SN2002ap (middle), and SN2008ax (right). For each temperature, we take a \citet{1993Kurucz} solar metallicity stellar model and scale it to a range of luminosities between $\log_{10}(L/L_{\odot})= 2-7$. Then, after correcting for distance and extinction appropriate for each object, we perform synthetic photometry for comparison with observed upper limits. For a majority of the grid we utilize the $\log_{10}(g) = 4$ stellar models, with the exception of models with $T_{\rm{eff}} > 36,000$ for which we use $\log_{10}(g) = 5$. For reference, we also show a set of solar metallicity single star evolution tracks from \texttt{POSYDON} (colored lines) and the location of the MS based on the observational calibration of \citet[][dotted black line]{Pecaut2013}.
\label{fig:rule-out}
}
  \end{center}
\end{figure*}

%``Modern Mean Dwarf Stellar Color and Effective Temperature Sequence'' distributed by E. Mamajek\footnote{\url{https://www.pas.rochester.edu/~emamajek/EEM_dwarf_UBVIJHK_colors_Teff.txt}} \citep{Pecaut2013}

{\bf SN2019yvr:} The Type Ib SN2019yvr is the only SN in our sample where constraints on the companion properties are inferred from \emph{pre-}SN imaging. It is also notable in that the SN originally presented as a H-free Type Ib SN, before interacting with a H-rich shell of dense circumstellar material approximately 100 days after the explosion \citep{Ferrari2024}. However, modelling of the early time light curve showed that its explosion properties were consistent with the range typically found for Type Ib SN. Namely, it had an ejecta mass of $\sim$1.4-2.6 \Msun\ \citep{Sun+2022}. By performing hydrodynamical modeling, \cite{Ferrari2024} favor a pre-explosion progenitor mass of $\sim3-4$ \Msun (we adopt this range in Table~\ref{tab:comp_sample}). 

\cite{Kilpatrick+2021} identified a red source at the location of SN2019yvr in HST imaging obtained 2.6 years prior to explosion. They demonstrate that the source is consistent with a $\log_{10}(L/L_{\odot}) = 5.3\pm0.2$ and $T_{\rm{eff}} = 6800^{+400}_{-200}$K. This is usually cool for a H-free SN progenitor. Alternatively, by comparing to BPASS \citep{Eldridge+2017} spectral models, \cite{Kilpatrick+2021} also find that the pre-SN detection could be consistent with the combined light from a binary system where the progenitor underwent common envelope evolution (although these results are also seemingly at odds with the H-free nature of SN2019yvr). Subsequently, \cite{Sun+2022} reanalyze the pre-SN photometry of the progenitor system. In particular, they perform binary SED fitting where they fix the contribution progenitor star to be consistent with expectations for a Type Ib SN progenitor that can reproduce the explosion properties found above (i.e., a hot/compact star). 
With this set of assumptions they infer the presence of a cool/inflated companion star that dominates the optical light of the progenitor system and has the following properties: $\log_{10}(L/L_{\odot}) = 4.91\pm0.14$ and $\log_{10}(T_{\rm{eff}}/K) = 3.80^{ +0.02}_{ - 0.03}$. We adopt these values from \cite{Sun+2022}.

%They find that the photometry can be reproduced by the sum of a hot/compact star (as expected for a Type Ib SN progenitor) and a cool/inflated companion star. 

%Finally, for the Type Ib SN~2019yvr a red source with $\log L/L_\odot\approx4.9$ and $\log T_{\rm eff}/{\rm K}\simeq3.80$ has been inferred from the composite pre‑explosion image dominated by the progenitor's flux\citep{Sun+2022}.We do not re-analyze SN2019yvr as the detections come from composite images %… 

\subsection{Companion Upper Limits}

The remaining six \sesne\ in our sample have post-explosion HST photometry that yield only upper limits on any putative companion (these are listed in the lower portion of Table~\ref{tab:comp_sample}).  In general, non‑detections imply three possibilities: (i) the companion is present at the \sesn\ but intrinsically faint, whether a low‑mass main‑sequence star or a more evolved one; (ii) the companion is a compact object (white dwarf, neutron star, or black hole) that is optically invisible with \emph{HST}; or (iii) no companion exists at the SN site, either because the progenitor was born single, the binary merged prior to collapse, or the exploding star was expelled by its binary system before explosion (due to a previous SN event). As noted in Section~\ref{sec:obs_sample}, for all systems that have only upper limits available, we re-assess the types of companions that would be allowed by the data by comparing to a grid of \cite{Kurucz1992} model atmospheres with a range of temperature and luminosity.

{\bf SN1994I:} Late‑time \emph{HST} UV photometry (F275W and F336W) of the explosion site of the canonical Type Ic SN1994I was reported by \cite{van-Dyk+2016}. Using these upper limits obtained they infer that stars hot stars ($\log_{10}(T_{\rm{eff}}/K) \gtrsim 4.36$) with luminosities greater than $\log_{10}(L/L_\odot) \gtrsim 3.6$ would have been detected. As a result, they rule out MS companions with masses $\gtrsim10 M_\odot$. In the left panel of Figure~\ref{fig:rule-out} we plot results from our reanalysis of this data. Using the photometry, distance, and extinction measurements from \cite{van-Dyk+2016} we find an upper luminosity limit of $\log_{10}(L/L_\odot) \lesssim 3.8$ for a MS star (where we define the location of the MS based on the data from ). These are similar to the result of \cite{van-Dyk+2016}. Notably, because the late-time imaging was obtained in the UV, these luminosity limits are relatively flat for hotter stars, but are weaker at cooler temperatures (i.e., $\log_{10}(T_{\rm{eff}}/K) \lesssim 4.$) .

The light curve of SN\,1994I declines relatively quickly in the context of Type Ib/c SN \citep[e.g.]{Drout+2011,Lyman+2016,Prentice+2016}. Estimates of its ejecta mass are therefore correspondingly small, with typical values $\lesssim$1 \Msun \citep[e.g.]{Young+1995,Lyman+2016}. By comparing the bolometric light curve of SN1994I with a set of explosion models, \citet{Nomoto+1994,Iwamoto+1994} favor the explosion of a 2.1 \Msun\ C+O progenitor (with an ejected mass of 0.88 \Msun). We adopt this value in Table~\ref{tab:comp_sample}.

{\bf SN2002ap:} Comparable or deeper limits have been obtained for the Ic‑BL SN2002ap. \cite{Crockett+2007} analyzed both pre-explosion (ground-based imaging) and post-explosion images (HST) for Type Ic-BL SN2002ap, and they found no detection in either epochs. Their analysis placed a limit on the companion as either a MS star with mass $\lesssim$20 Msun or a compact object. Later, \cite{Zapartas+2017b} obtained deeper limits of luminosity $\log_{10}(L/L_\odot) \lesssim 3.35$, which corresponds to a MS companion star with mass $\lesssim$8–10 \Msun. Results from our reanalysis are shown in the middle panel of Figure~\ref{fig:rule-out}. Using the data, distance, and extinction measurements of \cite{Zapartas+2017b}, we also find a luminosity upper limit of $\log_{10}(L/L_\odot) \lesssim 3.35$ at the location of the MS. As these data was also obtained in the UV, luminosity limits are again relatively flat until quite low temperatures ($\log_{10}(T_{\rm{eff}}/K) \lesssim 3.9$) are reached.

\citet{Mazzali+2002,Mazzali+2007} perform a joint analysis of the spectra and light curves of SN2002ap, and find an ejecta mass of $\sim$2.5--5 \Msun, kinetic energy of $\sim$(4-10)$\times$10$^{51}$ erg, and nickel mass of $\sim$0.07 \Msun. Comparing these to set of model explosions of C+O cores, they find that they most closely match expectations for the explosion of an $\sim$5\Msun\ progenitor that forms a 2.5\Msun\ remnant. We therefore adopt a final progenitor mass of 5$\pm$1 \Msun\ in Table~\ref{tab:comp_sample}.
%Our reanalysis shows a consistent upper limit. 

%Less restrictive, but still relatively informative, are the shallow constraints for the IIb SN2008ax and SN2015G. 

{\bf SN2008ax:} Post-explosion imaging of the explosion site of the Type IIb SN2008ax was presented by \cite{Folatelli+2015}. Using this data, they argue that any companion of SN2008ax must be less luminous than a O9–B0 MS star. Using the photometry, distance, and extinction values from \cite{Folatelli+2015}, we obtain the results shown in the right panel Figure~\ref{fig:rule-out}. We find we can exclude a MS star with $\log_{10}(L/L_\odot) < 4.6$ at  $\log_{10}(T_{\textrm{eff}}/K)\sim4.50$. While post-explosion UV data was obtained at the location of SN2008ax, it was relatively shallow compared to the limits obtained in the optical. As a result, the luminosity limits on any surviving companion are deeper for cooler temperatures, reaching a minimum of $\log_{10}(L/L_\odot) < 3.5$ at  $\log_{10}(T_{\textrm{eff}}/K)\sim3.7$.
%This corresponds to $\log_{10}(L/L_\odot) < 4.88$ [de koter notes, add Manos for more description]. 
%With the upper limits reported, we obtained the constraint of . 

\citet{Folatelli+2014} also perform hydrodynamical modelling of the light curve of SN2008ax, and find that it is consistent with the explosion of star with a final mass of $\sim$4--5 \Msun (assuming a final NS remnant of 1.4 \Msun). This is consistent with other estimates of the ejecta mass from semi-analytic models \citep[e.g.][]{Lyman+2016,Prentice+2016} and we adopt a final progenitor mass of 4.5$\pm$0.5 \Msun in Table~\ref{tab:comp_sample}. 

{\bf SN2015G:} \cite{Sun+2020} report that no source was detected in late-time UV and optical observations of the explosion site of the Type Ibn SN2015G. Notably,   \cite{Sun+2020} perform a similar analysis as we describe above to constraint the types of possible surviving companions to the progentior of SN2015G as a function of effective temperature. We therefore do not repeat this analysis here, but instead refer readers to Figure 7 in \cite{Sun+2020}. Based on this analysis, they exclude MS companions with $\log_{10}(L/L_\odot)<5$ at $\log_{10}(T_{\textrm{eff}}/K) = 4.4$. The corresponds to an upper mass limit of of $60_{-25}^{+35} M_\odot$. Due to the depth of their upper limits in the optical vs ultraviolet, constraints on the luminosity of any companion are deepest in the Hertzsprung Gap, reaching a minimum at $\log_{10}(L/L_\odot)<4$ at $\log_{10}(T_{\textrm{eff}}/K) = 3.9$ before rising steeply at both hotter and cooler temperatures.

As mentioned above for SN2006jc, constraining the ejecta mass for Type Ibn SN in complex due to the impacts of CSM interaction. As far as we are aware, no ejecta mass constraints for SN2015G have been made. However, \citet{Shivvers+2017b} present pre-explosion constraints on the progenitor system that rule out massive WR stars (requiring that M$_{F555W}$ $\gtrsim -6.4$ mag). We therefore very broadly adopt that the pre-SN mass of the progenitor was $\lesssim$9 \Msun in Table~\ref{tab:comp_sample}.

{\bf SN2012fh:} Deep post-explosion HST images of the explosion site of the Type Ib/c SN2012fh were presented in \citet{Williams+2025}. Upper limits from these images constrain the luminosity of any surviving companion star to be  $\log_{10}(L/L_{\odot}) \lesssim 3.35$ at the location of the MS (at $T_{\textrm{eff}} \sim 20$kK). Overall, a temperature dependent analysis finds that a companion with luminosity  $\log_{10}(L/L_\odot) \gtrsim 3.45$ can be ruled out for the entire temperature range of 6000--26000 K. As a plot equivalent to those presented in Figure~\ref{fig:rule-out} is presented in \citet{Williams+2025}, we do not reproduce it here.

While the peak of the light curve of SN2012fh was not observed, \citet{Williams+2025} use several other methods to constrain the final mass of the progenitor star. Specifically, they find that the final progenitor mass was likely in the range of 5.6--10.6 \Msun through a combination of nebular spectra analysis and the non-detection of a Wolf-Rayet progenitor in pre-explosion imaging. We adopt this range in our analysis (see Table~\ref{tab:comp_sample}). 

{\bf SN2017gax:} SN2017gax is Type Ib/c SN \citep{Jha2017} for which both deep pre- and post-explosion imaging from HST are available. This data is analyzed in detail in an upcoming publication (Su et al., in prep.)---yielding only upper limits either the SN progenitor or any surviving companion. Using the same procedure as described in Section~\ref{sec:obs_sample} (see also \citealt{Williams+2025} for further details), Su et al.  (in prep.) exclude any MS companion with a luminosity greater than  $\log_{10}(L/L_{\odot}) = 3.6$ (measured $\log_{10}(T_{\textrm{eff}}/K) \sim 4.3$). As a plot equivalent to  those presented in Figure~\ref{fig:rule-out} will be presented in Su et al.  (in prep.), we do not reproduce it here.

A paper examining the explosion properties of SN2017gax is also forthcoming (Hoang et al., in prep.). While we refer to that manuscript for details, overall the ejecta mass is found to fall within the range that is typical for Type Ib/c SN (e.g. $\sim$2.5--4 $\Msun$). Assuming a NS remnant of 1.4 \Msun, we therefore adopt a final progenitor mass of 4.7$\pm$0.8 \Msun in Table~\ref{tab:comp_sample}.

\section{Rotationally-limited mass accretion efficiency of \posydon}\label{sec:beta}

 \begin{figure} 
% \vspace*{-2.0 cm}
\begin{center}
\includegraphics[width=0.99\linewidth]{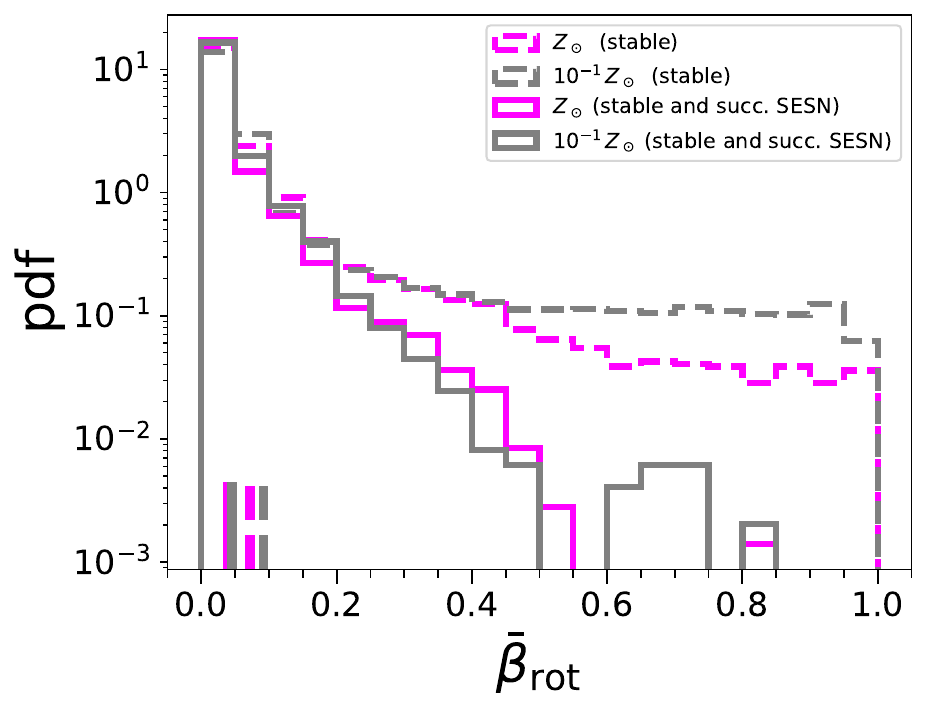}
\caption{Normalized probability density function of the mass-weighted, rotationally-limited mass accretion efficiency, $\bar{\beta}_{\rm rot}$, for all stable mass transfer sequences (dashed) and for only the ones that lead to to successful \sesne\ (solid) for our default \posydon\ population, for two metallicities. Vertical lines at the bottom indicate the mean $\langle \bar{\beta}_\mathrm{rot} \rangle$ for each distribution.%, with dashed lines for the full stable sample and solid lines for the \SESN-producing subset.  
\label{fig:beta}
}
  \end{center}
\end{figure}

The detailed binary \mesa\ models implemented in \posydon\ do not explicitly prescribe the mass transfer efficiency a priori, but instead adopt a rotationally-limited model which allows time-variable accretion depending on the accretor's rotation \citep[assuming conservative mass transfer until accretor reaches critical rotation; discussed further in ][]{Fragos+2023}. To quantify the impact of this modeling choice we calculate an effective mass transfer efficiency which arises from our physical model. 

During a single mass transfer phase we define the mass transfer efficiency as $\beta_\mathrm{rot} \equiv \Delta M_\mathrm{acc} / \Delta{M}_\mathrm{d,RLO}$, the ratio of mass accreted to mass transferred by the donor through Roche lobe overflow. % (excluding wind mass loss in our calculations). 
Note that during this phase, the instantaneous accreted ratio is not constant, so the effect of accretion may not be directly comparable to a fixed $\beta$ model assumption of the same value \citep[e.g.,][]{Renzo+Gotberg2021}, but can be used for a rough comparison.  
Binaries with a multiple ($n$) phases of mass transfer during their evolution will then have $n$ values of $\beta_\mathrm{rot}$ for a single binary sequence (e.g., a Case A phase with a Case B mass transfer phase later). 
Following the calculation of \citet{Rocha+2024}, we perform a mass-weighted average over all mass transfer phases for a given binary sequence to compute the effective mass-weighted rotationally-limited mass transfer efficiency: %$\bar{\beta}_\mathrm{rot} = \frac{1}{\sum^{n}_{i=1}\Delta{M}_{{\rm d,RLO},i}} \sum^{n}_{i=1} \beta_{\mathrm{rot},i} \cdot \Delta M_{\mathrm{acc},i}$. 
\begin{equation}
\bar{\beta}_\mathrm{rot} = 
\frac{
  \sum_{i=1}^{n} \beta_{\mathrm{rot},i} \cdot \Delta M_{\mathrm{acc},i}
}{
  \sum_{i=1}^{n} \Delta M_{{\rm acc},i}
}
\end{equation}

We focus only on stable mass transfer sequences between two non-compact-object stars (i.e., what is called in \posydon\ \emph{HMS-HMS} grid), excluding any isolated evolution, any non-mass transfer systems, any reverse mass transfer (from the initially less massive star towards the primary), any binary sequences that ended up eventually to unstable mass transfer and triggered CE, and also any mass transfer onto a compact object \citep[where Eddington-limited accretion is assumed instead][]{Fragos+2023}.  
%is not necessarily directly comparable to standard BPS implementations of $\beta$, which are often segmented through the canonical MT definitions (e.g., Case A vs. Case B). 

% Then through the \posydon{} framework, we use the same initial-final interpolation schemes used to simulate our binary populations, and use them to predict values for $\bar{\beta}_\mathrm{rot}$ for our entire astrophysical binary population of SN progenitors (undefined for single stars, unstable mass transfer and any mass transfer onto a compact object).

 In \fig{fig:beta} we show the normalized distribution of $\bar{\beta}_\mathrm{rot}$ for an astrophysical population with the same initial conditions as described Sec.~\ref{sec:pop_synth} (i.e.,  following the same initial mass function, initial period distribution, etc). We see a peak at very low $\bar{\beta}_\mathrm{rot}$ and a low tail to higher values. The average mass transfer for all stable mass transfer sequences (dashed) is found $\langle \bar{\beta}_{\rm rot} \rangle \sim 7\%$ and $9\%$ for $Z_\odot$ and $0.1Z_\odot$, respectively. This rotationally-limited efficiency is only very weakly dependent to the initial rotational distribution of the accretors (in our simulations all star are assumed initially synchronized with the orbit). If we only select the binary sequences that lead to eventual successful \sesne\ (our default \posydon\ model in this study; Sec.~\ref{sec:pop_synth}) the distribution skewers to the even lower efficiencies of $\langle \bar{\beta}_{\rm rot} \rangle \sim  4\%$ for both metallicities. High-mass donors in close orbits that typically experience more conservative mass transfer (due to stronger tidal spin-down caused by their larger radii \citealt{Sen+2022, Pauli+2022}), do not successfully explode. In addition, partial stripping in very wide orbits is found more conservative in some cases, but does not strip the donor fully \citep[e.g.,][]{Ercolino+2024} to contribute to the \sesn\ population. 
 %However, they do not successfully explode due to their larger masses.  high-mass donors, as well as slow case A systems of lower transfer rates, which tend to be more conservative predominantly due to tidal spin-down of the companion in tight orbits \citep{Sen+2022, Pauli+2022}, either do not strip the donors fully or they not eventually manage to explode to contribute to the \sesn\ population.  
 
 % The difference has to do with the excluding high mass donors that tend to implode and at the same time tend to higher relative higher mass transfer efficienceies. We also  case A phase that in the end do not stip the donor to produce a \sesn. These tends to be more conservative, both due to lower mass transfer rates as well as because they occur in tight orbital systems where tides contribute in spinning down the accreting companion which can thus accrete more. 

%%%%%%%%%%%%%%%%%%%%%%%%%%%%%%%%%%%%%%%%%%%%%%%%%%%%%%%%%%%%%%%%%%%%%%%%%
\section{Companion mass distribution}\label{sec:mass_distr}

\begin{figure}  
% \vspace*{-2.0 cm}
\begin{center}
 \includegraphics[width=0.99\linewidth]{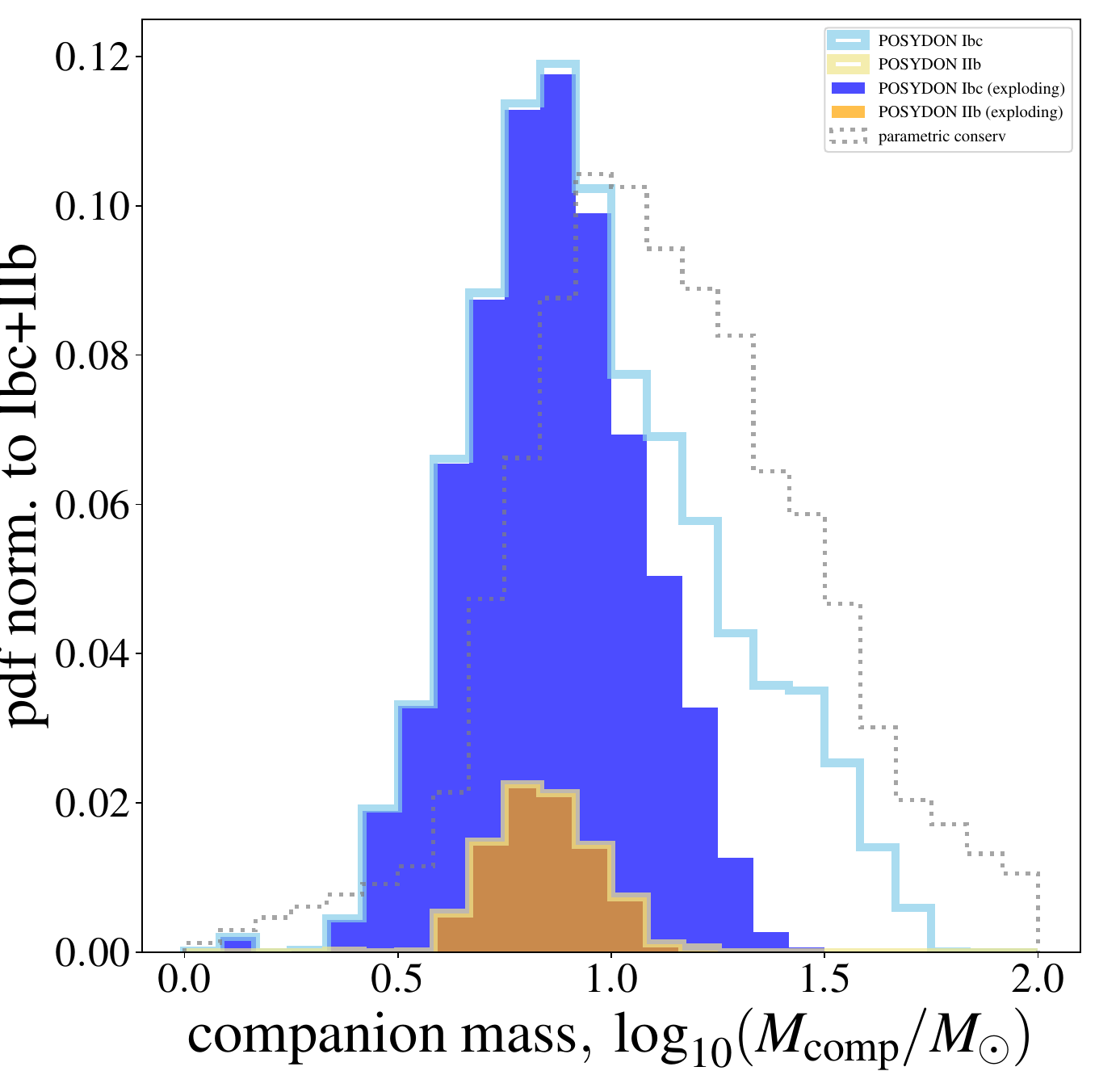}
 %{plots/cumulative_both3.pdf} 
% \vspace*{-1.0 cm}
 \caption{Mass distribution of stellar companions with \posydon. Step cyan (yellow) histograms correspond to all core-collapse progenitors that are candidates for Type \Ibc\ (\IIb), while the filled blue (orange) histograms show only those that successfully explode as \Ibc\ (\IIb). We also show the prediction from \Zt\ for %fully conservative
 thermally-limited mass transfer and assuming all massive stars explode (dotted). 
 \label{fig:mass_distr}
 }
 \end{center}
\end{figure}

We present the mass estimates here for completeness, although we refrain from emphasizing them given their strong dependence on stellar models when inferred from luminosity and HRD position (typically of a MS companion). 

Focusing on the most commonly expected MS companions, in \fig{fig:mass_distr} we show the predicted mass distribution for solar metallicity of stellar companions. We find a distribution  spanning from a few to several tens of \Msun, sharply peaked at $\sim7\Msun$, slightly lower than the $\sim9\Msun$ obtained with the default assumptions of \Zt, for the same reasons discussed in Sec.~\ref{sec:luminosity_cumulative}. 
From the high mass side, the distribution is constrained firstly by how conservative the mass transfer is assumed. For example, it extends much less to higher masses compared to the case of fully conservative assumption of a variation in \Zt\ (dotted line). The number of expected massive companions is also significantly influenced by which progenitors explode (cyan line for the case that everything explodes). %If we exclude the non-exploding ones according to \citet{Patton+2020} we exclude the high mass companions, because these tend to occur next to massive progenitors prone to implode. This is because for a fixed initial mass ratio range, a more massive primary has more massive companions on average, but also because the progenitor transfers a larger amount of mass towards their companion in case of stripping by stable mass transfer.
In the low-mass companion side, the distribution is populated by progenitors of low initial mass in binary systems of extreme mass ratio that trigger and survive CE. This is because low-mass companions next to stars massive enough to collapse are prone to unstable mass transfer and will not accrete significant mass). So this regime is limited by assumptions of unstable mass transfer and common envelope evolution. Stable mass transfer occurs in almost all cases, with CE playing a minor role of $\sim 3\%$ in stripping of progenitors from primary stars in binaries. %More lower mass companions are expected for higher $\alpha_{\rm CE}$ parameter (variation?).
%\MZ{Add in the figure the $\alpha_{CE}=5$ variation?}

% The detailed modeling of the binary mass transfer phase in \posydon\  shows that the parameter space for unstable mass transfer is less extended than 
% %when the $q_{\rm crit}$ values are assumed 
% what is assumed 
% in parametric population synthesis studies \citep{Gallegos-Garcia+2021}.  On top of that, in most occurrences of CE, when we take into account the binding energy profile of the detailed model of donor, we find that most systems would not be able to eject the envelope, leading to a merger and thus not have a binary companion next to them (and in most cases not even resulting in a \sesn). In principle, this regime would be affected by our lower mass ratio considered, but an even more extreme mass ratio would have an lower higher probability of surviving the CE.  

Recent observations have identified binary companions next to the first empirical detection of stripped stars, with typical companion masses in the range of 1–4~M$_\odot$ \citep{Drout+2023,Gotberg+2023}, lower than our resulting mass distribution (\fig{fig:mass_distr}). However, their findings are largely biased toward low-mass stripped stars that are not expected to explode as core-collapse \sesne. In addition, their observational selection favors systems with low-mass, faint companions, otherwise the stripped star would not be observable as the companion would dominate the system’s luminosity \citep{Gotberg+2018}.

\section{Non-detections for different assumptions in mass transfer efficiency}\label{sec:prob_nondetection_var}

 \begin{figure*}
% \vspace*{-2.0 cm}
\begin{center}
\includegraphics[width=0.45\linewidth]{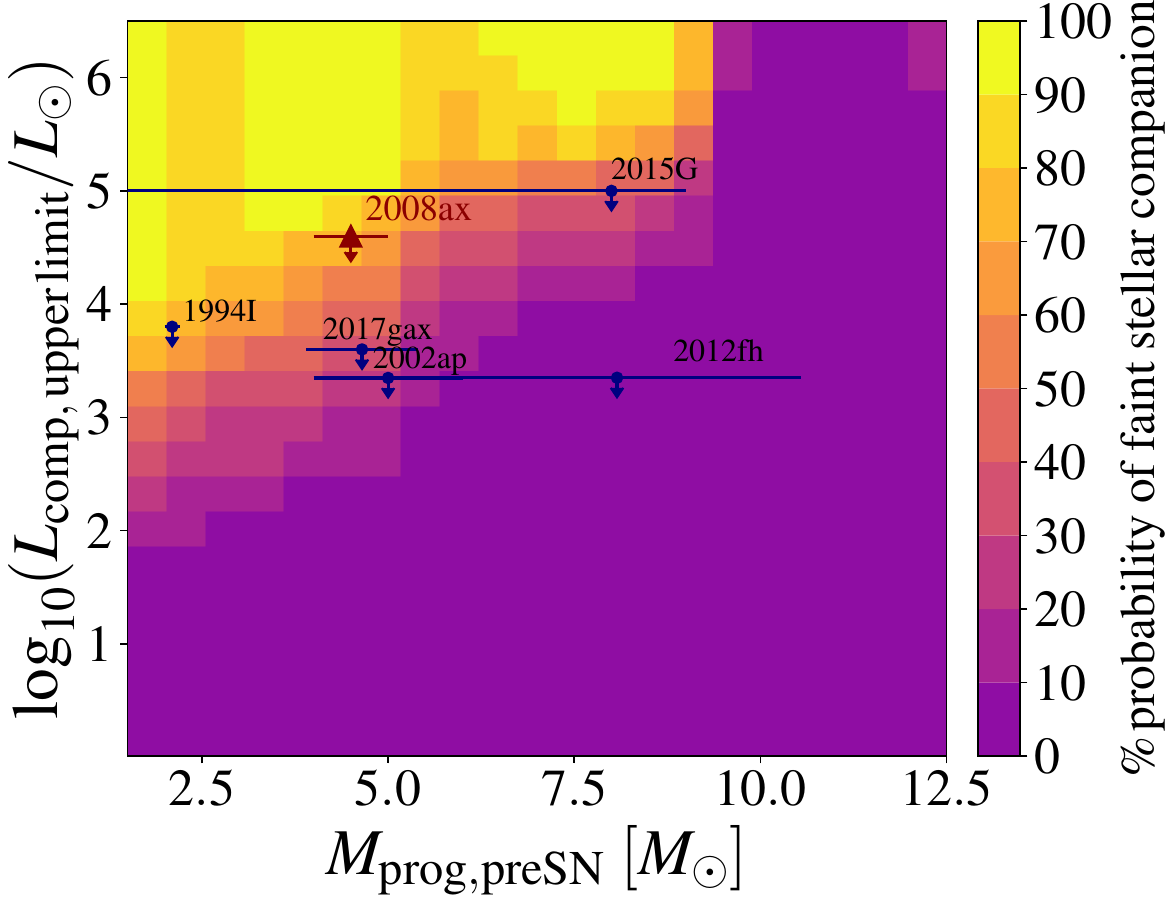}
\includegraphics[width=0.45\linewidth]{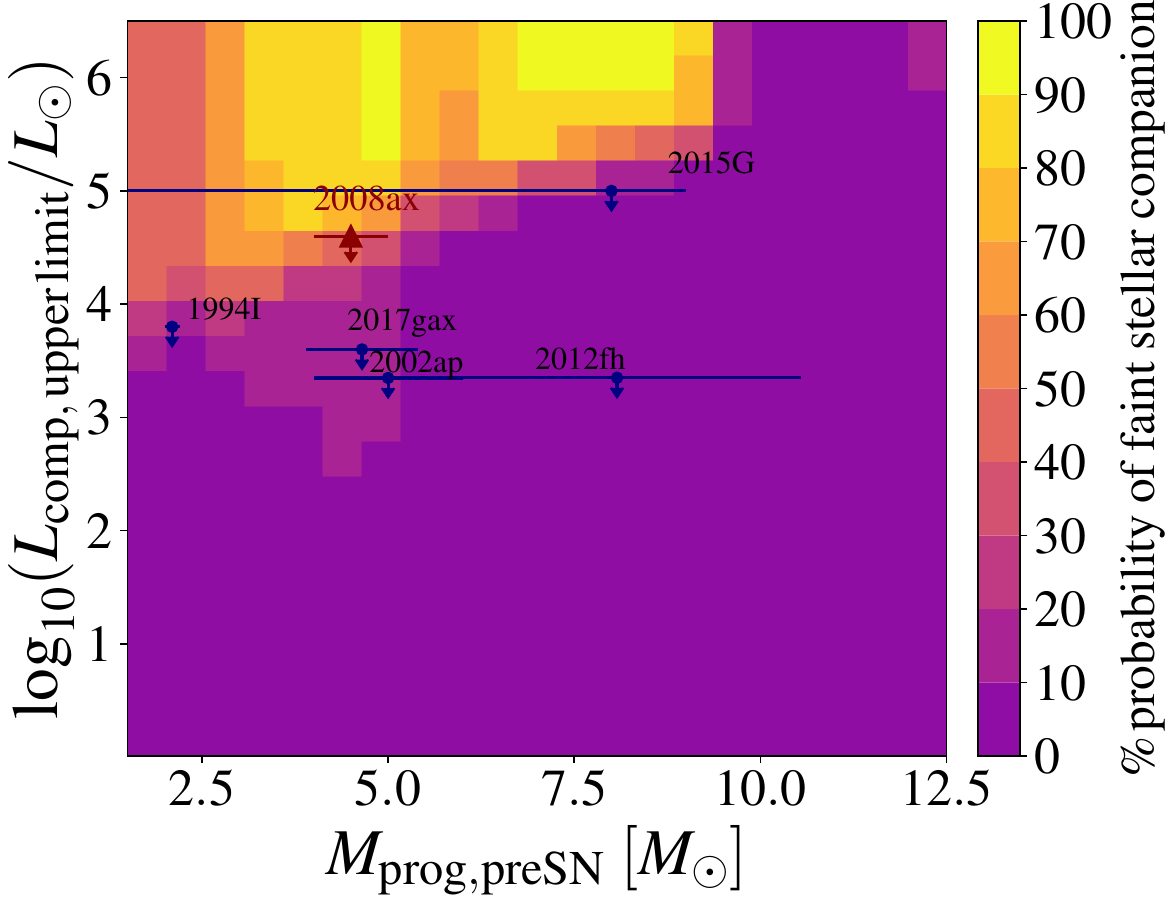}
 \caption{
Same as \fig{fig:prob_nondetection} but with models of \Zt\ for moderate, mass accretion efficiency, according to the thermal timescale of the mass gainer (left), and fully conservative mass transfer (right). 
\label{fig:prob_nondetection_var}
}
  \end{center}
\end{figure*}

In \fig{fig:prob_nondetection_var} we show the probability of a faint stellar companion in case of non-detections, for different mass accretion efficiencies, $\beta$. Higher efficiencies imply an even stronger possibility of a companion's absence for the same upper limit on each companion. Equivalently, highly conservative mass transfer would require no stellar companion in most non-detections, even in cases of low inferred progenitor masses (such as SN1994I), which seems highly improbable.

%%%%%%%%%%%%%%%%%%%%%%%%%%%%%%%%%%%%%%%%%%%%%%%%%%

% Don't change these lines
%\bsp	% typesetting comment
\label{lastpage}
\end{document}

%% file: summary_comp_detected2.tex
\begin{table*}
\centering
\caption{Summary of the observed companions sample, ordered by their luminosity constraint, with detections on top.} \label{tab:comp_sample}
\begin{tabular}{lcccccccc}
\hline
SN & SN Type &
\shortstack[c]{Epoch of image\\(yr after SN)} &
$\log_{10}(L_{\rm comp}/L_\odot)^*$ &
$\log_{10}(T_{\rm eff,comp}/\mathrm{K})^{*}$  & \shortstack[c]{Distance\\ (Mpc)}  & \shortstack[c]{Extinction, $A_{V}$ \\ (mag.)} &
\shortstack[c]{$M_{\rm prog}$\\ ($M_\odot$)} & 
\shortstack[c]{Metallicity \\ ($Z_\odot$)} \\
\hline
2019yvr & Ib         & $-1$ & $4.91\pm0.14^{(a_1)}$                        & $3.80^{+0.02}_{-0.03}{}^{(a_1)}$ & $14.4\pm1.3^{(a_2)}$ & 2.4$^{(a_3)}$ &$3.5\pm0.5^{(a_1,a_4)}$              & 0.40--0.91$^{(a_1)}$ \\
2013ge  & Ib/c       & 7 & $4.85\pm0.05^{(\#,b_1)}$                        & $4.20^{+0.03}_{-0.06}{}^{(\#,b_1)}$ & 23.7$^{(b_1,b_2)}$ & 0.2077$^{(b_2)}$ & $3.9\pm0.5^{(b_2)}$              & $0.5^{(b_1)}$ \\
1993J   & IIb        & 18.9   & $\sim$$4.71^{(\#,c_1)}$                        & $4.38^{+0.05}_{-0.10}{}^{(\#,c_1)}$ & 3.6$^{(\#,c_1)}$ & 0.52$^{(\#,c_1)}$ &  $4.0\pm0.5^{(c_2)}$              & $\sim1^{(c_1)}$ \\
2006jc  & Ibn        & 10.4  & $4.52\pm0.13^{(d_1)}$                        & $4.09^{+0.05}_{-0.04}{}^{(d_1)}$ & 27.8$^{(d_2)}$ & 0.155$^{(d_1)}$ & $4.4$--$6.9^{(d_3,d_4)}$    & $\sim0.5^{(d_1)}$ \\
2001ig  & IIb        & 14.4  & $3.92\pm0.14^{(e_1)}$                        & $4.31^{+0.03}_{-0.03}{}^{(e_1)}$ & $11.2\pm0.3^{(e_2)}$ & 0.06$^{(e_1)}$ & $3.5\pm1.0^{(e_1,e_3)}$ & $0.3$--$0.7^{(e_1)}$ \\
2011dh  & IIb       & 6.3 & $3.94 \pm0.13{}^{(f_1)}$              & $4.30^{+0.06}_{-0.06}{}^{(f_1)}$ & $7.78_{-0.92}^{+1.09}{}^{(f_2)}$ & 0.217$^{(f_3)}$ & $3.5\pm0.5^{(f_4)}$    & $0.83^{(f_5)}$ \\
\hline
2015G   & Ibn        & 1.8  & $<5.00^{(d_1)}$                              & $\sim 4.40^{(d_1)}$ & 23.2$^{(g_1)}$ & 1.19$^{(g_1)}$ &$\lesssim9^{(g_2)}$    & $\sim0.5^{(d_1)}$ \\
2008ax  & IIb        & 7    & $<4.60^{(\#,h_1)}$                              & $\sim 4.50 ^{(\#,h_1)}$ & 7.77$^{(\#,h_1)}$ & 1.30$^{(h_1)}$ &$4.5\pm0.5^{(h_1)}$              & $\sim 1^{(h_2)}$ \\
1994I   & Ic         & 20   & $<3.80^{(\#,i_1)}$                              & $\sim 4.38^{(\#,i_1)}$                  & 6.7$^{(i_1)}$ & 0.775$^{(i_1)}$ &$2.1\pm0.1^{(i_2)}$           & $0.83^{(f_5)}$ \\
2017gax & Ib/c       & 6    & $<3.60^{(j_1)}$                              & $\sim 4.35^{(j_1)}$                   & 15.9$^{(j_1)}$  & 0.0403$^{(j_1)}$  & $4.7\pm0.8^{(j_2)}$              & $\sim1^{(j_3)}$ \\
2002ap  & Ic-BL      & 23   & $<3.35^{(\#,k_1)}$                              & $\sim 4.30^{(\#,k_1)}$ & 10.19$^{(k_2)}$ & 0.279$^{(k_3)}$ & $5.0\pm1.0^{(k_3)}$              & $\sim$ 0.3-1$^{(k_1)}$ \\
2012fh  & Ib/c       & 14   & $<3.35^{(l_1)}$                              & $\sim4.30^{(l_1)}$ & 9.8$^{(l_2)}$ & 0.09$^{(l_3)}$ & $5.6$--$10.6^{(l_1)}$                      & 0.5--2$^{(l_4)}$ \\
\hline
\end{tabular}

\medskip
\begin{flushleft}
\textbf{Notes.}
$^{(a_1)}$\,\citet{Sun+2022};\,
$^{(a_2)}$\,\citet{Shappee+2016};\,
$^{(a_3)}$\,\citet{Kilpatrick+2021};\,
$^{(a_4)}$\,\citet{Ferrari2024};\,
$^{(b_1)}$\,\citet{Fox+2022};\,
$^{(b_2)}$\,\citet{Drout2016};\,
$^{(c_1)}$\,\citet{Maund+2004},\,\citet{Fox+2014};\,
$^{(c_2)}$\,\citet{Woosley+1994};\,
$^{(d_1)}$\,\citet{Sun+2020};\,
$^{(d_2)}$\,\citet{Maund+2016};\,
$^{(d_3)}$\,\citet{Tominaga2008};\,
$^{(d_4)}$\,\citet{Maeda2022};\,
$^{(e_1)}$\,\citet{Ryder+2006},\,\citet{Ryder+2018};\,
$^{(e_2)}$\,\citet{Böker+2002}, \citet{Tully1988}, \citet{Soria+2006};\,
$^{(e_3)}$\,\citet{Silverman2009};\,
$^{(e_4)}$\,\citet{Modjaz+2011};\,
$^{(f_1)}$\,\citet{Maund2019};\,
$^{(f_2)}$\,\citet{Ergon+2015};\,
$^{(f_3)}$\,\citet{Schlafly+2011};\,
$^{(f_4)}$\,\citet{Bersten+2012};\,
$^{(f_5)}$\,\citet{Kuncarayakti+2013};\,
$^{(g_1)}$\,\citet{Shivvers+2017};\,
$^{(g_2)}$\,\citet{Shivvers+2017b};\,
$^{(h_1)}$\,\citet{Folatelli+2015};\,
$^{(h_2)}$\,\citet{Crockett+2008};\,
$^{(i_1)}$\,\citet{van-Dyk+2016};\,
$^{(i_2)}$\,\citet{Nomoto+1994}, \citet{Iwamoto+1994}, \citet{Young+1995};\,
%$^{(j_1)}$\,\citet{Su+2025};\,
$^{(j_1)}$\, Su et al. (in prep.);\,
%$^{(j_2)}$\,\citet{Hoang+2025};\,
$^{(j_2)}$\, Hoang et al. (in prep.);\,
$^{(j_3)}$\,\citet{Schmitt+2006};\,
$^{(k_1)}$\,\citet{Zapartas+2017b};\,
$^{(k_2)}$\,\citet{Jang+2014};\,
$^{(k_3)}$\,\citet{Mazzali+2002}, \citet{Mazzali+2007};\,
%$^{(k_3)}$\,\citet{Mazzali+2007}, \citet{Maurer+2010};\,
$^{(l_1)}$\,\citet{Williams+2025};\,
$^{(l_2)}$\,\citet{Jacobs+2009}, \citet{Anand+2021};\,
$^{(l_3)}$\,\citet{Schlafly+2011};\,
$^{(l_4)}$\,\citet{Moumen+2019}\\
$^{(*)}$\, For non-detections (lower portion of the table), we list  the luminosity upper limit and temperature corresponding to a MS star (see Appendix~\ref{sec:obs_sample_app}). \\
$^{(\#)}$\,This work (see Section~\ref{sec:obs_sample} and Appendix~\ref{sec:obs_sample_app} for further details).
\end{flushleft}
\end{table*}

%% file: likelihoods_table3.tex
\begin{table*}
\centering
\caption{Log--likelihood $\ln\mathcal L$ of each model as a function of the assumed stripped--SN binary fraction $f_{\rm bin}$.\;
In parentheses we give the Bayes factor
$\mathcal B=\exp\!\bigl[\ln\mathcal L-\ln\mathcal L_{\rm ref}\bigr]$
relative to the reference model
(\posydon{} default at $f_{\rm bin}=1.0$, for which
$\ln\mathcal L_{\rm ref}=-11.360$ and $\mathcal B=1$ by definition, shown in bold).}
\label{tab:likelihoods}
\begin{tabular}{c
                r@{\;\,(}l
                r@{\;\,(}l
                r@{\;\,(}l
                r@{\;\,(}l
                r@{\;\,(}l}
\multicolumn{1}{c}{} &
\multicolumn{2}{c}{This work, POSYDON} &
\multicolumn{2}{c}{Z17 -- model} &
\multicolumn{2}{c}{Z17 -- default} &
\multicolumn{2}{c}{Z17 -- model} &
\multicolumn{2}{c}{Z17 -- model} \\
\multicolumn{1}{c}{} &
\multicolumn{2}{c}{rot.-limit accr.} &
\multicolumn{2}{c}{0\% accr.} &
\multicolumn{2}{c}{therm.-limit accr.} &
\multicolumn{2}{c}{30\% accr.} &
\multicolumn{2}{c}{100\% accr.} \\
 & 
\multicolumn{2}{c}{only succ. SESNe} &
\multicolumn{2}{c}{only succ. SESNe} &
\multicolumn{2}{c}{only succ. SESNe} &
\multicolumn{2}{c}{only succ. SESNe} &
\multicolumn{2}{c}{only succ. SESNe} \\
\hline\hline
$f_{\rm bin}$ &
\multicolumn{2}{c}{$\ln\mathcal L\;(\mathcal B)$} &
\multicolumn{2}{c}{$\ln\mathcal L\;(\mathcal B)$} &
\multicolumn{2}{c}{$\ln\mathcal L\;(\mathcal B)$} &
\multicolumn{2}{c}{$\ln\mathcal L\;(\mathcal B)$} &
\multicolumn{2}{c}{$\ln\mathcal L\;(\mathcal B)$} \\
\hline
1.00 & $\mathbf{-11.360}$ & $\mathbf{1.00}$) &
       $-11.458$ & 0.91) &
       $-12.040$ & 0.51) &
       $-12.183$ & 0.44) &
       $-16.151$ & $8\cdot10^{-3}$) \\[2pt]

0.95 & $-11.397$ & 0.96) &
       $-11.415$ & 0.95) &
       $-11.726$ & 0.69) &
       $-11.747$ & 0.68) &
       $-14.894$ & 0.029) \\

0.90 & $-11.468$ & 0.90) &
       $-11.418$ & 0.94) &
       $-11.510$ & 0.86) &
       $-11.450$ & 0.91) &
       $-14.075$ & 0.066) \\

0.85 & $-11.575$ & 0.81) &
       $-11.464$ & 0.90) &
       $-11.374$ & 0.99) &
       $-11.258$ & 1.11) &
       $-13.511$ & 0.12) \\

0.70 & $-12.108$ & 0.47) &
       $-11.852$ & 0.61) &
       $-11.361$ & 1.00) &
       $-11.142$ & 1.24) &
       $-12.711$ & 0.26) \\

0.50 & $-13.418$ & 0.13) &
       $-13.023$ & 0.19) &
       $-12.158$ & 0.45) &
       $-11.899$ & 0.58) &
       $-12.966$ & 0.20) \\

0.25 & $-16.840$ & $4\cdot10^{-3}$) &
       $-16.318$ & $7\cdot10^{-3}$) &
       $-15.138$ & 0.023) &
       $-14.864$ & 0.030) &
       $-15.553$ & 0.015) \\

0.10 & $-21.953$ & $3\cdot10^{-5}$) &
       $-21.371$ & $4\cdot10^{-5}$) &
       $-20.045$ & $2\cdot10^{-4}$) &
       $-19.772$ & $2.2\cdot10^{-4}$) &
       $-20.298$ & $1\cdot10^{-4}$) \\
\end{tabular}
\end{table*}